\documentclass[review]{elsarticle}

\usepackage{lineno}
\modulolinenumbers[5]
\usepackage{hyperref}
\usepackage{bm}
\usepackage{amsmath}
\usepackage{subfigure}
\usepackage{color}

\journal{Journal of Computational Physics}

\begin{document}

\begin{frontmatter}

\title{An Extension of Godunov SPH I\hspace{-.1em}I: Application to Elastic Dynamics}

%% include affiliations in footnotes:
\author[mymainaddress]{Keisuke Sugiura\corref{mycorrespondingauthor}}
\cortext[mycorrespondingauthor]{Corresponding author}
\ead{sugiura.keisuke@a.mbox.nagoya-u.ac.jp}

\author[mymainaddress]{Shu-ichiro Inutsuka}

\address[mymainaddress]{Department of Physics, Nagoya University, Aichi 464-8602, Japan}

\begin{abstract}
Godunov Smoothed Particle Hydrodynamics (Godunov SPH) method is a computational fluid dynamics method that utilizes a Riemann solver and achieves the second-order accuracy in space. In this paper, we extend the Godunov SPH method to elastic dynamics by incorporating deviatoric stress tensor that represents the stress for shear deformation or anisotropic compression. Analogously to the formulation of the original Godunov SPH method, we formulate the equation of motion, the equation of energy, and the time evolution equation of deviatoric stress tensor so that the resulting discretized system achieves the second-order accuracy in space. The standard SPH method tends to suffer from the tensile instability that results in unphysical clustering of particles especially in tension-dominated region. We find that the tensile instability can be suppressed by selecting appropriate interpolation for density distribution in the equation of motion for the Godunov SPH method even in the case of elastic dynamics. Several test calculations for elastic dynamics are performed, and the accuracy and versatility of the present method are shown.
\end{abstract}

\begin{keyword}
Smoothed Particle Hydrodynamics \sep Elastic dynamics \sep Tensile instability \sep Linear stability analysis \sep Godunov's method
\end{keyword}

\end{frontmatter}

\linenumbers

\section{Introduction}
Smoothed Particle Hydrodynamics (SPH) is one of the computational fluid dynamics methods using particles that mimic fluid elements (e.g.\,\cite{Lucy1977},\cite{Gingold-and-Monaghan1977},\cite{Monaghan1992}). Recently the standard SPH method, i.e., the most popular form of SPH method, is developed to elastic dynamics and applied to calculations of planetesimal collisions (e.g.\,\cite{Benz-and-Asphaug1995},\cite{Benz-and-Asphaug1999}). The SPH method does not require a Eulerian mesh. Thus it is favourable for simulations with large deformation, and we can easily track information accompanying to particles such as clack history. Therefore, the SPH method is suited for calculations of disruptive collisions.

However, the standard SPH method for elastic dynamics has a serious problem that results in unphysical clustering of particles especially in tension-dominated region. This problem is called the tensile instability\cite{Swegle-et-al1995}. The property of the tensile instability for the case of the Nyquist wavelength is analyzed in \cite{Morris1996} for hydrodynamics, and in \cite{Iwasaki2015} for magnetohydrodynamics. The tensile instability occurs also in positive pressure region that represents compressed material or usual fluid. According to \cite{Dehnen-and-Aly2012}, B-spline kernels produce the tensile instability even in the positive pressure regime if the number of neighbor particles is too large. The simple test calculation of oscillating plate in \cite{Gray-et-al2001} demonstrates that the standard SPH method suffers from unphysical fracture caused by the tensile instability. Thus the mitigation of the tensile instability is required when we use the SPH method for elastic dynamics.

Some researches have tried to prevent the tensile instability (e.g.\,\cite{Randles-and-Libersky1996},\cite{Johnson-and-Beissel1996}). For example, in \cite{Monaghan1999} and \cite{Gray-et-al2001}, Monaghan and Gray et al. introduce artificial stress term that provides a strong repulsive force only when particles become too close to each other, and try to prevent the tensile instability. They conducted the linear stability analysis and found that this method suppresses the instability at short wavelengths and does not strongly affect the perturbations of long wavelengths. However, this method includes the artificial stress term that does not exist in the original equations. Moreover, according to \cite{Mehra-et-al2012}, this method does not seem to suppress the tensile instability in simulations of hypervelocity impacts. Sugiura and Inutsuka \cite{Sugiura-and-Inutsuka2016} mitigate the tensile instability using the Godunov SPH method \cite{Inutsuka2002} that utilizes a Riemann solver and achieves the second-order accuracy in space. They conduct the linear stability analysis for the equations of the Godunov SPH method, and find that the tensile instability can be suppressed by selecting appropriate interpolation for $V_{ij}^{2}$ (i.e., weighted average of $\rho^{-2}$) depending on the sign of pressure. However, they conduct the linear stability analysis only for the equations of hydrodynamics and it is not obvious that their approach works for those of elastic dynamics that uses the deviatoric stress tensor.

The accuracy of the standard SPH method is below the first-order in the case of disordered particle distribution. This means very slow convergence for the increase of spatial resolution. For example Genda et al. \cite{Genda-et-al2015} conducted simulations of planetesimal collision using the standard SPH method, and evaluated critical kinetic energy $Q_{D}^{\ast}$, which is required to disrupt planetesimals while increasing the number of particles. As a result, they found that at least five million particles are required to obtain converged $Q_{D}^{\ast}$, and convergence is the first order with respect to mean particle spacing. They claim that this first-order convergence is due to the effect of shock waves because the spatial accuracy of physical quantities becomes the first order at shock surface. The Godunov SPH method can resolve shock surface with much small number of particles thanks to the utilization of the Riemann solver, and thus much fast convergence is expected.

In this study, we extend the Godunov SPH method, which can achieve the second-order accuracy in space, to elastic dynamics. The equation of motion and the equation of energy for elastic dynamics include deviatoric stress tensor. We formulate the equation of motion, the equation of energy, and the evolution equation of deviatoric stress tensor itself so that formulated equations can achieve the second-order accuracy in space. Moreover, we develop a method to treat the Riemann solver for general equation of state (hereafter, EoS) for elastic dynamics, and enable calculations of elastic dynamics using the Godunov SPH method. We perform several test calculations of elastic dynamics, and show that even in elastic dynamics the tensile instability can be suppressed just by selecting appropriate interpolation for $V_{ij}^{2}$ depending on the sign of pressure.

The structure of this paper is as follows: in Section 2 we extend the Godunov SPH method to elastic dynamics. The detailed method for the implementation is described in Section 3, which includes the treatment of the Riemann solver for non-ideal gas EoS or the method to mitigate the tensile instability. In Section 4 we perform several test calculations of elastic dynamics. Section 5 is for summary. 

\section{Godunov SPH method for elastic dynamics}
In this section, we introduce fundamental equations for elastic dynamics and formulate the Godunov SPH method for these equations  to achieve the second-order accuracy in space.

\subsection{Fundamental equations for elastic dynamics}
Fundamental equations for elastic dynamics can be found e.g., in \cite{Benz-and-Asphaug1995}. The equation of continuity is,

\begin{equation}
\frac{d\rho}{dt}=-\rho \frac{\partial}{\partial x^{\alpha}}v^{\alpha},
\label{EoC}
\end{equation}

\noindent where $d/dt$ means Lagrangian time derivative, $\rho$ is the density, $v^{\alpha}$ is the $\alpha$-th component of the velocity $\bm{v}$, and $x^{\alpha}$ is the $\alpha$-th component of the position $\bm{r}$. We also assume the summation rule over repeated indices of Greek letter. Hereafter, a superscript of Greek letter means component of vector or tensor, a subscript of Roman letter means particle number. 

The equation of motion is,

\begin{equation}
\frac{d v^{\alpha}}{dt}=\frac{1}{\rho}\frac{\partial}{\partial x^{\beta}}\sigma ^{\alpha \beta},
\label{EoM}
\end{equation}

\noindent where $\sigma^{\alpha \beta}$ is the stress tensor. The stress tensor can be decomposed to pressure $P$ that represents the diagonal part and deviatoric stress tensor $S^{\alpha \beta}$ that corresponds to the non-diagonal part,

\begin{equation}
\sigma^{\alpha \beta}=-P\delta^{\alpha \beta}+S^{\alpha \beta},
\label{stress-tensor}
\end{equation}

\noindent where $\delta^{\alpha \beta}$ is Kronecker delta. $P$ can be expressed by appropriate EoS for the solid.

The equation of energy is,

\begin{equation}
\frac{du}{dt}=\frac{1}{\rho}\sigma^{\alpha \beta} \dot{\epsilon}^{\alpha \beta},
\label{EoE}
\end{equation}

\noindent where $u$ is the specific internal energy, $\dot{\epsilon}^{\alpha \beta}$ is the strain rate tensor, 

\begin{equation}
\dot{\epsilon}^{\alpha \beta}=\frac{1}{2}\Bigl( \frac{\partial}{\partial x^{\beta}}v^{\alpha}+\frac{\partial}{\partial x^{\alpha}}v^{\beta} \Bigr).
\label{strain-rate-tensor}
\end{equation}

\noindent $\sigma^{\alpha \beta}$ is a symmetric tensor. Thus Eq.\,(\ref{EoE}) can be expressed by simpler form as,

\begin{equation}
\frac{du}{dt}=\frac{1}{\rho}\sigma^{\alpha \beta} \frac{\partial}{\partial x^{\beta}}v^{\alpha}.
\label{EoE-2}
\end{equation}

In addition to these equations, a equation that determines the deviatoric stress tensor $S^{\alpha \beta}$ is necessary. We use the time evolution equation of the deviatoric stress tensor that assumes Hook's law,

\begin{equation}
\frac{dS^{\alpha \beta}}{dt}=2\mu \Bigl(\dot{\epsilon}^{\alpha \beta}-\frac{1}{3}\delta^{\alpha \beta}\dot{\epsilon}^{\gamma \gamma} \Bigr) + S^{\alpha \gamma}R^{\beta \gamma} + S^{\beta \gamma}R^{\alpha \gamma},
\label{evol-of-Sab}
\end{equation}

\noindent where $\mu$ is the shear modulus, $R^{\alpha \beta}$ is the rotational rate tensor,

\begin{equation}
R^{\alpha \beta}=\frac{1}{2}\Bigl( \frac{\partial}{\partial x^{\beta}}v^{\alpha} - \frac{\partial}{\partial x^{\alpha}}v^{\beta} \Bigr).
\label{rotation-rate-tensor}
\end{equation}

If we use the EoS $P=P(\rho ,u)$, we can describe the motion of elastic body.

\subsection{Equations for Godunov SPH method}

In the SPH method, we define the density at arbitrary position $\bm{r}$ as,

\begin{equation}
\rho(\bm{r})=\sum_{j}m_{j}W(\bm{r}-\bm{r}_{j},h),
\label{SPH-density}
\end{equation}

\noindent where $W(\bm{r},h)$ is a kernel function and $h$ is a parameter called the smoothing length. In Section 2, we treat this smoothing length as constant in space. The kernel function has various forms. Throughout this paper, we use Gaussian kernel,

\begin{equation}
W(\bm{r},h)=\Bigl[ \frac{1}{h\sqrt{\pi}} \Bigr]^{d} e^{-\bm{r}^{2}/h^{2}},
\label{kernel-function}
\end{equation}

\noindent where $d$ represents the number of dimensions.

The equation of motion and the equation of energy for the Godunov SPH method are defined by the convolution of Eq.\,(\ref{EoM}) and Eq.\,(\ref{EoE-2}) respectively. The acceleration of the $i$-th particle is expressed as,

\begin{equation}
\dot{v}_{i}^{\alpha} \equiv \int \frac{dv^{\alpha}(\bm{r})}{dt}W(\bm{r}-\bm{r}_{i},h)d\bm{r}=\int \frac{1}{\rho (\bm{r})}\frac{\partial}{\partial x^{\beta}}\sigma^{\alpha \beta}(\bm{r})W(\bm{r}-\bm{r}_{i},h)d\bm{r}, 
\label{Godunov-EoM-1} 
\end{equation}

\noindent where the overdot represents time derivative. Similarly, time derivative of the internal energy of the $i$-th particle is,

\begin{equation}
\dot{u}_{i} \equiv \int \frac{du(\bm{r})}{dt}W(\bm{r}-\bm{r}_{i},h)d\bm{r}=\int \frac{1}{\rho (\bm{r})}\sigma^{\alpha \beta}(\bm{r})\frac{\partial}{\partial x^{\beta}}v^{\alpha}(\bm{r})W(\bm{r}-\bm{r}_{i},h)d\bm{r}.
\label{Godunov-EoE-1}
\end{equation}

We can formulate the equation of motion (\ref{Godunov-EoM-1}) in almost the same way as for hydrodynamics in \cite{Inutsuka2002}. What we should do is just replacing $-P(\bm{r})$ in \cite{Inutsuka2002} with $\sigma^{\alpha \beta}(\bm{r})$. Finally the equation of motion for the Godunov SPH method for elastic dynamics becomes,

\begin{align}
& \dot{v}_{i}^{\alpha}=\sum_{j}2m_{j}\sigma^{\alpha \beta \ast}_{ij}V_{ij}^{2}(h)\frac{\partial}{\partial x_{i}^{\beta}}W(\bm{r}_{i}-\bm{r}_{j},\sqrt{2}h), \label{Godunov-EoM-2} \\ & V_{ij}^{2}(h)=\int \Bigl[ \frac{\sqrt{2}}{h\sqrt{\pi}} \Bigr]^{d} \frac{1}{\rho^{2}(\bm{r})} \exp \Bigl[ -\frac{2(\bm{r}-(\bm{r}_{i}+\bm{r}_{j})/2)^{2}}{h^{2}}\Bigr]d\bm{r}, \label{Vij2}
\end{align}

\noindent where,

\begin{equation}
\sigma^{\alpha \beta \ast}_{ij}=-P_{ij}^{\ast}\delta^{\alpha \beta}+S_{ij}^{\alpha \beta \ast},
\label{sigma-ast}
\end{equation}

\noindent $P_{ij}^{\ast}$ is resultant pressure of the Riemann problem that uses the physical quantities of the $i$-th and $j$-th particles as initial condition, and

\begin{align}
& S_{ij}^{\alpha \beta \ast}=\frac{S_{i}^{\alpha \beta}+S_{j}^{\alpha \beta}}{2}+s_{ij}^{\ast}\frac{S_{i}^{\alpha \beta}-S_{j}^{\alpha \beta}}{|\bm{r}_{i}-\bm{r}_{j}|}, \label{Sabij-ast} \\ & s_{ij}^{\ast}=\int \Bigl[ \frac{\sqrt{2}}{h\sqrt{\pi}} \Bigr]^{d} \frac{\bm{r}-(\bm{r}_{i}+\bm{r}_{j})/2}{\rho^{2}(\bm{r})}\cdot \frac{\bm{r}_{i}-\bm{r}_{j}}{|\bm{r}_{i}-\bm{r}_{j}|} \exp \Bigl[ -\frac{2(\bm{r}-(\bm{r}_{i}+\bm{r}_{j})/2)^{2}}{h^{2}}\Bigr]d\bm{r}, \label{sijast}
\end{align}

\noindent If we define the $s$-axis, which is along $\bm{r}_{i}-\bm{r}_{j}$ and has its origin at $(\bm{r}_{i}+\bm{r}_{j})/2$, and expand $\rho^{-2}(\bm{r})$ linearly in the direction perpendicular to the $s$-axis, $V_{ij}^{2}(h)$ and $s_{ij}^{\ast}$ become simpler form,

\begin{align}
& V_{ij}^{2}(h)=\int_{-\infty}^{\infty}\frac{\sqrt{2}}{h\sqrt{\pi}}\frac{1}{\rho^{2}(s)}\exp\Bigl( -\frac{2s^{2}}{h^{2}} \Bigr)ds, \label{Vij2-simpler} \\ & s_{ij}^{\ast}=\int_{-\infty}^{\infty}\frac{\sqrt{2}}{h\sqrt{\pi}}\frac{s}{\rho^{2}(s)}\exp\Bigl(-\frac{2s^{2}}{h^{2}} \Bigr)ds. \label{sijast-simpler}
\end{align}

\noindent Equation (\ref{Vij2-simpler}) is also written in \cite{Sugiura-and-Inutsuka2016}. To calculate $V_{ij}^{2}(h)$ and $s_{ij}^{\ast}$, we need to interpolate $1/\rho (s)$ along $s$-axis. In this paper we use linear interpolation and cubic spline interpolation. The formula of $V_{ij}^{2}(h)$ and $s_{ij}^{\ast}$ in the case of linear interpolation and cubic spline interpolation are written in \cite{Inutsuka2002}. Note that $V_{ij}^{2}(h)$ is also a function of smoothing length.

If we use cubic spline interpolation when the particles become much closer to each other than the smoothing length, $V_{ij}^{2}$ diverges due to the interpolation. $V_{ij}^{2}$ is originally weighted average of $1/\rho^{2}(\bm{r})$. Thus its value should be about $1/\rho^{2}(\bm{r})$. Therefore, if $V_{ij}^{2}$ calculated by cubic spline interpolation is much larger than $1/\rho^{2}(\bm{r})$, we should use linear interpolation. In this study, we use linear interpolation when $V_{ij}^{2}$ becomes larger than $V_{ij,{\rm crit}}^{2}$,

\begin{equation}
V_{ij,{\rm crit}}^{2}=10\Bigl( \frac{1}{\rho_{ij}^{2}} \Bigr),
\label{Vij2-crit}
\end{equation}

\noindent where $\rho_{ij}=(\rho_{i}+\rho_{j})/2$.

As we use the result of Riemann problem for $P_{ij}^{\ast}$, we can use the result of the Riemann problem in elastic dynamics for $S_{ij}^{\alpha \beta \ast}$. However, in the Godunov method we utilize the Riemann solver to describe the shock wave accurately, and for this purpose it is enough to use the result of Riemann problem for pressure. Thus we use simple weighted average of deviatoric stress tensor expressed in Eq.\,(\ref{Sabij-ast}) for $S_{ij}^{\alpha \beta \ast}$. 

We can also transform the equation of energy in almost the same way as in \cite{Inutsuka2002}. Finally the equation of energy becomes,

\begin{equation}
\dot{u}_{i}=\sum_{j}2m_{j}\sigma_{ij}^{\alpha \beta \ast}V_{ij}^{2}(h)(v_{ij}^{\alpha \ast}-v_{i}^{\alpha \ast})\frac{\partial}{\partial x_{i}^{\beta}}W(\bm{r}_{i}-\bm{r}_{j},\sqrt{2}h),
\label{Godunov-EoE-2}
\end{equation}

\noindent where we use time centered velocity for $v_{i}^{\alpha \ast}$ to achieve the conservation of total energy. 

\begin{equation}
v_{i}^{\alpha \ast}=v_{i}^{\alpha}+\frac{1}{2}\Delta t \dot{v}_{i}^{\alpha},
\label{time-centered-vel}
\end{equation}

\noindent where $\Delta t$ is the time step. The reason why the total energy is conserved is written in \cite{Inutsuka2002} in the case of hydrodynamics. For the same reason, the total energy can be conserved exactly in our formulation. In \cite{Inutsuka2002}, Inutsuka uses the result of Riemann problem for $v_{ij}^{\alpha \ast}$, but this treatment can cause a problem if the EoS is not for ideal gas. In the case of positive pressure, resultant velocity of the Riemann problem causes effective energy transfer from high-pressure particle to low-pressure particle. For example, in the case of collision between aluminum sphere and plate (test calculation in Section 4.4), collisional surface becomes contact discontinuity. The pressure should be constant across contact discontinuity, but SPH calculation makes ``pressure wiggle'' at contact discontinuity due to discretization error. If the EoS is for ideal gas, energy transfer stops when the pressure becomes constant even when pressure wiggle exists. However, stiffened gas EoS (e.g.\,\cite{Mehra-and-Chaturvedi2006}) or Tillotson EoS (e.g.\,\cite{Tillotson1962}) has terms that represent the elastic body such as $P=C_{s}^{2}(\rho - \rho_{0})$. Thus if the ``density wiggle'' exists the pressure wiggle also exists irrespective of the internal energy, and energy can be transferred from high-pressure particle continuously. Eventually the internal energy of high-pressure particle becomes largely negative even though this particle is located in a compressed region. To prevent this problem, in this study we use simple average value for $v_{ij}^{\alpha \ast}$ expressed as,

\begin{equation}
v_{ij}^{\alpha \ast}=\frac{v_{i}^{\alpha}+v_{j}^{\alpha}}{2}+s_{ij}^{\ast}\frac{v_{i}^{\alpha}-v_{j}^{\alpha}}{|\bm{r}_{i}-\bm{r}_{j}|},
\label{vij-ast}
\end{equation}

\noindent and the result of Riemann problem is used only for pressure. 

Finally, we formulate the time evolution equation of deviatoric stress tensor. Following the formulation of the induction equation in \cite{Iwasaki-and-Inutsuka2011}, we formulate the time derivative of $S^{\alpha \beta}/\rho$. We simply differentiate $S^{\alpha \beta}/\rho$ and obtain,

\begin{equation}
\frac{d}{dt}\Bigl( \frac{S^{\alpha \beta}}{\rho} \Bigr) = \frac{1}{\rho}\frac{dS^{\alpha \beta}}{dt} - \frac{S^{\alpha \beta}}{\rho^{2}}\frac{d\rho}{dt}.
\label{dSab_rho_dt-1}
\end{equation}

\noindent Substituting Eqs.\,(\ref{EoC}) and (\ref{evol-of-Sab}) into Eq.\,(\ref{dSab_rho_dt-1}), we can obtain,

\begin{equation}
\frac{d}{dt}\Bigl( \frac{S^{\alpha \beta}}{\rho} \Bigr) = 2\mu \Bigl( \frac{\dot{\epsilon}^{\alpha \beta}}{\rho} - \frac{1}{3}\delta^{\alpha \beta}\frac{\dot{\epsilon}^{\gamma \gamma}}{\rho} \Bigr) + \frac{S^{\alpha \gamma}}{\rho}R^{\beta \gamma} + \frac{S^{\beta \gamma}}{\rho}R^{\alpha \gamma} + \frac{S^{\alpha \beta}}{\rho}\dot{\epsilon}^{\gamma \gamma}.
\label{dSab_rho_dt-2}
\end{equation}

\noindent Note that $(\partial / \partial x^{\alpha})v^{\alpha}=\dot{\epsilon}^{\gamma \gamma}$. As with the equation of motion or the equation of energy, we define the time derivative of $S^{\alpha \beta}/\rho$ of the $i$-th particle as the convolution of Eq.\,(\ref{dSab_rho_dt-2}).

\begin{align}
\frac{d}{dt}\Bigl( \frac{S^{\alpha \beta}}{\rho} \Bigr)_{i} & \equiv \int \frac{d}{dt}\Bigl( \frac{S^{\alpha \beta}}{\rho} \Bigr)(\bm{r})W(\bm{r}-\bm{r}_{i},h)d\bm{r} \nonumber \\ & = \int \Bigl[ 2\mu \Bigl( \frac{\dot{\epsilon}^{\alpha \beta}}{\rho} - \frac{1}{3}\delta^{\alpha \beta}\frac{\dot{\epsilon}^{\gamma \gamma}}{\rho} \Bigr) + \frac{S^{\alpha \gamma}}{\rho}R^{\beta \gamma} + \frac{S^{\beta \gamma}}{\rho}R^{\alpha \gamma} + \frac{S^{\alpha \beta}}{\rho}\dot{\epsilon}^{\gamma \gamma} \Bigr]d\bm{r}. \label{Godunov-dSab_rho_dt-1}
\end{align}

This equation includes the following terms (Note that $\dot{\epsilon}^{\alpha \beta}$ and $R^{\alpha \beta}$ are the sums of velocity gradient):

\begin{align}
& \int \frac{1}{\rho (\bm{r})}\frac{\partial v^{\alpha}(\bm{r})}{\partial x^{\beta}}W(\bm{r}-\bm{r}_{i},h)d\bm{r}, \label{dSab_rho_dt-integrations-1} \\ & \int \frac{S^{\alpha^{'} \beta^{'}}(\bm{r})}{\rho(\bm{r})}\frac{\partial v^{\alpha}(\bm{r})}{\partial x^{\beta}}W(\bm{r}-\bm{r}_{i},h)d\bm{r}, \label{dSab_rho_dt-integrations-2}
\end{align}

\noindent where $\alpha ,\beta ,\alpha^{'} ,\beta^{'}$ change depending on the subscript of each term of Eq.\,(\ref{Godunov-dSab_rho_dt-1}). Regarding Eq.\,(\ref{dSab_rho_dt-integrations-2}), we can transform it in almost the same way as in \cite{Inutsuka2002} and obtain,

\begin{equation}
2\sum_{j}m_{j}V_{ij}^{2}(h)S_{ij}^{\alpha^{'} \beta^{'} \ast}(v_{ij}^{\alpha \ast}-v_{i}^{\alpha})\frac{\partial}{\partial x_{i}^{\beta}}W(\bm{r}_{i}-\bm{r}_{j},\sqrt{2}h).
\label{dSab_rho_dt-integrations-2-1}
\end{equation}

\noindent Using Eq.\,(\ref{SPH-density}), the identity $\sum_{j}\frac{m_{j}}{\rho (\bm{r})}W(\bm{r}-\bm{r}_{j},h)=1$, $\frac{\partial v_{i}^{\alpha}}{\partial x^{\beta}}=0$ and the partial integration, we can transform Eq.\,(\ref{dSab_rho_dt-integrations-1}) to,

\begin{align}
& \int \frac{1}{\rho (\bm{r})}\frac{\partial v^{\alpha}(\bm{r})}{\partial x^{\beta}}W(\bm{r}-\bm{r}_{i},h)d\bm{r} \nonumber \\ & = \int \frac{1}{\rho (\bm{r})}\frac{\partial}{\partial x^{\beta}}(v^{\alpha}(\bm{r})-v_{i}^{\alpha})W(\bm{r}-\bm{r}_{i},h)d\bm{r} \nonumber \\ & = - \int (v^{\alpha}(\bm{r})-v_{i}^{\alpha})\frac{\partial}{\partial x^{\beta}}\Bigl(\frac{W(\bm{r}-\bm{r}_{i},h)}{\rho (\bm{r})}\Bigr)d\bm{r} \nonumber \\ & = \sum_{j}m_{j}\int \frac{1}{\rho ^{2}(\bm{r})}(v^{\alpha}(\bm{r})-v_{i}^{\alpha})\Bigl[ \frac{\partial}{\partial x_{i}^{\beta}} - \frac{\partial}{\partial x_{j}^{\beta}} \Bigr] W(\bm{r}-\bm{r}_{i},h)W(\bm{r}-\bm{r}_{j},h)d\bm{r}. \label{dSab_rho_dt-integrations-1-2}
\end{align}

\noindent Finally, we calculate the integral using interpolation as in \cite{Inutsuka2002}, and Eq.\,(\ref{dSab_rho_dt-integrations-1-2}) becomes,

\begin{equation}
2\sum_{j}m_{j}V_{ij}^{2}(h)(v_{ij}^{\alpha \ast}-v_{i}^{\alpha})\frac{\partial}{\partial x_{i}^{\beta}}W(\bm{r}_{i}-\bm{r}_{j},\sqrt{2}h).
\label{dSab_rho_dt-integrations-1-3}
\end{equation}

We can transform Eq.\,(\ref{Godunov-dSab_rho_dt-1}) using Eqs.\,(\ref{dSab_rho_dt-integrations-2-1}) and (\ref{dSab_rho_dt-integrations-1-3}), the time derivative of $S^{\alpha \beta}/\rho$ of the $i$-th particle becomes,

\begin{equation}
\frac{d}{dt}\Bigl( \frac{S^{\alpha \beta}}{\rho} \Bigr)_{i} = \sum_{j}2\mu \Bigl( \dot{\epsilon}_{\rho ,ij}^{\alpha \beta} - \frac{1}{3}\delta^{\alpha \beta}\dot{\epsilon}_{\rho ,ij}^{\gamma \gamma} \Bigr) + S_{ij}^{\alpha \gamma \ast}R_{\rho ,ij}^{\beta \gamma} + S_{ij}^{\beta \gamma \ast}R_{\rho ,ij}^{\alpha \gamma} + S_{ij}^{\alpha \beta \ast}\dot{\epsilon}_{\rho ,ij}^{\gamma \gamma},
\label{Godunov-dSab_rho_dt-2}
\end{equation}

\noindent where,

\begin{align}
&\dot{\epsilon}_{\rho ,ij}^{\alpha \beta} \equiv m_{j}V_{ij}^{2}(h) \Bigl[ (v_{ij}^{\alpha \ast}-v_{i}^{\alpha}) \frac{\partial}{\partial x_{i}^{\beta}}+(v_{ij}^{\beta \ast}-v_{i}^{\beta})\frac{\partial}{\partial x_{i}^{\alpha}} \Bigr] W(\bm{r}_{i}-\bm{r}_{j},\sqrt{2}h), \label{dot-eps-rho-ij-ab} \\ & R_{\rho ,ij}^{\alpha \beta} \equiv m_{j}V_{ij}^{2}(h) \Bigl[ (v_{ij}^{\alpha \ast}-v_{i}^{\alpha}) \frac{\partial}{\partial x_{i}^{\beta}}-(v_{ij}^{\beta \ast}-v_{i}^{\beta})\frac{\partial}{\partial x_{i}^{\alpha}} \Bigr] W(\bm{r}_{i}-\bm{r}_{j},\sqrt{2}h). \label{R-rho-ij-ab}
\end{align}

In actual calculation, we follow the time evolution of $S^{\alpha \beta}/\rho$ using Eq.\,(\ref{Godunov-dSab_rho_dt-2}), and then we can obtain $S_{i}^{\alpha \beta}$ at each time step using,

\begin{equation}
S^{\alpha \beta}_{i}=\Bigl( \frac{S^{\alpha \beta}}{\rho} \Bigr)_{i} \rho_{i}.
\label{way-to-express-Sab}
\end{equation}

Our formulation of the equation of motion, the equation of energy and the time evolution equation of deviatoric stress tensor essentially follows \cite{Inutsuka2002}. Therefore, these equations are expected to achieve the second-order accuracy. We confirm this fact in the convergence test in Section 4.1.

The density can be calculated by Eq.\,(\ref{SPH-density}). However, it is known that this equation causes a problem in a surface of solid body. Density calculated by Eq.\,(\ref{SPH-density}) becomes small nearby the free surface, and pressure also becomes small via EoS. Thus the solid body tend to be deformed by unphysical gradient of pressure nearby a free surface \cite{Monaghan1988}. We can prevent this problem by calculating the time evolution of the density using the equation of continuity. In this study, we use simple Lagrangian derivative of Eq.\,(\ref{SPH-density}) as the equation of continuity,

\begin{equation}
\dot{\rho}_{i}=\sum_{j}m_{j}(v_{i}^{\alpha}-v_{j}^{\alpha})\frac{\partial}{\partial x_{i}^{\alpha}}W(\bm{r}_{i}-\bm{r}_{j},h).
\label{Godunov-EoC}
\end{equation}

\noindent Eq.\,(\ref{Godunov-EoC}) is used, e.g., in \cite{Benz-and-Asphaug1995}.

Linear momentum is conserved exactly in our method because the equation of motion (\ref{Godunov-EoM-2}) is written in the anti-symmetric form. However, as is usually the case with SPH methods for elastic dynamics or magnetohydrodynamics, angular momentum of our method is not conserved exactly in our method because of the existence of non-central forces. This problem is stated in \cite{Price-and-Monaghan2004}, and \cite{Bonet-and-Lok1999} proposed modification of the gradient of the kernel function to recover angular momentum conservation. This aspect will be studied in our next paper.

\section{Implementation}
In this section, we describe detailed implementation of our Godunov SPH method for elastic dynamics. In Section 3.1, the method to use the Riemann solver for non-ideal gas EoS is described. In Section 3.2, we explain the mitigation of the tensile instability in our formulation. In Section 3.3, we explain how to use the variable smoothing length.

\subsection{Riemann solver for non-ideal gas equation of state}
The Riemann solver is a method to solve the Riemann problem (the shock tube problem). In the Godunov scheme, we can describe the shock wave accurately using the Riemann solver. We have semi-analytic formula of the Riemann solver in the case of ideal gas EoS or simple EoS for elastic body ($P=C_{s}^{2}(\rho -\rho_{0})$), and we can solve it using iteration. The Riemann solver for ideal gas EoS is introduced in \cite{Leer1978}, and for EoS of elastic body is written in \cite{Sugiura-and-Inutsuka2016}. However, general EoS such as Tillotson EoS is complicated in contrast to that for ideal gas or elastic body. At present analytical solutions of the Riemann problems for such EoS are not available. The Riemann solver is a tool to treat the shock wave, and we do not necessarily use the analytical solution. Therefore, in this study, we propose the method to obtain numerical solutions of the Riemann problems for general EoS. 

The EoS that represents solids such as Tillotson EoS or stiffened gas EoS behaves like elastic body at low temperature and like ideal gas at very high temperature because of sublimation. Therefore, it is expected that we may use the Riemann solver for EoS of elastic body at low temperature, and that for ideal gas EoS at high temperature. 

First, we consider the case that EoS behaves like ideal gas at high temperature. The specific heat ratio $\gamma$ is a good indicator to measure the property of ideal gas. In adiabatic change, polytropic relation $P=K\rho^{\gamma}$ holds, and the specific heat ratio shows the power of the density. Similarly we can evaluate effective specific heat ratio $\gamma_{{\rm eff}}$ for general EoS by calculating the exponent of the density,

\begin{equation}
\gamma_{{\rm eff}} \equiv \frac{d\ln P}{d\ln \rho} = \frac{\rho}{P}\Bigl[ \frac{\partial P}{\partial \rho} + \frac{\partial P}{\partial u}\frac{du}{d\rho} \Bigr],
\label{eff-gamma}
\end{equation}

\noindent where we can express $du/d\rho$ using the first law of thermodynamics $du=-PdV=(P/\rho^{2})d\rho$ as,

\begin{equation}
\frac{du}{d\rho}=\frac{P}{\rho^{2}}.
\label{du_drho}
\end{equation}

\noindent We can calculate the formula of $\partial P/\partial \rho$ and $\partial P/\partial u$ easily once EoS is obtained. We solve the Riemann solver at high temperature by approximating it as the Riemann solver for ideal gas with the specific heat ratio of

\begin{equation}
\gamma = \frac{\gamma_{{\rm eff,L}}+\gamma_{{\rm eff,R}}}{2},
\label{gamma-for-riemann-solver}
\end{equation}

\noindent where $\gamma_{{\rm eff,L}}$ is effective specific heat ratio of left hand side of the Riemann problem, $\gamma_{{\rm eff,R}}$ is that of right hand side. Hereafter, subscript of ${\rm L}$ denotes the value of left hand side of the Riemann problem, and ${\rm R}$ denotes that of right hand side. It is assumed that this approximation is valid when $\gamma_{{\rm eff,L}}$ and $\gamma_{{\rm eff,R}}$ are comparable, because in that case this EoS behaves like ideal gas EoS locally, but becomes poor when $\gamma_{{\rm eff,L}}$ and $\gamma_{{\rm eff,R}}$ are largely different. 

Next, we consider the case that EoS behaves like elastic body at low temperature. We can describe EoS of elastic body $P=C_{s}^{2}(\rho -\rho_{0})$ once we determine the bulk sound speed $C_{s}$ and the reference density $\rho_{0}$. We approximate the bulk sound speed as,

\begin{equation}
C_{s}=\frac{C_{s,{\rm L}}+C_{s,{\rm R}}}{2}.
\label{Cs-for-riemann-solver}
\end{equation}

\noindent We can express the reference density using $C_{s}$ as $\rho_{0}=\rho -P/C_{s}^{2}$ in the case of EoS of elastic body. Thus we approximate $\rho_{0}$ used for the Riemann solver as,

\begin{equation}
\rho_{0}=\frac{1}{2}[(\rho_{{\rm L}}-P_{{\rm L}}/C_{s}^{2})+(\rho_{{\rm R}}-P_{{\rm R}}/C_{s}^{2})].
\label{rho0-for-riemann-solver}
\end{equation}

\noindent Using Eqs.\,(\ref{Cs-for-riemann-solver}) and (\ref{rho0-for-riemann-solver}) to the Riemann solver for EoS of elastic body, we can approximately obtain the result of Riemann problem at low temperature. 

In the Godunov SPH method, we use the resultant pressure of Riemann problem for $P_{ij}^{\ast}$, which is defined for each pair of particle $i$ and $j$. When we calculate $P_{ij}^{\ast}$, we use physical quantities of the $i$-th and $j$-th particle for the values of left and right hand side of the Riemann problem. Thus the values with subscript of L or R in Eqs.\,(\ref{gamma-for-riemann-solver}), (\ref{Cs-for-riemann-solver}) and (\ref{rho0-for-riemann-solver}) are variables depending on particles, and $\gamma, C_{s}$ and $\rho_{0}$ are the appropriate values that are valid nearby each pair of the $i$-th and $j$-th particle and used for the Riemann solver of ideal gas or elastic body EoS. 

We should have the criterion for which approximation we should use appropriately, and this criterion will depend on the EoS. For example, in the case of stiffened gas EoS,

\begin{equation}
P=C_{0}^{2}(\rho -\rho_{0})+(\gamma_{0}-1)\rho u,
\label{stiffened-gas-equation-of-state}
\end{equation}

\noindent a possible criterion that uses sound speed for solid $C_{0}^{2}$ and that for gas $\gamma_{0}P/\rho$ is,

\begin{equation}
C_{0}^{2}>\gamma_{0}\Bigl( \frac{P_{i}}{\rho_{i}} + \frac{P_{j}}{\rho_{j}} \Bigr) /2.
\label{criterion-for-riemann-solver}
\end{equation}

\noindent If Eq.\,(\ref{criterion-for-riemann-solver}) is satisfied, we use the Riemann solver for EoS of elastic body, and elsewhere we use one for ideal gas EoS, for each pair of the $i$-th and $j$-th particle. In the calculation of collision between aluminum sphere and aluminum plate in Section 4.4, we use this EoS and criterion, and we can calculate without any problem. For Tillotson EoS, a possible criterion is the internal energy of complete vaporization $E_{{\rm cv}}$, which is one of the parameters for Tillotson EoS. If the internal energy of the $i$-th or $j$-th particle is greater than $E_{{\rm cv}}$, we can utilize the Riemann solver for EoS of ideal gas, and elsewhere we use one for elastic body EoS.

As stated in \cite{Inutsuka2002}, the gradients of density, pressure and velocity are necessary if we utilize the second-order Riemann solver. The gradients can be calculated by standard method \cite{Monaghan1992},

\begin{equation}
\nabla f_{i}=\sum_{j}\frac{m_{j}f_{j}}{\rho_{j}} \nabla_{i} W(\bm{r}_{i}-\bm{r}_{j},h).
\label{gradient-f}
\end{equation}

\noindent However, this method produces unphysical gradient nearby the free surface because there is no particle outside of the free surface. To prevent this problem, we modify Eq.\,(\ref{gradient-f}) as follows:

\begin{equation}
\nabla f_{i}=\sum_{j}\frac{m_{j}(f_{j}-f_{i})}{\rho_{j}} \nabla_{i}W(\bm{r}_{i}-\bm{r}_{j},h).
\label{gradient-f-modified}
\end{equation}

\noindent Eq.\,(\ref{gradient-f-modified}) is also used in \cite{Puri-and-Ramachandran2014}. 

As pointed out by \cite{Sugiura-and-Inutsuka2016}, the gradient of pressure that is calculated by Eq.\,(\ref{gradient-f}) helps instability of Nyquist frequency perturbation in the negative pressure region. In the case of the perturbation of Nyquist frequency, the density and pressure of particles become constant, and if the pressure is negative gradients of pressure and density are anti-parallel. In that case we tend to estimate the resultant pressure of the Riemann problem mistakenly smaller. That's why Nyquist frequency perturbation can be unstable. However, the gradient of pressure calculated by Eq.\,(\ref{gradient-f-modified}) becomes zero for the perturbation of Nyquist frequency because the pressure of particles is constant. Therefore, if we use Eq.\,(\ref{gradient-f-modified}), the problem pointed out in \cite{Sugiura-and-Inutsuka2016} does not occur. In this study, we calculate the gradients of density, pressure and velocity for the second-order Riemann solver using Eq.\,(\ref{gradient-f-modified}).

\cite{Puri-and-Ramachandran2014} introduces approximate Riemann solver into the Godunov SPH method. In principle, it can be used for any EoS with relatively smaller computational cost (See \cite{Einfeldt-et-al1991} for cares required in some cases).

\subsection{Mitigation of the tensile instability using the Godunov SPH method}
In \cite{Sugiura-and-Inutsuka2016}, Sugiura and Inutsuka conduct the linear stability analysis of the Godunov SPH method for hydrodynamics equations, and evaluate the stability against the tensile instability. They find that if we choose the interpolation method for $V_{ij}^{2}$ appropriately depending on the sign of pressure and the number of dimensions, we can calculate stably. In two or three dimensions, linear interpolation is stable for positive pressure, and cubic spline interpolation is stable for negative pressure. Therefore, the equation of motion of the Godunov SPH method for hydrodynamics is,

\begin{align}
&\dot{v}^{\alpha}_{i}=-2\sum_{j}m_{j}P_{ij}^{\ast}V_{ij}^{2}\frac{\partial}{\partial x^{\alpha}_{i}}W(\bm{r}_{i}-\bm{r}_{j},\sqrt{2}h), \nonumber \\ &V_{ij}^{2}=\left\{ \begin{array}{ll} V_{ij,{\rm linear}}^{2} & {\rm if} \ (P_{i}+P_{j})>0, \\ V_{ij,{\rm cubic}}^{2} & {\rm if} \ (P_{i}+P_{j})<0. \\ \end{array} \right.\label{EoM-for-2-3D}
\end{align}

\noindent To achieve conservation of total energy, we should use the same type of $V_{ij}^{2}$ for the equation of energy. 

This result is for the equations of hydrodynamics, and it is not obvious that the same method is valid for elastic dynamics. However, in usual calculations, if two particles approach each other, the deviatoric stress tensor becomes repulsive force, and this can stabilize the tensile instability. Thus we can assume that the same method as in \cite{Sugiura-and-Inutsuka2016} is sufficient. Indeed the test calculations of Section 4 show that we can calculate stably by this method. We describe the linear stability analysis of the Godunov SPH method for elastic dynamics in Appendix A, and the result of the linear stability analysis also supports our conclusion.

Therefore, in this paper, we use Eq.\,(\ref{elastic-EoM-for-2-3D}) as the equation of motion of the Godunov SPH method for elastic dynamics,

\begin{align}
&\dot{v}^{\alpha}_{i}=2\sum_{j}m_{j}\sigma_{ij}^{\alpha \beta \ast}V_{ij}^{2}\frac{\partial}{\partial x^{\beta}_{i}}W(\bm{r}_{i}-\bm{r}_{j},\sqrt{2}h), \nonumber \\ &V_{ij}^{2}=\left\{ \begin{array}{ll} V_{ij,{\rm linear}}^{2} & {\rm if} \ (P_{i}+P_{j})>0, \\ V_{ij,{\rm cubic}}^{2} & {\rm if} \ (P_{i}+P_{j})<0. \\ \end{array} \right. \label{elastic-EoM-for-2-3D}
\end{align}

$V_{ij}^{2}$ in the time evolution equation of the deviatoric stress tensor does not contribute to the stability, thus we can use any type of $V_{ij}^{2}$ for it. However, using the same type of $V_{ij}^{2}$ is favourable in terms of computational cost.

Cubic spline interpolation needs the gradient of specific volume. As discussed in Section 3.1, if we calculate the gradient of specific volume as,

\begin{equation}
\frac{\partial V_{i}}{\partial x_{i}^{\alpha}}=-\frac{1}{\rho_{i}^{2}}\sum_{j}m_{j}\frac{\partial}{\partial x_{i}^{\alpha}} W(\bm{r}_{i}-\bm{r}_{j},h),
\label{specific-volume-gradient}
\end{equation}

\noindent undesirable gradient is produced nearby free surface. Eq.\,(\ref{specific-volume-gradient}) is suggested in \cite{Inutsuka2002}. The gradient of specific volume calculated by Eq.\,(\ref{gradient-f-modified}) does not cause such a problem. However, Eq.\,(\ref{gradient-f-modified}) totally changes the stability of the Godunov SPH method against the tensile instability, which is prominent for perturbation of Nyquist frequency. As mentioned above, density of particles is constant for perturbation of Nyquist frequency, so that the gradient of specific volume calculated by Eq.\,(\ref{gradient-f-modified}) becomes zero. Cubic spline interpolation is stable for negative pressure because the gradient of specific volume calculated by Eq.\,(\ref{specific-volume-gradient}) does not become zero even for Nyquist frequency perturbation. If we use Eq.\,(\ref{gradient-f-modified}) for the gradient of specific volume, all interpolations are unstable for negative pressure. Thus, in this study, we calculate the gradient of specific volume for cubic spline interpolation using Eq.\,(\ref{specific-volume-gradient}). Surely this equation produces undesirable gradient nearby free surface, but it does not affect the result of simulations as shown in test calculations.

\subsection{Variable smoothing length}
We have so far treated the smoothing length as constant in space. However, the smoothing length should be close to the average particle spacing. Thus in calculations where the density largely varies in space, the smoothing length should also vary. In \cite{Inutsuka2002}, the smoothing length of the $i$-th particle is defined as,

\begin{align}
&h_{i}=\eta \Bigl[ \frac{m_{i}}{\rho^{\ast}_{i}} \Bigr]^{1/d},\nonumber \\ &\rho_{i}^{\ast}=\sum_{j}m_{j}W(\bm{r}_{i}-\bm{r}_{j},h_{i}^{\ast}), \ \ h_{i}^{\ast}=h_{i}C_{{\rm smooth}}, \label{variable-h}
\end{align}

\noindent where $\eta$ is a constant and corresponds to the ratio between the smoothing length and the average particle spacing, and $C_{{\rm smooth}}$ is a constant to determine the distribution of the smoothing length. $\eta$ should be about 1, and throughout this paper we use $\eta =1$. If $C_{{\rm smooth}}$ is larger than 1, the distribution of the smoothing length becomes smoother than the distribution of density. 

If the smoothing length is represented by spatial variable $h(\bm{r})$, we can not integrate Eq.\,(\ref{Godunov-EoM-1}) analytically even if polynomial approximation of $\rho^{-1}(\bm{r})$ is used. In \cite{Inutsuka2002}, Inutsuka conducts integration analytically assuming that the smoothing length is $h_{i}$ for the half of the integration space that includes the $i$-th particle, and $h_{j}$ for the other half. Also in this study we adopt the same procedure. The equation of motion and the equation of energy for the variable smoothing length are,

\begin{align}
\dot{v}_{i}^{\alpha}=\sum_{j}m_{j}\sigma^{\alpha \beta \ast}_{ij}\Bigl[ & V_{ij}^{2}(h_{i})\frac{\partial}{\partial x_{i}^{\beta}}W(\bm{r}_{i}-\bm{r}_{j},\sqrt{2}h_{i}) \nonumber \\ &+V_{ij}^{2}(h_{j})\frac{\partial}{\partial x_{i}^{\beta}}W(\bm{r}_{i}-\bm{r}_{j},\sqrt{2}h_{j})\Bigr], \label{EoM-Godunov-variable-h} \\ \dot{u}_{i}=\sum_{j}m_{j}\sigma_{ij}^{\alpha \beta \ast}(v_{ij}^{\alpha \ast}-v_{i}^{\alpha \ast})\Bigl[ & V_{ij}^{2}(h_{i})\frac{\partial}{\partial x_{i}^{\beta}}W(\bm{r}_{i}-\bm{r}_{j},\sqrt{2}h_{i}) \nonumber \\ & +V_{ij}^{2}(h_{j})\frac{\partial}{\partial x_{i}^{\beta}}W(\bm{r}_{i}-\bm{r}_{j},\sqrt{2}h_{j})\Bigr]. \label{EoE-Godunov-variable-h}
\end{align}

\noindent Eqs.\,(\ref{dot-eps-rho-ij-ab}) and (\ref{R-rho-ij-ab}) for the variable smoothing length are,

\begin{align}
\dot{\epsilon}_{\rho ,ij}^{\alpha \beta} \equiv &\frac{1}{2}m_{j}\Bigl(V_{ij}^{2}(h_{i}) \Bigl[ (v_{ij}^{\alpha \ast}-v_{i}^{\alpha}) \frac{\partial}{\partial x_{i}^{\beta}}+(v_{ij}^{\beta \ast}-v_{i}^{\beta})\frac{\partial}{\partial x_{i}^{\alpha}} \Bigr] W(\bm{r}_{i}-\bm{r}_{j},\sqrt{2}h_{i}) \nonumber \\ & +V_{ij}^{2}(h_{j}) \Bigl[ (v_{ij}^{\alpha \ast}-v_{i}^{\alpha}) \frac{\partial}{\partial x_{i}^{\beta}}+(v_{ij}^{\beta \ast}-v_{i}^{\beta})\frac{\partial}{\partial x_{i}^{\alpha}} \Bigr] W(\bm{r}_{i}-\bm{r}_{j},\sqrt{2}h_{j})\Bigr), \label{dot-eps-rho-ij-ab-variable-h} \\ R_{\rho ,ij}^{\alpha \beta} \equiv &\frac{1}{2}m_{j}\Bigl(V_{ij}^{2}(h_{i}) \Bigl[ (v_{ij}^{\alpha \ast}-v_{i}^{\alpha}) \frac{\partial}{\partial x_{i}^{\beta}}-(v_{ij}^{\beta \ast}-v_{i}^{\beta})\frac{\partial}{\partial x_{i}^{\alpha}} \Bigr] W(\bm{r}_{i}-\bm{r}_{j},\sqrt{2}h_{i}) \nonumber \\ & +V_{ij}^{2}(h_{j}) \Bigl[ (v_{ij}^{\alpha \ast}-v_{i}^{\alpha}) \frac{\partial}{\partial x_{i}^{\beta}}-(v_{ij}^{\beta \ast}-v_{i}^{\beta})\frac{\partial}{\partial x_{i}^{\alpha}} \Bigr] W(\bm{r}_{i}-\bm{r}_{j},\sqrt{2}h_{j})\Bigr). \label{R-rho-ij-ab-variable-h}
\end{align}

Also in the case of variable smoothing length, we should use appropriate interpolation method for $V_{ij}^{2}$ depending on the sign of $P_{i}+P_{j}$ to suppress the tensile instability. 

We define the density for the variable smoothing length as so-called ``gather'' formulation \cite{Hernquist-and-Katz1989}.

\begin{equation}
\rho_{i}=\sum_{j}m_{j}W(\bm{r}_{i}-\bm{r}_{j},h_{i}).
\label{SPH-density-variable-h}
\end{equation}

\noindent In the case of the variable smoothing length, we have to take into account the gradient of smoothing length to derive the equation of continuity. According to \cite{Price2012}, the proposed equation of continuity for the variable smoothing length is as follows:

\begin{align}
& \dot{\rho}_{i}=\frac{1}{\Omega_{i}}\sum_{j}m_{j}(v_{i}^{\alpha}-v_{j}^{\alpha})\frac{\partial}{\partial x_{i}^{\alpha}}W(\bm{r}_{i}-\bm{r}_{j},h_{i}), \nonumber \\ & \Omega_{i}=1+\frac{h_{i}}{\rho_{i}d}\sum_{j}m_{j}\frac{\partial}{\partial h_{i}}W(\bm{r}_{i}-\bm{r}_{j},h_{i}). \label{EoC-Godunov-variable-h}
\end{align}

\noindent In this study, we use Eq.\,(\ref{EoC-Godunov-variable-h}) as the equation of continuity for the variable smoothing length.

The use of the variable smoothing length tends to enhance the tensile instability for negative pressure. If particles approach each other, the smoothing length becomes short and it makes the shape of the kernel function sharp. Thus in the negative pressure region, the attractive force becomes strong, and this strengthens the tensile instability. The tensile instability that is caused by the extension to the variable smoothing length can not be suppressed by just selecting interpolation explained in Section 3.2. Instead, if $C_{{\rm smooth}}$ is large the smoothing length behaves like constant for short perturbation. Thus large $C_{{\rm smooth}}$ can suppress the tensile instability caused by the variable smoothing length. In Appendix B, we conduct the linear stability analysis for the equations of variable smoothing length, and derive how large $C_{{\rm smooth}}$ should be.

\section{Test Calculation}
In this section, to evaluate the validity of the Godunov SPH method for elastic dynamics, we conduct test calculations such as collision of rubber rings, oscillation of plate, and impact of aluminum sphere on aluminum plate. We show that the Godunov SPH method can suppress the tensile instability even in elastic dynamics.

In this study, we use simple predictor corrector method as a time integrator. This method is almost the same as second-order Runge-Kutta method. We follow time evolution of position, velocity, density, internal energy and $S^{\alpha \beta}/\rho$. First, we calculate time derivative of physical quantities at the $n$-th time step using values at the $n$-th time step, and then derive time-centered physical quantities as,

\begin{align}
&U_{i,n+1/2}=U_{i,n}+\dot{U}_{i,n}\frac{\Delta t}{2}, \nonumber \\ &\bm{r}_{i,n+1/2}=\bm{r}_{i,n}+\bm{v}_{i,n}\frac{\Delta t}{2}+\frac{1}{2}\dot{\bm{v}}_{i,n}\Bigl( \frac{\Delta t}{2} \Bigr)^{2}, \label{time-evolution-1}
\end{align}

\noindent where $U=\rho ,u,\bm{v} ,S^{\alpha \beta}/\rho$. Next, we calculate time-centered derivatives using time-centered physical quantities. Finally, physical quantities of next time step are calculated as,

\begin{align}
&U_{i,n+1}=U_{i,n}+\dot{U}_{i,n+1/2}\Delta t,\nonumber \\ &\bm{r}_{i,n+1}=\bm{r}_{i,n}+\bm{v}_{i,n}\Delta t + \frac{1}{2}\dot{\bm{v}}_{i,n+1/2}\Delta t^{2}. \label{time-evolution-2}
\end{align}

Time step $\Delta t$ is determined by the Courant condition as,

\begin{equation}
\Delta t=\min_{i}C_{{\rm CFL}}\Bigl( \frac{[m_{i}/\rho_{i}]^{1/d}}{C_{{s,i}}} \Bigr),
\label{delta-t}
\end{equation}

\noindent where $C_{s,i}$ is local sound speed at the position of the $i$-th particle. In this study, we use $C_{{\rm CFL}}=0.5$.

We use the second-order Riemann solver that is describe in \cite{Inutsuka2002} with the modified monotonicity constraint of \cite{Sugiura-and-Inutsuka2016}. This monotonicity constraint is that we use the first-order Riemann solver when there are some particles with opposite-sign gradients nearby their positions. This condition is written for a pair of the $i$-th and $j$-th particles as,

\begin{equation}
\Bigl( \frac{\partial f}{\partial s} \Bigr)_{i} \cdot \Bigl( \frac{\partial f}{\partial s} \Bigr)_{j} < 0,
\label{monotonicity-constraint}
\end{equation}

\noindent where,

\begin{align}
\Bigl( \frac{\partial f}{\partial s} \Bigr)_{i}=\Bigl( \frac{\bm{r}_{i}-\bm{r}_{j}}{|\bm{r}_{i}-\bm{r}_{j}|}\Bigr) \cdot \nabla f_{i}, \nonumber \\ \Bigl( \frac{\partial f}{\partial s} \Bigr)_{j}=\Bigl( \frac{\bm{r}_{i}-\bm{r}_{j}}{|\bm{r}_{i}-\bm{r}_{j}|}\Bigr) \cdot \nabla f_{j}, \label{dfds}
\end{align}

\noindent and $f$ represents $\rho$ or $P$. If there is any one particle $j$ that satisfies the condition of Eq.\,(\ref{monotonicity-constraint}) within $3h_{i}$ from the $i$-th particle, we use the first-order Riemann solver for the $i$-th particle. Here, the gradient of physical quantity $f$ is calculated by Eq.\,(\ref{gradient-f-modified}).

\subsection{Convergence test}
First, we conduct a convergence test to confirm that our Godunov SPH method for elastic dynamics really achieves the second-order accuracy in space. In elastic dynamics, longitudinal wave and tangential wave exist as linear waves. In this subsection, we conduct the calculations of longitudinal and tangential wave in two dimensions as a test problem for the convergence test. 

Here, we use simple EoS of elastic body,

\begin{equation}
P=C_{s}^{2}(\rho -\rho_{0}),
\label{EoS-elastic}
\end{equation}

\noindent where $C_{s}$ is bulk sound speed, $\rho_{0}$ is reference density of material. In this subsection, we set $C_{s}=1.0$ and $\rho_{0}=0.1$. The density in the unperturbed state is $\overline{\rho}=1.0$, and thus the pressure in the unperturbed state is $\overline{P}=0.9$. We set the shear modulus to $\mu = 1.0$. Simulations are performed in the square domain,\ $x,\ y \in [0.0,1.0]$, and we assume the periodic boundary condition. The positions of particles in the unperturbed state $(\overline{x_{i}},\overline{y_{i}})$ are given as a square lattice. The initial conditions for the longitudinal wave are,

\begin{align}
&x_{i}=\overline{x_{i}}+X\sin (k\overline{x_{i}}), \nonumber \\ & y_{i}=\overline{y_{i}}, \nonumber \\ & v_{i,x}=-\omega X\cos (k\overline{x_{i}}), \nonumber \\ & v_{i,y}=0, \nonumber \\ & \rho_{i}=\overline{\rho}-\overline{\rho}k X\cos (k\overline{x_{i}}), \nonumber \\ & S_{i}^{xx}=\frac{4}{3}\mu k X \cos (k\overline{x_{i}}), \nonumber \\ & S_{i}^{xy}=S_{i}^{yx}=S_{i}^{yy}=0, \label{initial-longitudinal-wave}
\end{align}

\noindent where $X=0.001/k$ and $k=2\pi$. In the case of the longitudinal wave, $\omega=\sqrt{C_{s}^{2}+(4\mu/3\overline{\rho})}k=2\pi\sqrt{7/3}$. The initial conditions for the tangential wave are,

\begin{align}
&x_{i}=\overline{x_{i}}, \nonumber \\ & y_{i}=\overline{y_{i}}+X\sin (k\overline{x_{i}}), \nonumber \\ & v_{i,x}=0, \nonumber \\ & v_{i,y}=-\omega X\cos (k\overline{x_{i}}), \nonumber \\ & \rho_{i}=\overline{\rho}, \nonumber \\ & S_{i}^{xy}=S_{i}^{yx}=\mu k X \cos (k\overline{x_{i}}), \nonumber \\ & S_{i}^{xx}=S_{i}^{yy}=0,
\end{align}

\noindent where, in the case of the tangential wave, $\omega=\sqrt{\mu /\overline{\rho}}k=2\pi$. We consider the variable smoothing length with $C_{{\rm smooth}}=1.0$. 

To measure the error, we calculate difference between the reference data as,

\begin{equation}
\epsilon = \frac{1}{N_{{\rm tot}}}\sum_{i=1}^{N_{{\rm tot}}}|U_{{\rm ref}}(\bm{r}_{i})-U_{i}|,
\label{error-definition}
\end{equation}

\noindent where $N_{{\rm tot}}$ is the total number of particles, $U_{{\rm ref}}(\bm{r}_{i})$ represents the reference data at position $\bm{r}_{i}$. We use $U=\rho$ for the longitudinal wave, and we use $U=S^{xy}$ for the tangential wave because the density remains constant in this case. In this convergence test, we use the result of $N_{{\rm tot}}=512 \times 512$ as the reference data. The tests are conducted with the total number of particles $N_{{\rm tot}}=16\times 16, 32\times 32,64\times 64,128\times 128,256\times 256$. The errors are evaluated after 100 time-steps. To reduce the error coming from time integration as much as possible, we set $\Delta t$ to be very small value $5.0\times 10^{-4}$. 

\begin{figure}[!htb]
 \begin{center}
 \includegraphics[height=8.0cm,width=12cm]{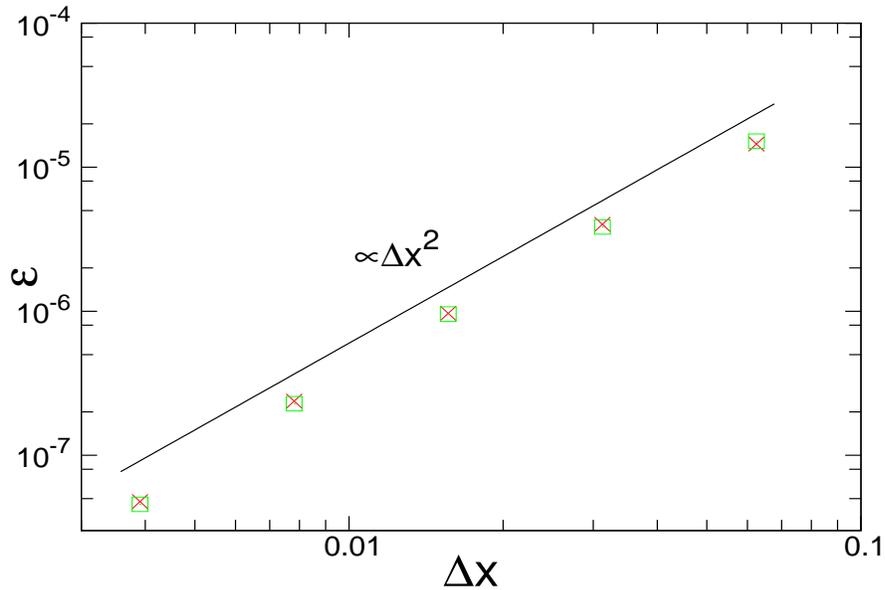}
 \caption{Result of the convergence test for the Godunov SPH method for elastic dynamics. The vertical axis shows relative error, the horizontal axis shows the average particle spacing $\Delta x$. Red crosses show the result of the longitudinal wave, and Green open squares show that of the tangential wave. Solid line shows the line $\propto \Delta x^{2}$.}
 \label{Godunov-SPE-convergence-test}
\end{center}
\end{figure}

In Fig.\,\ref{Godunov-SPE-convergence-test}, $\epsilon$ is plotted as a function of the average particle spacing $\Delta x$. As shown in Fig.\,\ref{Godunov-SPE-convergence-test}, the errors are proportional to $\Delta x^{2}$ for both cases of the longitudinal and the tangential wave. Therefore, the Godunov SPH method for elastic dynamics that we develop in this study shows second-order accuracy in space. 

\subsection{One-dimensional shock tube problem using Tillotson EoS}
To evaluate the validity of our approximation in the Riemann solver for non-ideal gas EoS, we calculate one-dimensional shock tube problem using Tillotson EoS. For simplicity, we use the equations for hydrodynamics. We use the parameters of Tillotson EoS for basalt \cite{Benz-and-Asphaug1999}, and the unit is cgs. For comparison, we also perform calculation by the standard SPH method using artificial viscosity \cite{Monaghan1992} with high resolution. The initial conditions for this shock tube problem are,

\begin{align}
&\rho_{{\rm L}}=2.72, \ \ \rho_{{\rm R}}=2.72, \nonumber \\ &u_{{\rm L}}=1.8\times 10^{12}, \ \ u_{{\rm R}}=1.8\times 10^{8}, \nonumber \\ & v_{{\rm L}}=0.0, \ \ v_{{\rm R}}=0.0. \label{initial-tillotson-shocktube}
\end{align}

\noindent We use $200$ particles for each side, and the mass of each particle is $m=0.0136$. In the case of calculation by the standard SPH method, we use $2000$ particles for each side. Wall boundary condition ($v(x=1)=v(x=-1)=0$) is applied at $x=\pm 1$. We adopt the variable smoothing length with $C_{{\rm smooth}}=1.0$. For simplicity, to derive the density we use Eq.\,(\ref{SPH-density-variable-h}) instead of the continuity equation (\ref{EoC-Godunov-variable-h}) in this shock tube test. Here, we use the Riemann solver for ideal gas EoS only because initial internal energy for left hand side is sufficiently high. $\gamma_{{\rm eff}}$ for each side is,

\begin{equation}
\gamma_{{\rm eff,L}}=2.7, \ \ \gamma_{{\rm eff,R}}=93. 
\label{gamma-eff-tillotson-shocktube}
\end{equation}

\begin{figure}[!htb]
 \begin{center}
 \includegraphics[width=10cm,height=7cm]{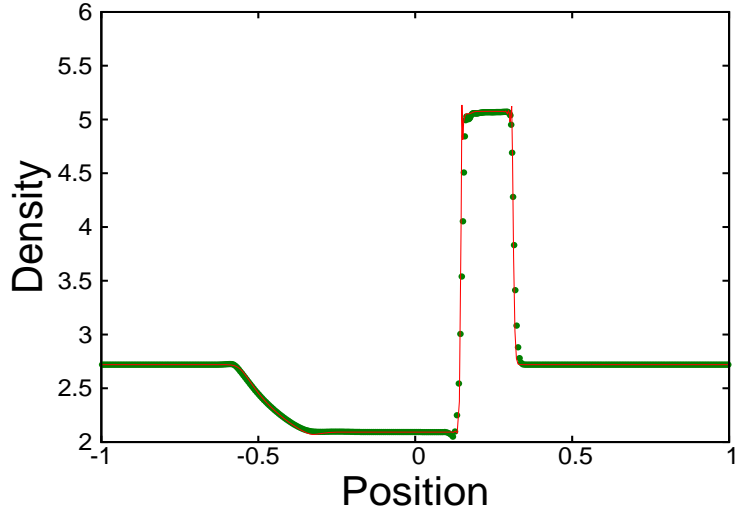}
 \caption{The density distribution of the shock tube problem using Tillotson EoS. The horizontal axis shows the position, and the vertical axis shows the density. The unit is cgs. Green filled circles show the result of the Godunov SPH method using the Riemann solver for ideal gas EoS, and red curve shows that of the standard SPH method.}
 \label{shocktube-tillotson-3-arith-stand}
 \end{center}
\end{figure}

\noindent $\gamma_{{\rm eff,L}}$ and $\gamma_{{\rm eff,R}}$ are largely different in this case, and thus this problem provides a severe test. Figure \ref{shocktube-tillotson-3-arith-stand} shows the result of this shock tube problem calculated by our Godunov SPH method and the standard SPH method.

As we can notice from Fig.\,\ref{shocktube-tillotson-3-arith-stand}, the results of the Godunov SPH method using the Riemann solver for ideal gas EoS and the standard SPH method are almost the same. Therefore, our approximation method can describe shock waves correctly even if EoS is for non-ideal gas. In particular, our Godunov SPH method is valid for hypervelocity impact because the Godunov scheme can treat extremely strong shock waves accurately. 

\subsection{Collision of rubber rings in two dimensions}
Gray et al. \cite{Gray-et-al2001} calculate collision and bounce off of two rubber rings to evaluate the effectiveness of their method against the tensile instability. If we conduct this calculation without any prescription against the tensile instability, numerical fragmentation occurs in the simulation and we can not calculate bounce off of rubber rings. They prevent the tensile instability by introducing artificial stress. In this subsection, we conduct the same simulation using the Godunov SPH method for elastic dynamics.

Also in this subsection, we use EoS of Eq.\,(\ref{EoS-elastic}). The density is scaled using $\rho_{0}$, the velocity is scaled using $C_{s}$ and the length is scaled using the width of ring $w$. We adopt constant smoothing length because in this simulation density is almost constant, and the Riemann solver for elastic EoS is used. 

 We place two rings with 1$w$ separation. The inner radius of rings is 3$w$, and the outer radius is 4$w$. These rings collide with the relative velocity of 0.118$C_{s}$. The particles are put on the square lattice with the side length of 0.1$w$ within two rings. The smoothing length is $h=0.1w$, and we set shear modulus to $\mu = 0.22C_{s}^{2}\rho_{0}$. Initial density of each particle is set to $\rho_{0}$, and all components of initial deviatoric stress tensor is set to $0$. The same condition for initial density and deviatoric stress tensor is adopted for subsequent test calculations. 

\begin{figure}[!htb]
  \begin{center}
    \leavevmode
    \subfigure
    {\includegraphics[width=5cm,height=3.75cm]{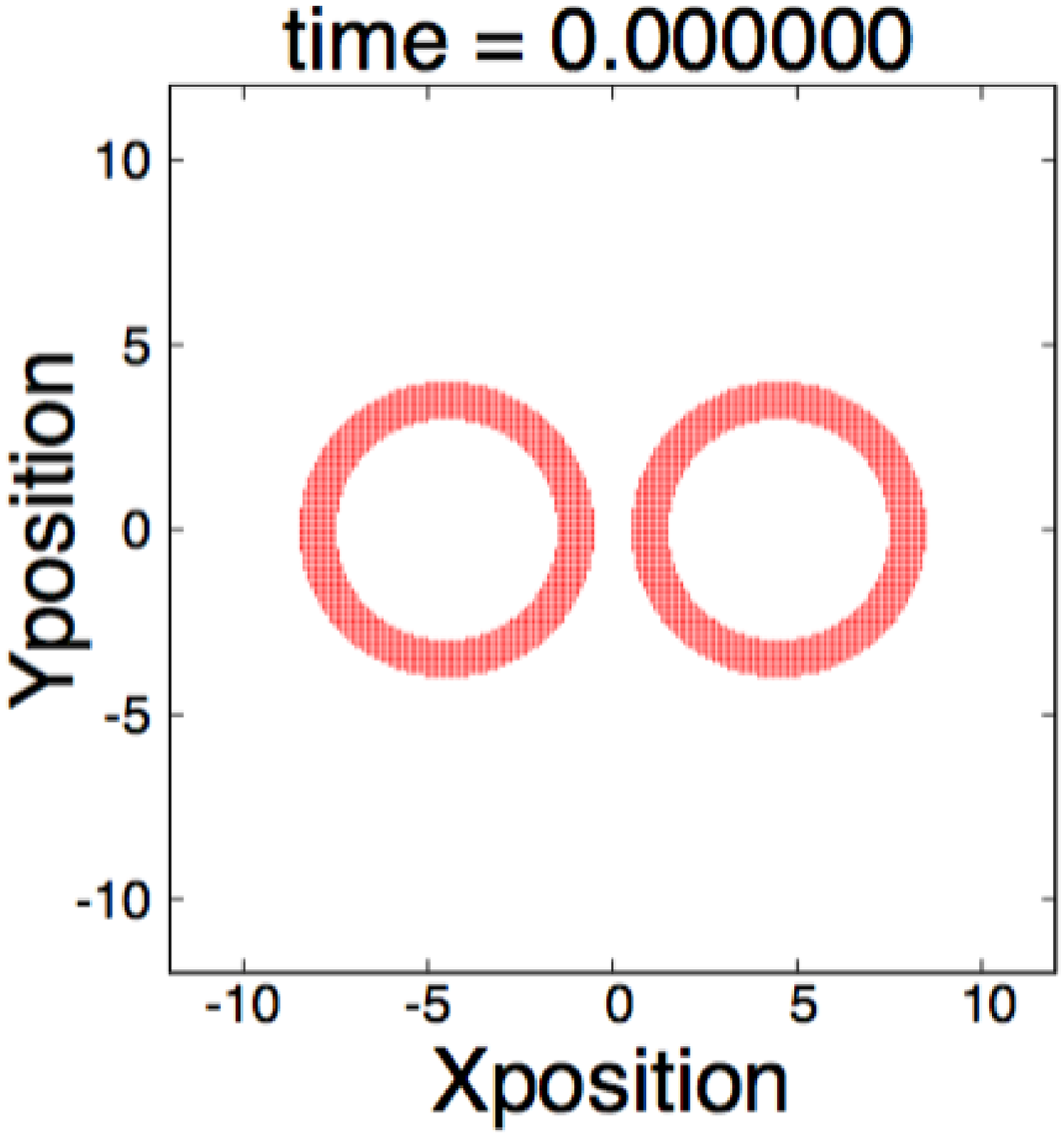}}
    \subfigure
    {\includegraphics[width=5cm,height=3.75cm]{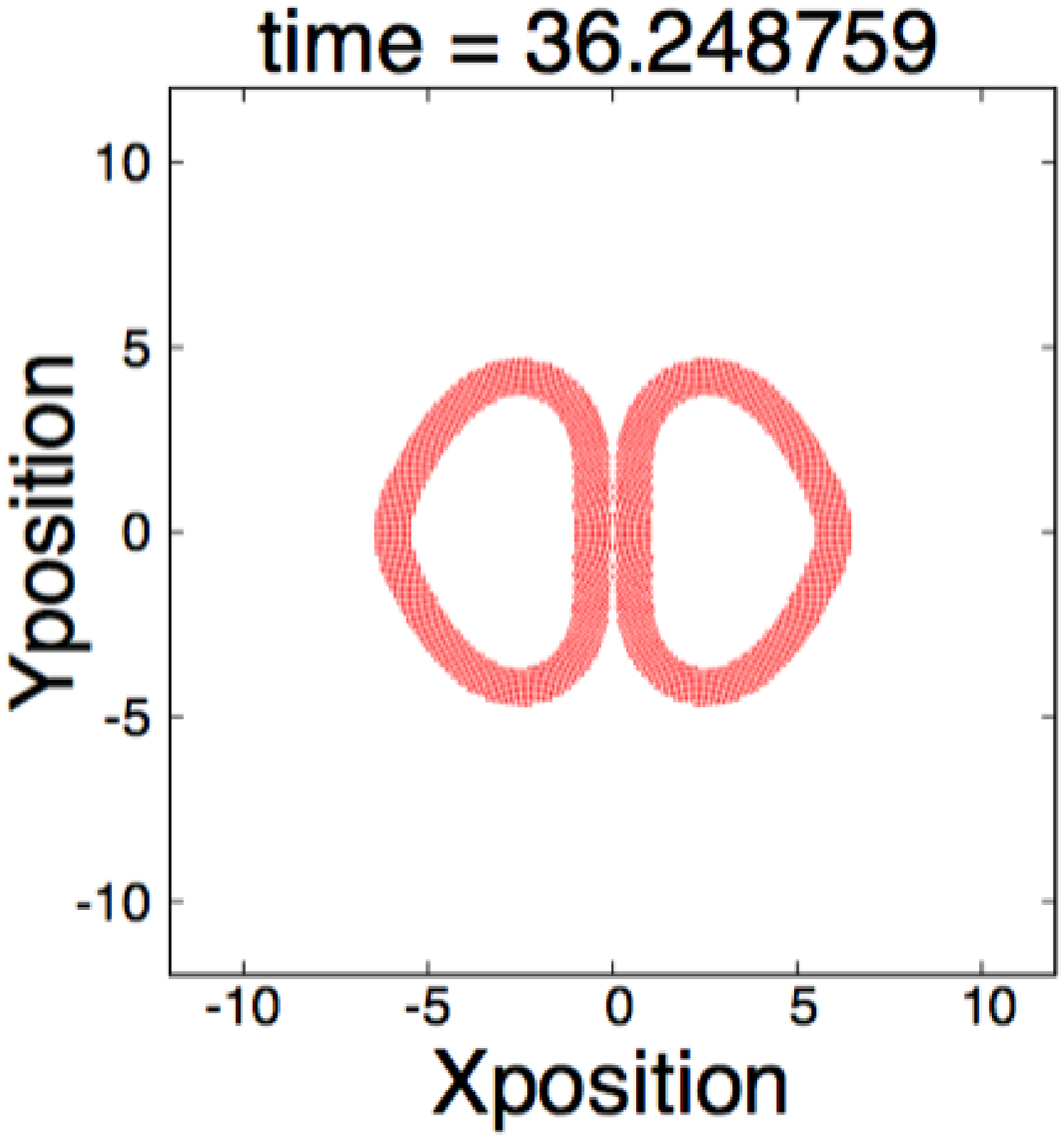}}
    \\
    \subfigure
    {\includegraphics[width=5cm,height=3.75cm]{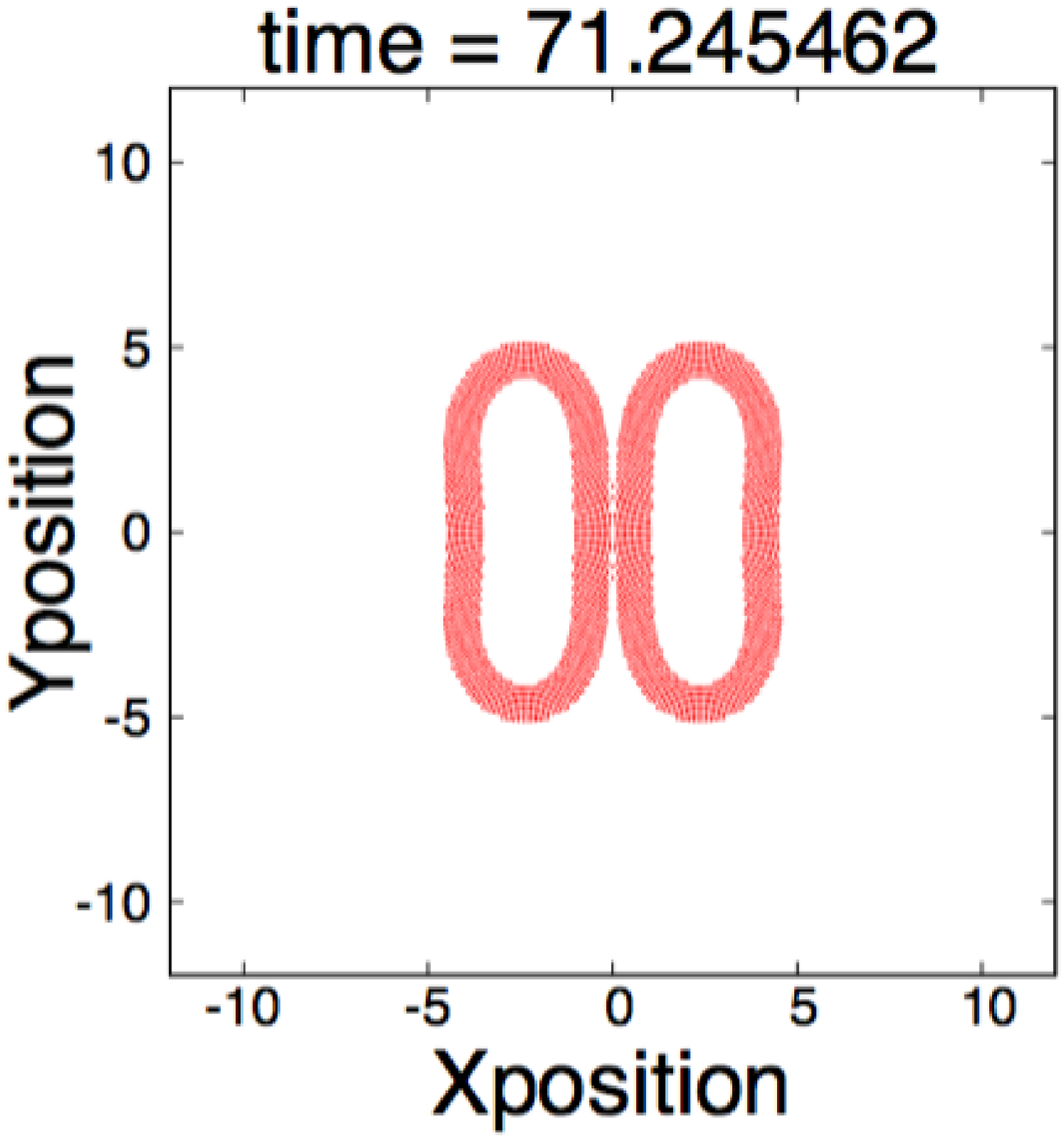}}
    \subfigure
    {\includegraphics[width=5cm,height=3.75cm]{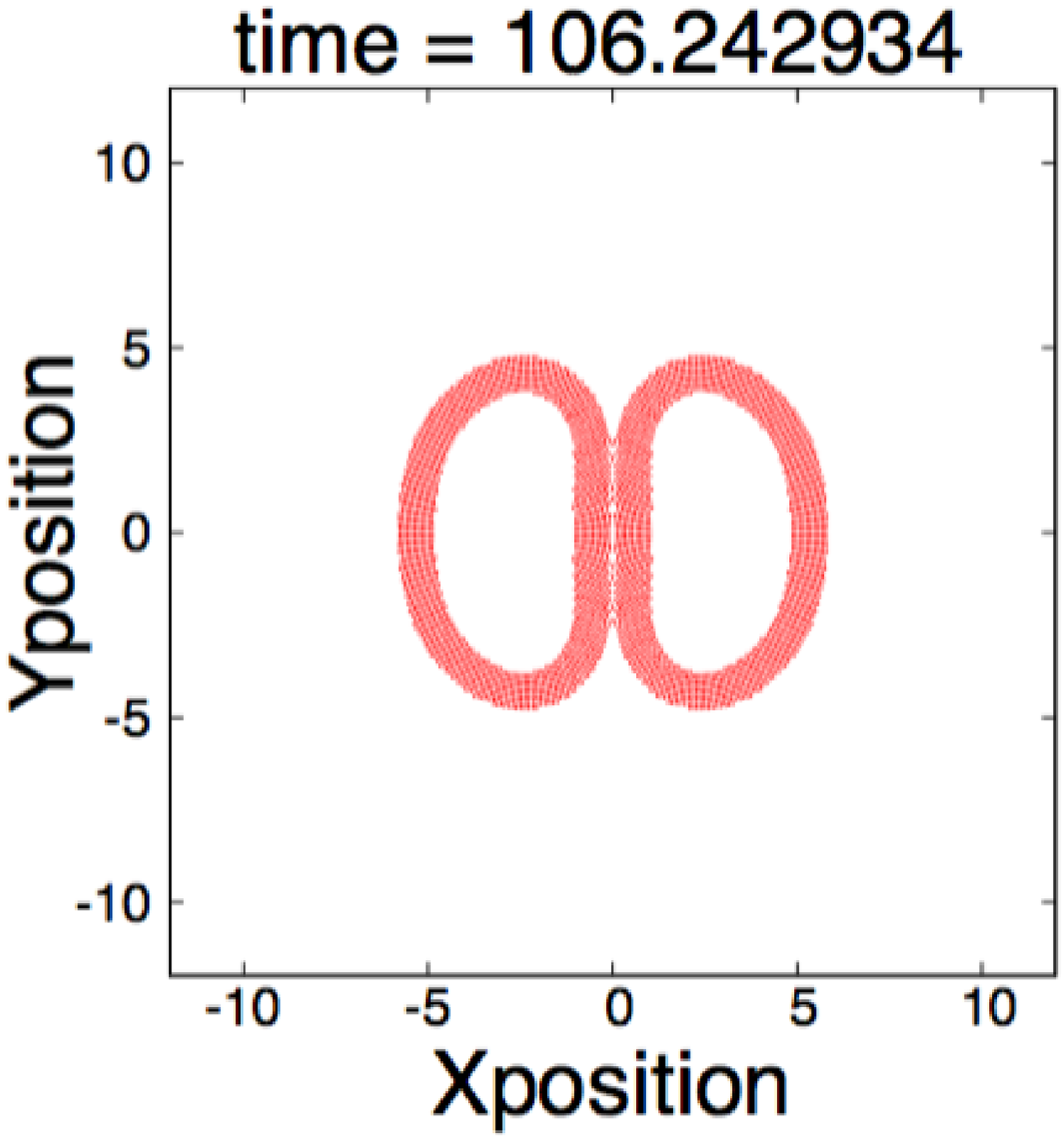}}
    \\
    \subfigure
    {\includegraphics[width=5cm,height=3.75cm]{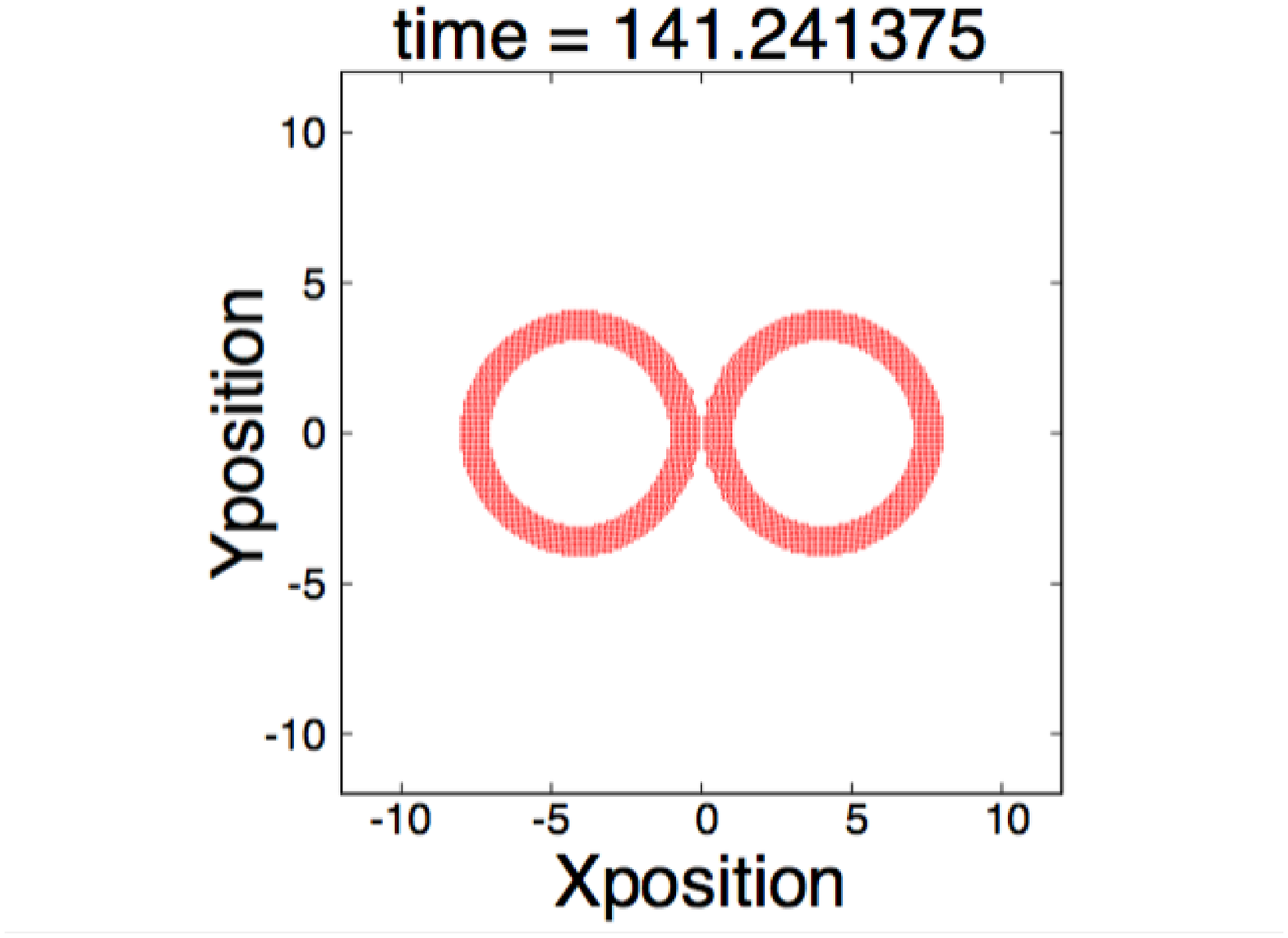}}
    \subfigure
    {\includegraphics[width=5cm,height=3.75cm]{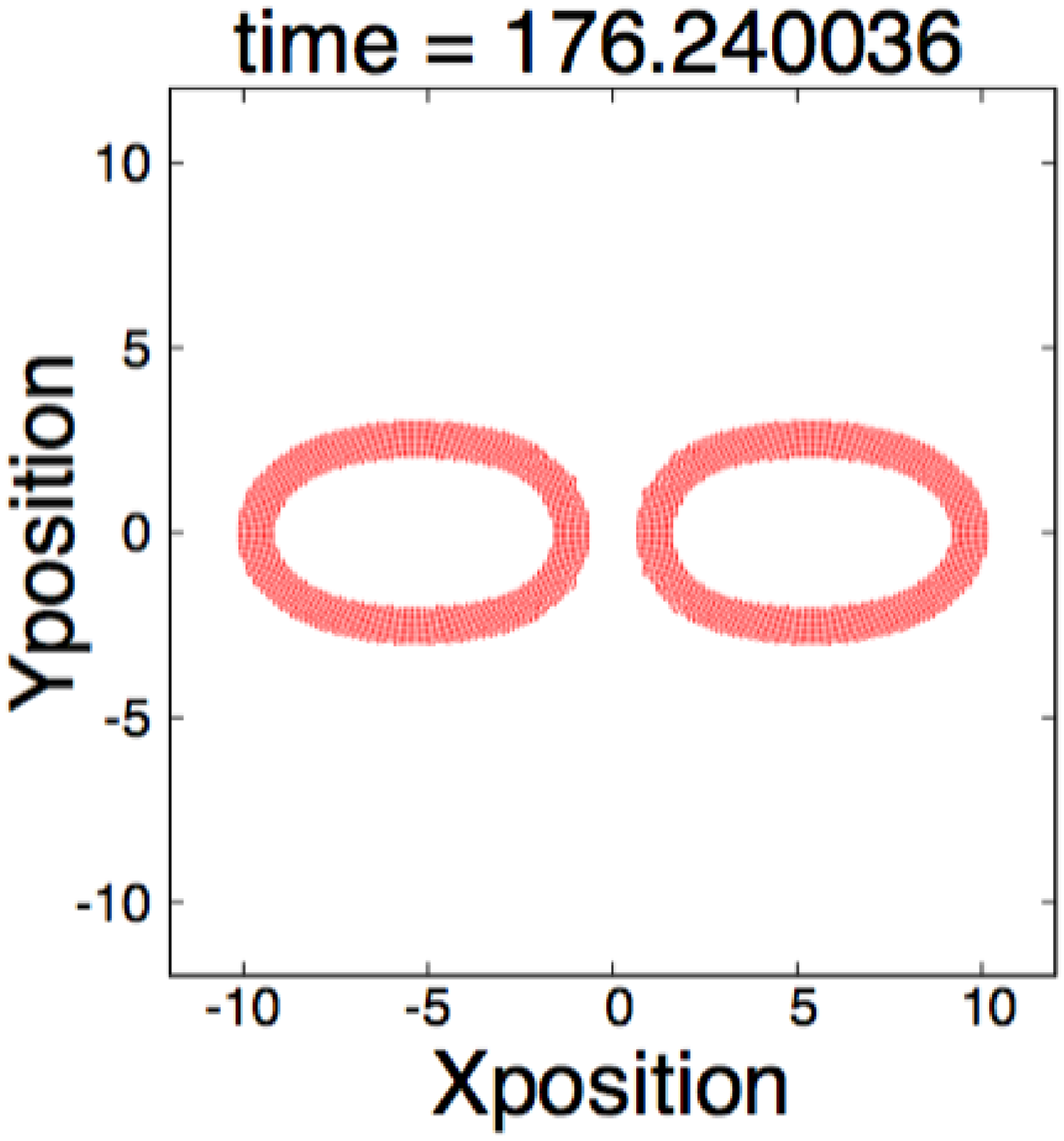}}
    \\
    \subfigure
    {\includegraphics[width=5cm,height=3.75cm]{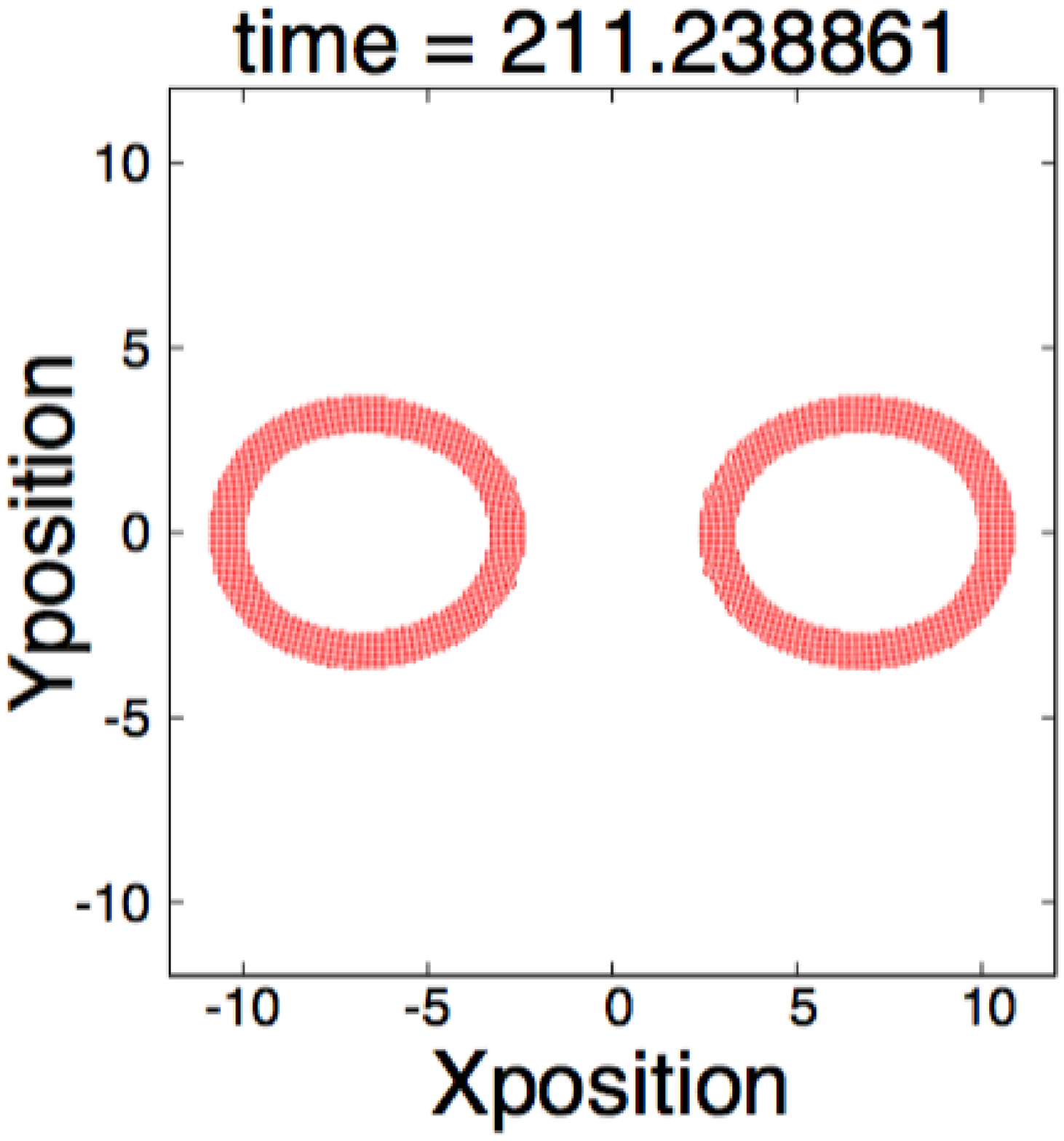}}
    \subfigure
    {\includegraphics[width=5cm,height=3.75cm]{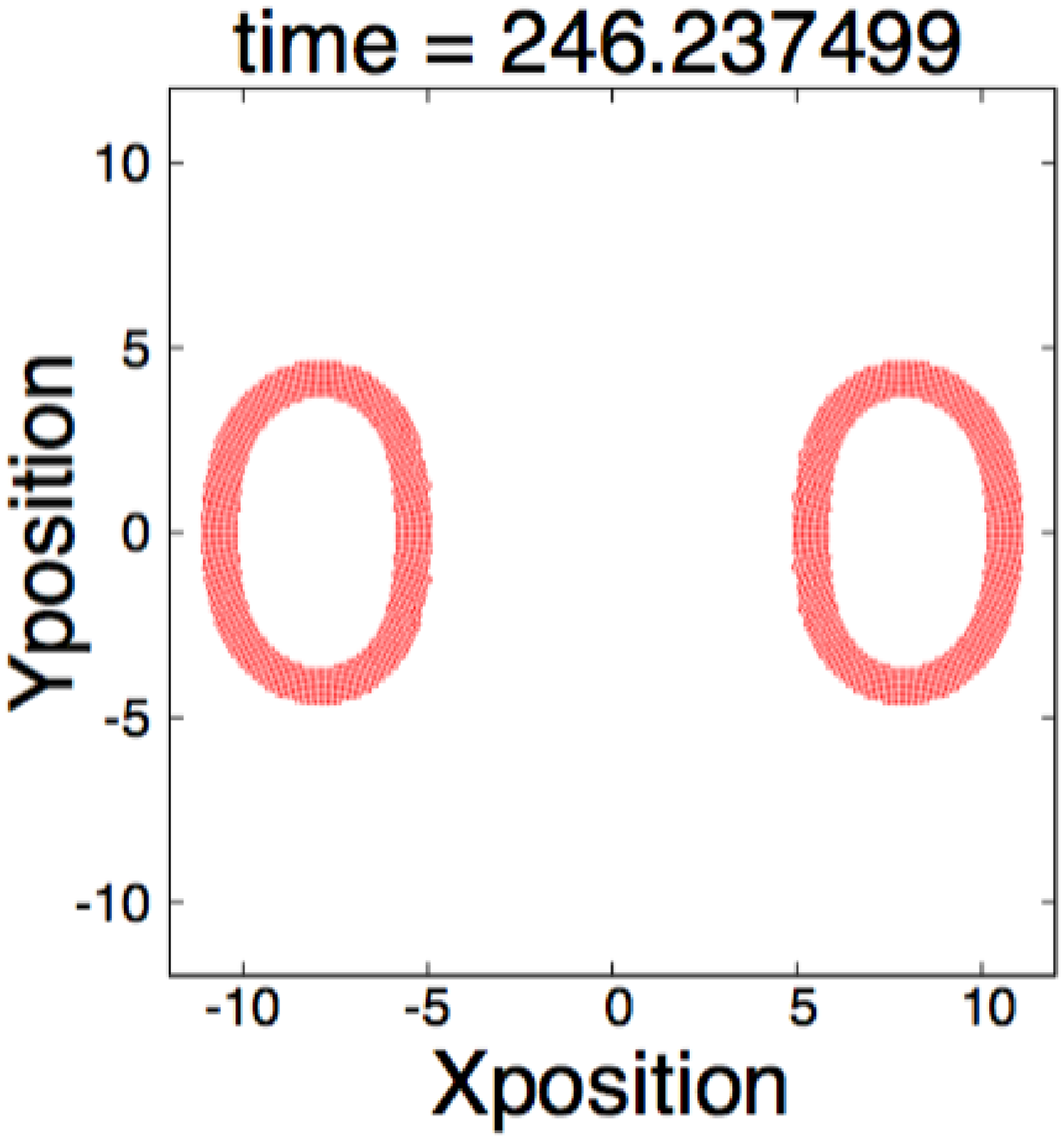}}
    \caption{Result of the calculation of rubber rings collision when we use appropriate interpolation method depending on the sign of pressure.}
    \label{rubber-ring-collision-2D-01234567}
  \end{center}
\end{figure}

\begin{figure}[!htb]
  \begin{center}
    \leavevmode
    \subfigure
    {\includegraphics[width=5cm,height=3.75cm]{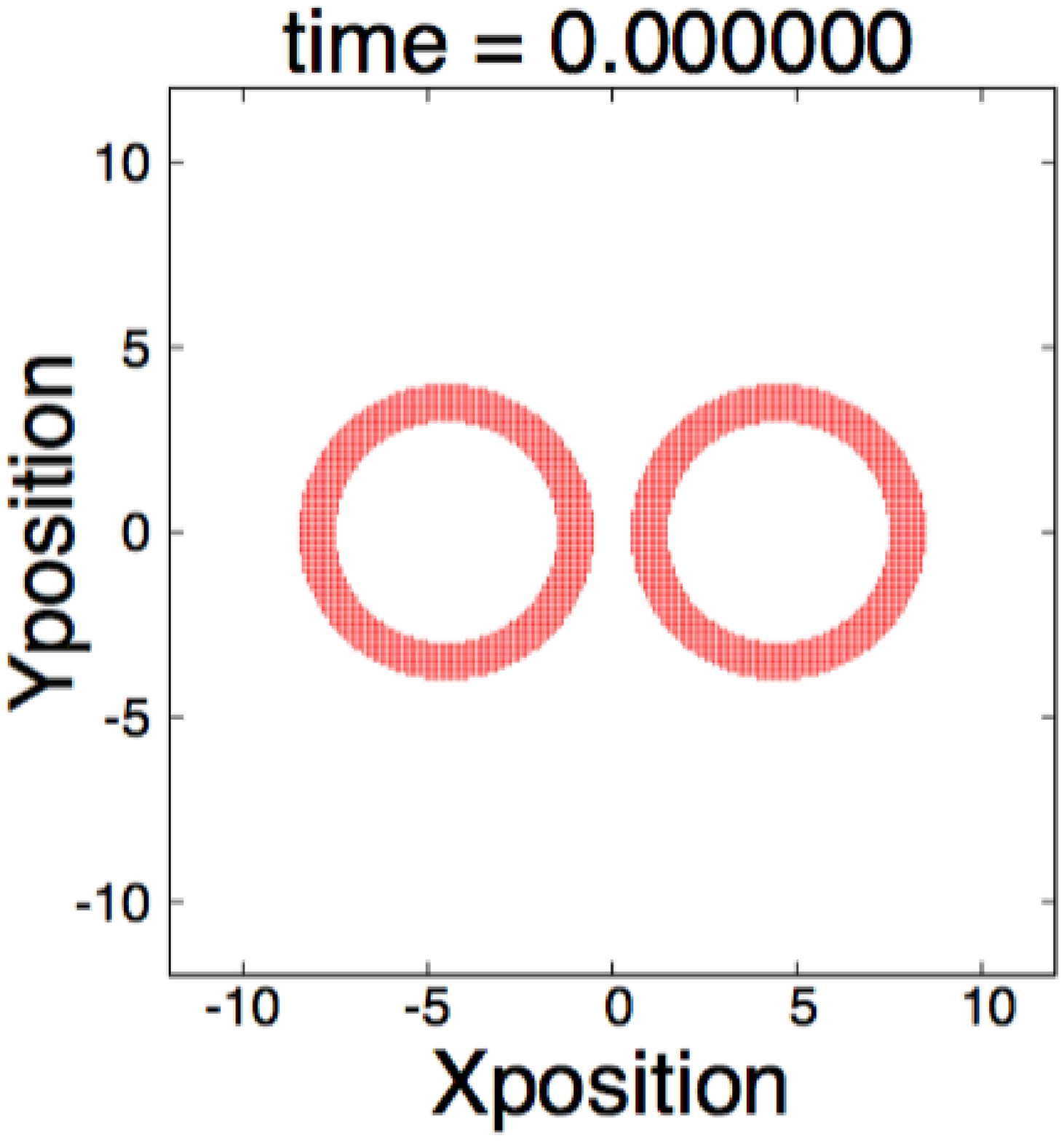}}
    \subfigure
    {\includegraphics[width=5cm,height=3.75cm]{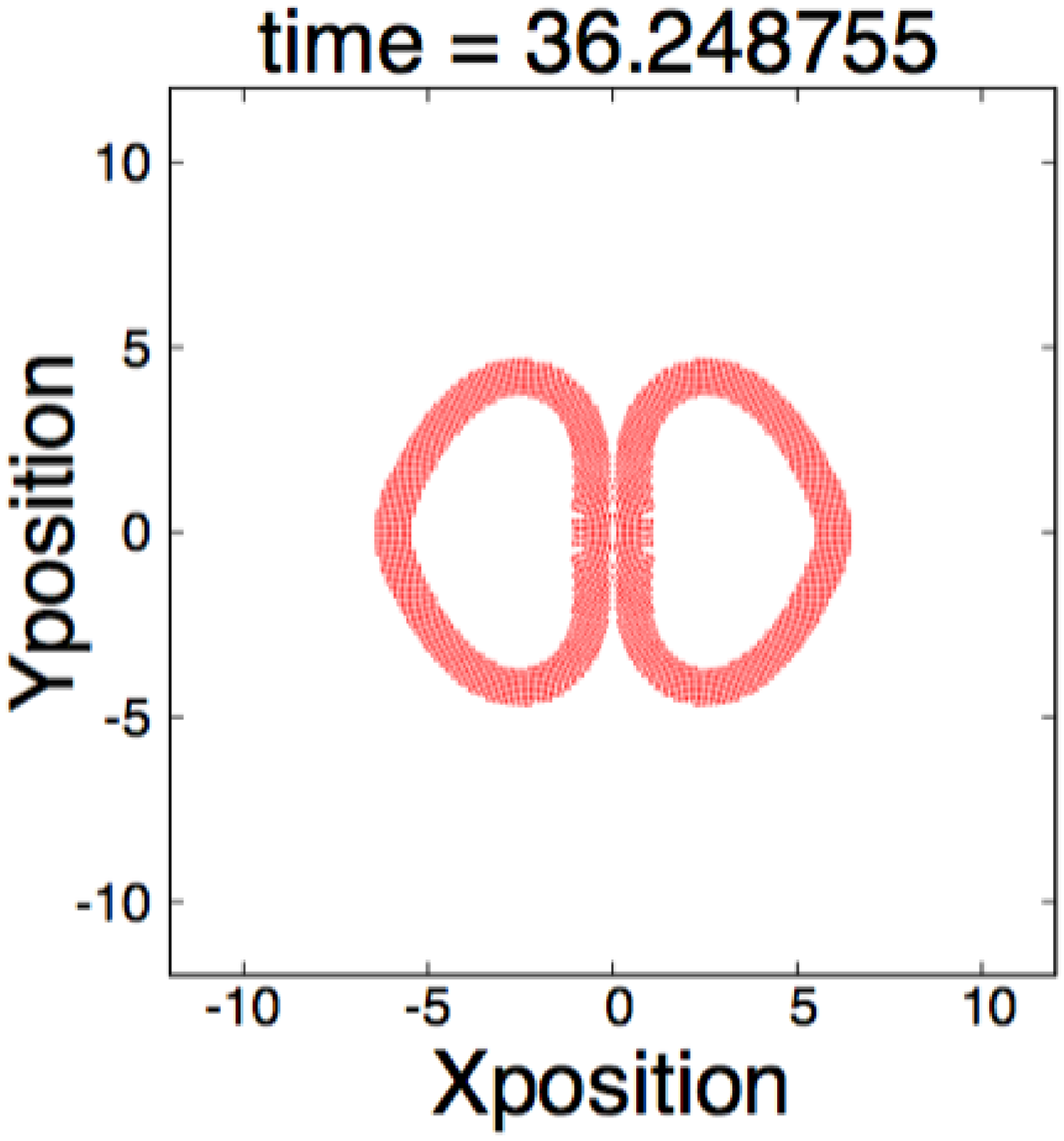}}
    \\
    \subfigure
    {\includegraphics[width=5cm,height=3.75cm]{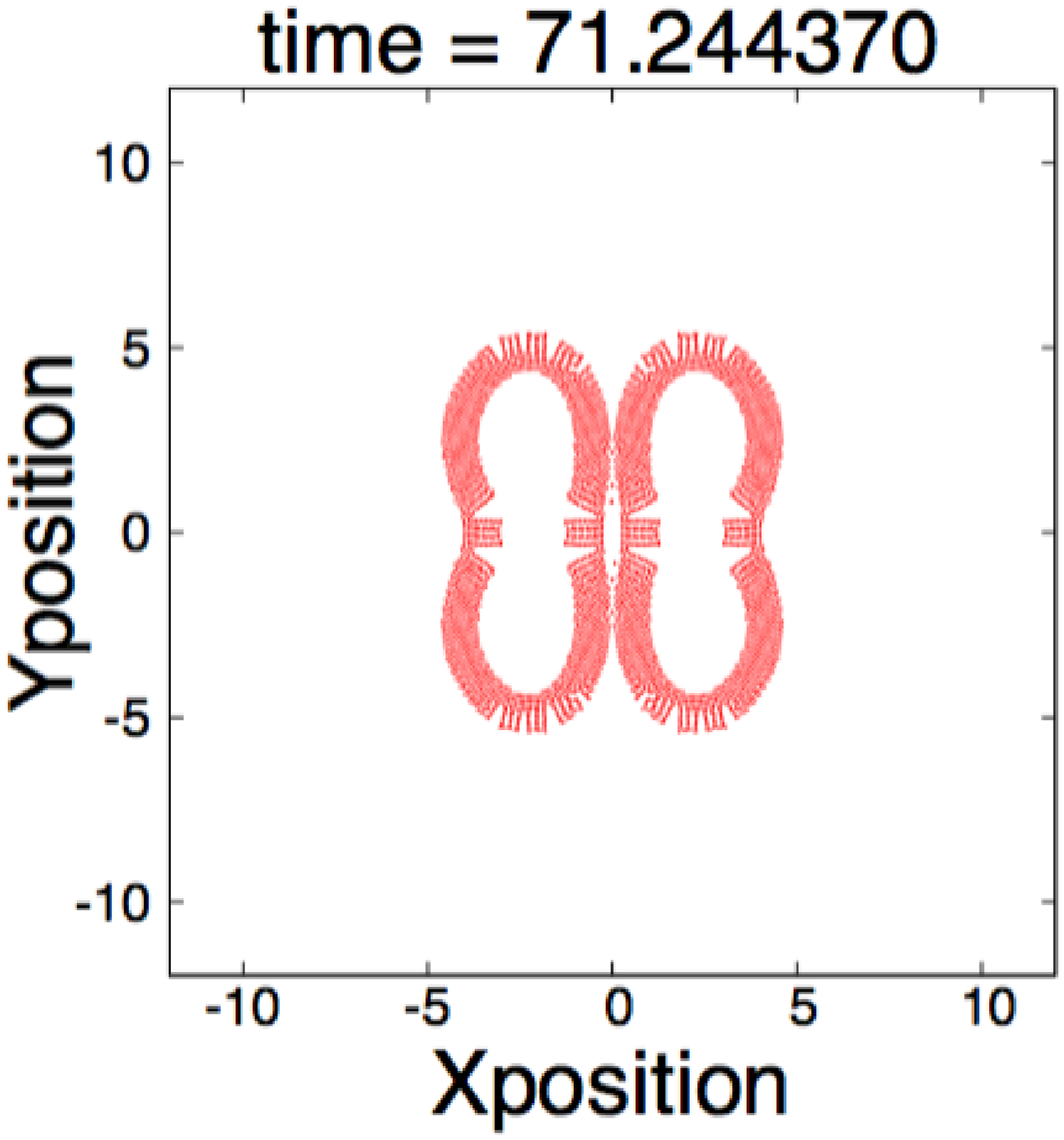}}
    \subfigure
    {\includegraphics[width=5cm,height=3.75cm]{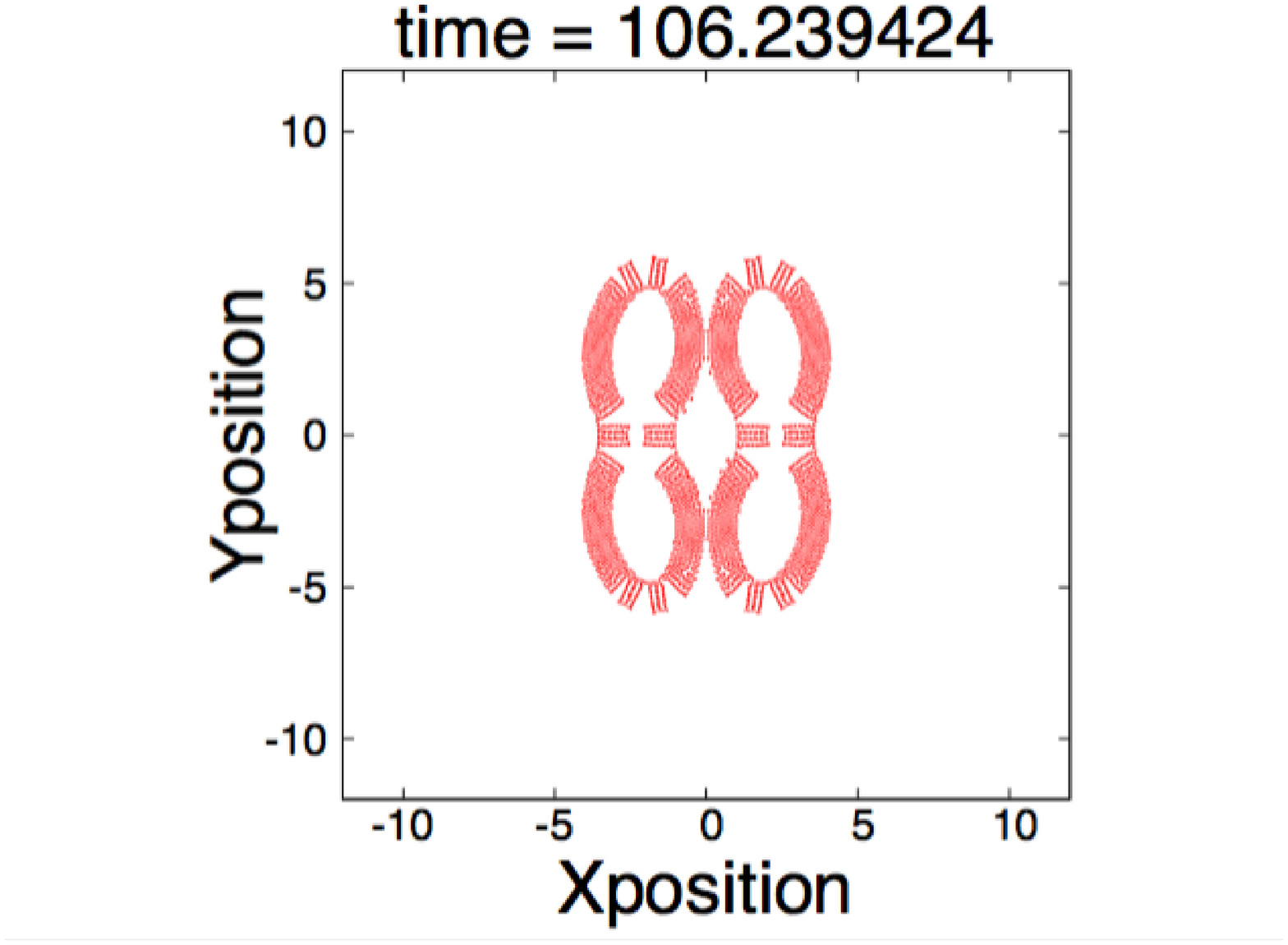}}
    \\
    \subfigure
    {\includegraphics[width=5cm,height=3.75cm]{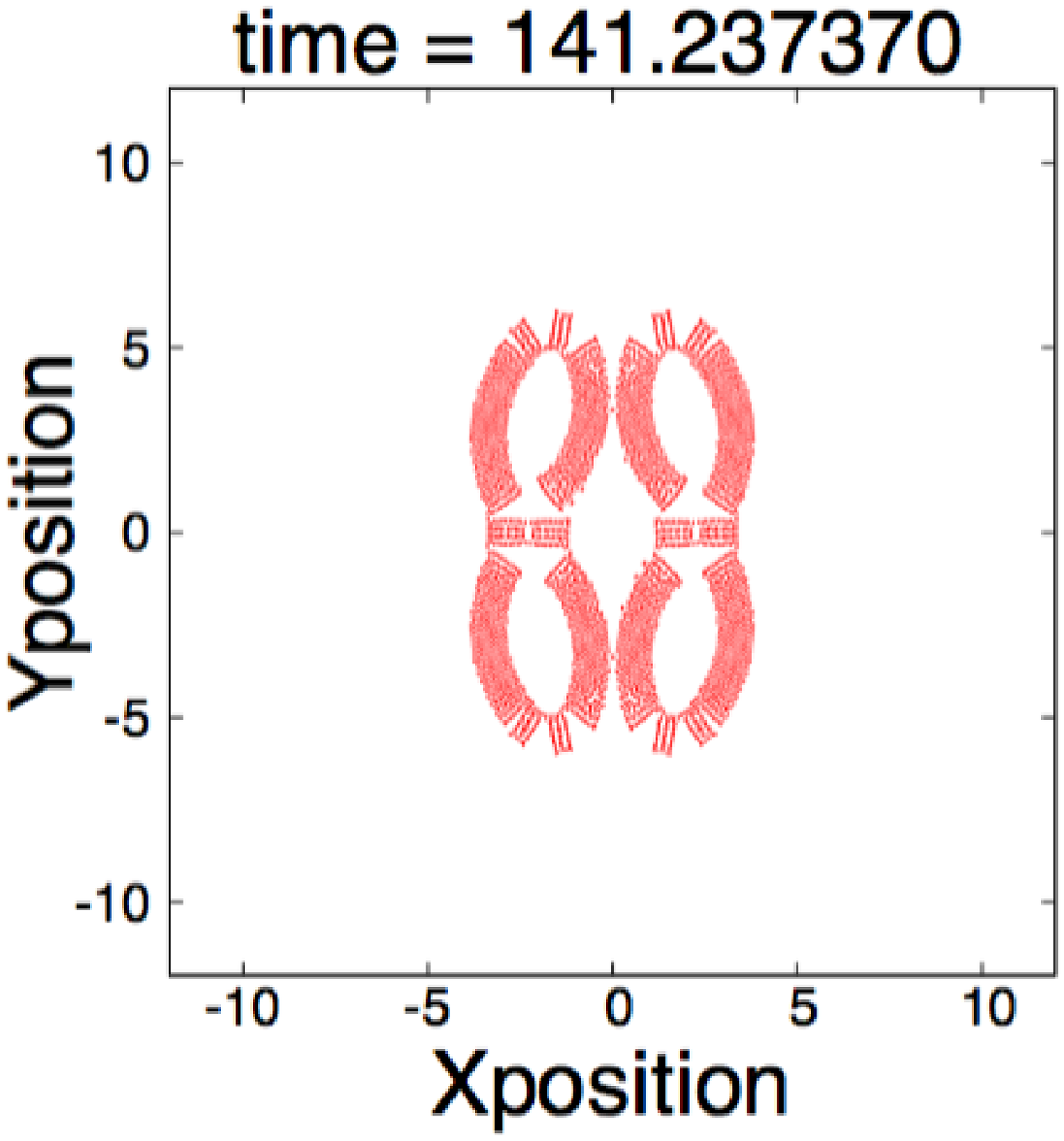}}
    \subfigure
    {\includegraphics[width=5cm,height=3.75cm]{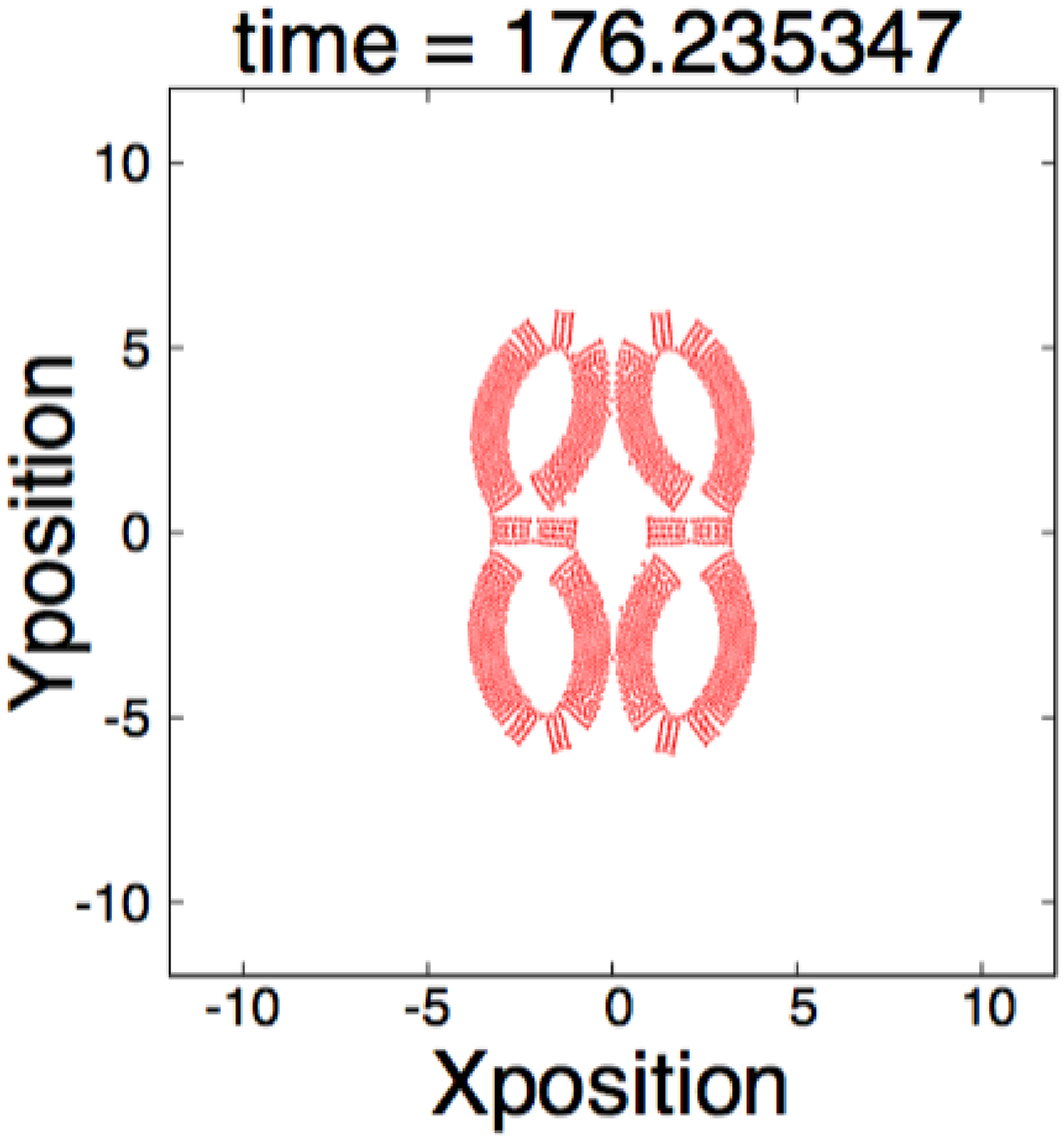}}
    \\
    \subfigure
    {\includegraphics[width=5cm,height=3.75cm]{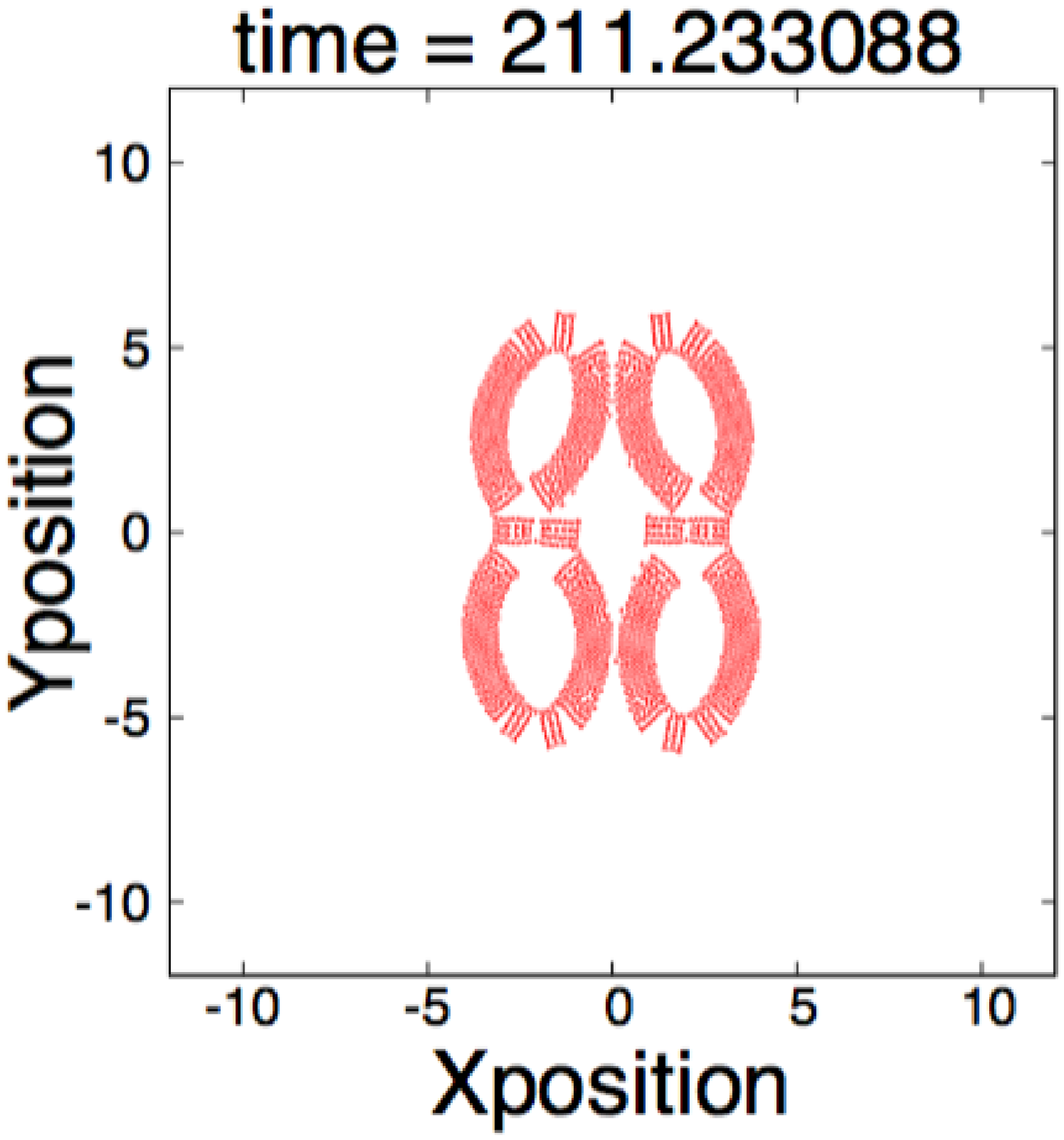}}
    \subfigure
    {\includegraphics[width=5cm,height=3.75cm]{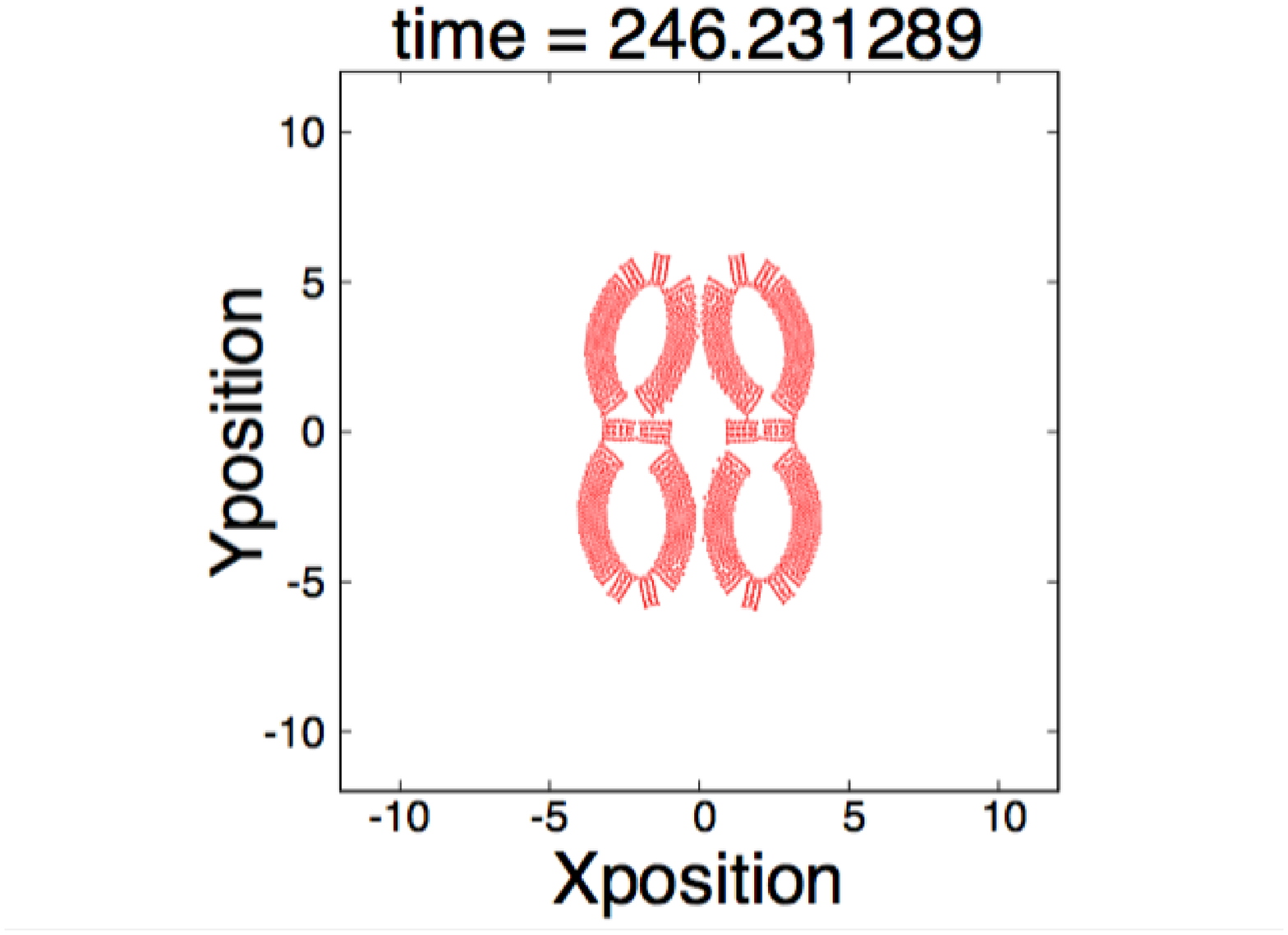}}
    \caption{Same as Fig.\,\ref{rubber-ring-collision-2D-linear-01234567}, but we use only lienar interpolation independent of the sign of pressure.}
    \label{rubber-ring-collision-2D-linear-01234567}
  \end{center}
\end{figure}

Figure \ref{rubber-ring-collision-2D-01234567} shows the configurations of rings when we select appropriate interpolation method depending on the sign of pressure as in Eq.\,(\ref{elastic-EoM-for-2-3D}), and Fig.\,\ref{rubber-ring-collision-2D-linear-01234567} shows the same configuration but we use only linear interpolation independent of the sign of pressure. 

As we can notice from Fig.\,\ref{rubber-ring-collision-2D-01234567}, if we use appropriate interpolation we can calculate the bounce off of two rings stably. However, as shown in Fig.\,\ref{rubber-ring-collision-2D-linear-01234567}, we can not calculate the bounce off due to unphysical fracture caused by the tensile instability at stretched part if we use only linear interpolation. The configurations of two rings shown in Fig.\,\ref{rubber-ring-collision-2D-01234567} agree well with that of Gray et al.

\subsection{Oscillation plate in three dimensions}
To evaluate the validity of our method in three dimensions, we calculate oscillation of elastic plate, one edge of which is fixed. The same test calculation is done by Gray et al. \cite{Gray-et-al2001}; however, this calculation is in two dimensions. Analytical solution of oscillation of extremely thin plate can be found in \cite{Landau-Lifshitz1970}.

We use the same EoS and unit system as those of Section 4.2 except for the unit of length. In this section the length is scaled using the thickness of plate $H$. We also consider the case of constant smoothing length that is the same as initial particle spacing. The length of plate $L$ is 11$H$ ($x$-direction) and the width is 2$H$ ($z$-direction). The particles are put on the square lattice with the side length of 0.1$H$ within this plate. The shear modulus $\mu$ is 0.5$C_{s}^{2}\rho_{0}$. Gray et al. expressed fixed edge by putting the plate between two layers of SPH particles that are not allowed to move. Here, for simplicity, we fix the particles that are located within 1$H$ from left end of the plate. The initial velocity distribution is the same as that of \cite{Gray-et-al2001}. The velocity of $y$-direction $v_{y}$ at the position of $x$-direction $x$ is given by,

\begin{equation}
\frac{v_{y}}{C_{s}}=V_{f}\frac{[M(\cos (kx)-\cosh (kx))-N(\sin (kx)-\sinh (kx))]}{Q},
\label{initial-velocity-oscillation-plate}
\end{equation}

\noindent where $V_{f}$ is the velocity at the free edge of the plate, 

\begin{align}
&M=\sin (kL)+\sinh (kL),\nonumber \\ &N=\cos (kL)+\cosh (kL), \nonumber \\ &Q=2(\cos (kL)\sinh (kL)-\sin (kL)\cosh (kL)), \label{M-N-Q-defs}
\end{align}

\noindent and $k$ is the solutions of

\begin{equation}
\cos (kL)\cosh (kL)=-1.
\label{k-def}
\end{equation}

\noindent In this subsection, $V_{f}$ is set to be 0.05. For the fundamental mode $kL=1.875$. 

\begin{figure}[!htb]
  \begin{center}
    \leavevmode
    \subfigure
    {\includegraphics[width=5cm,height=3.75cm]{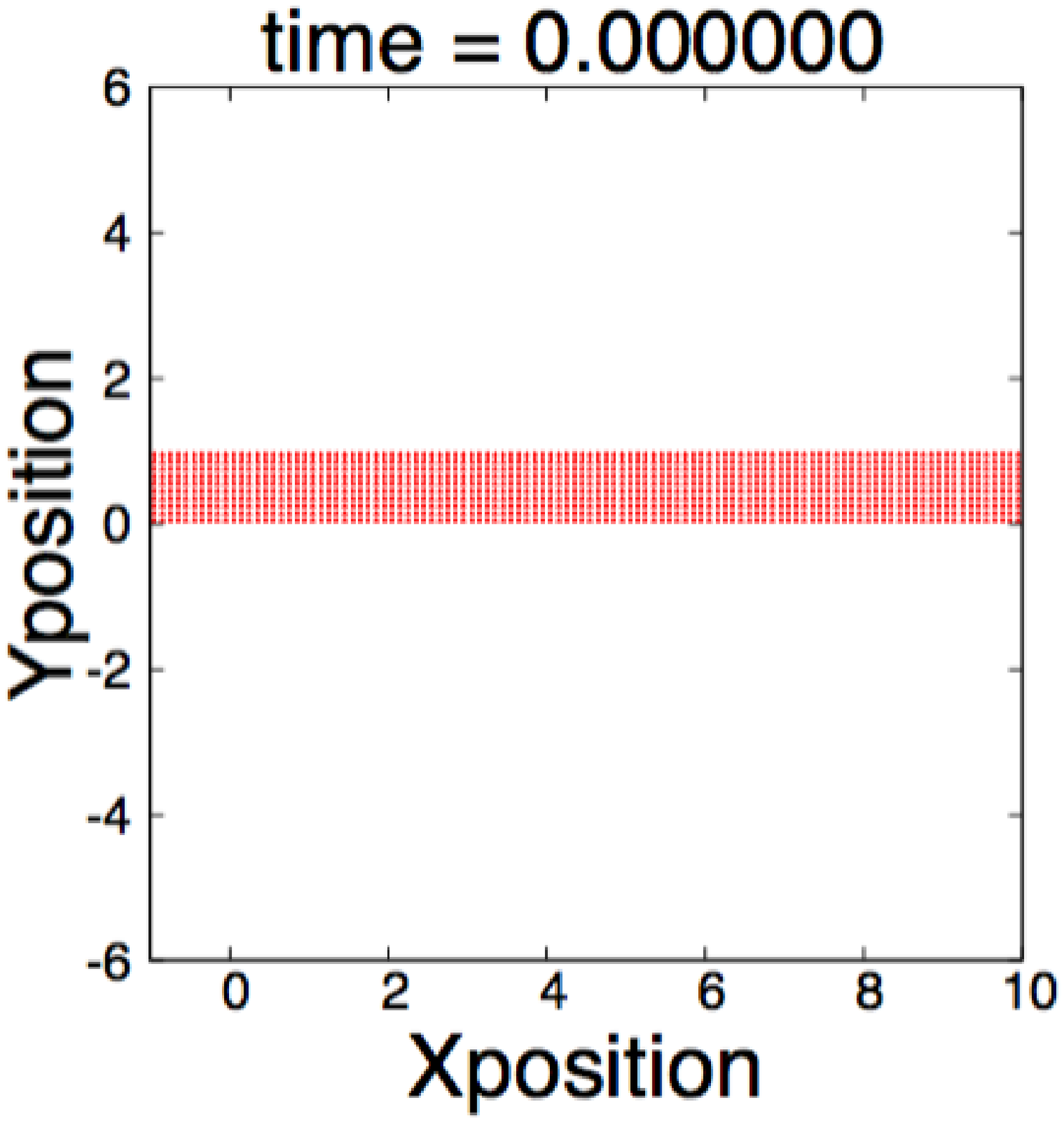}}
    \subfigure
    {\includegraphics[width=5cm,height=3.75cm]{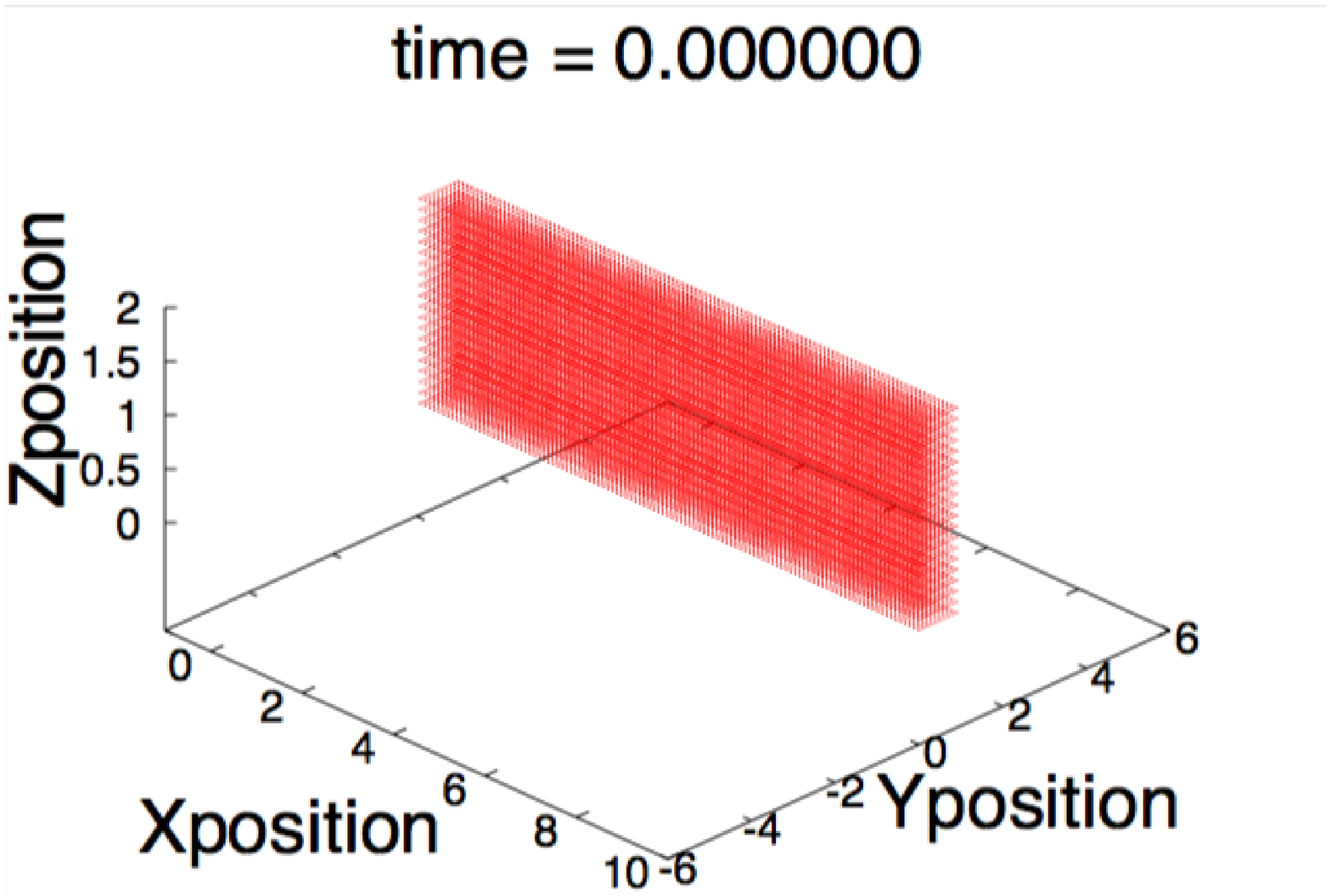}}
    \\
    \subfigure
    {\includegraphics[width=5cm,height=3.75cm]{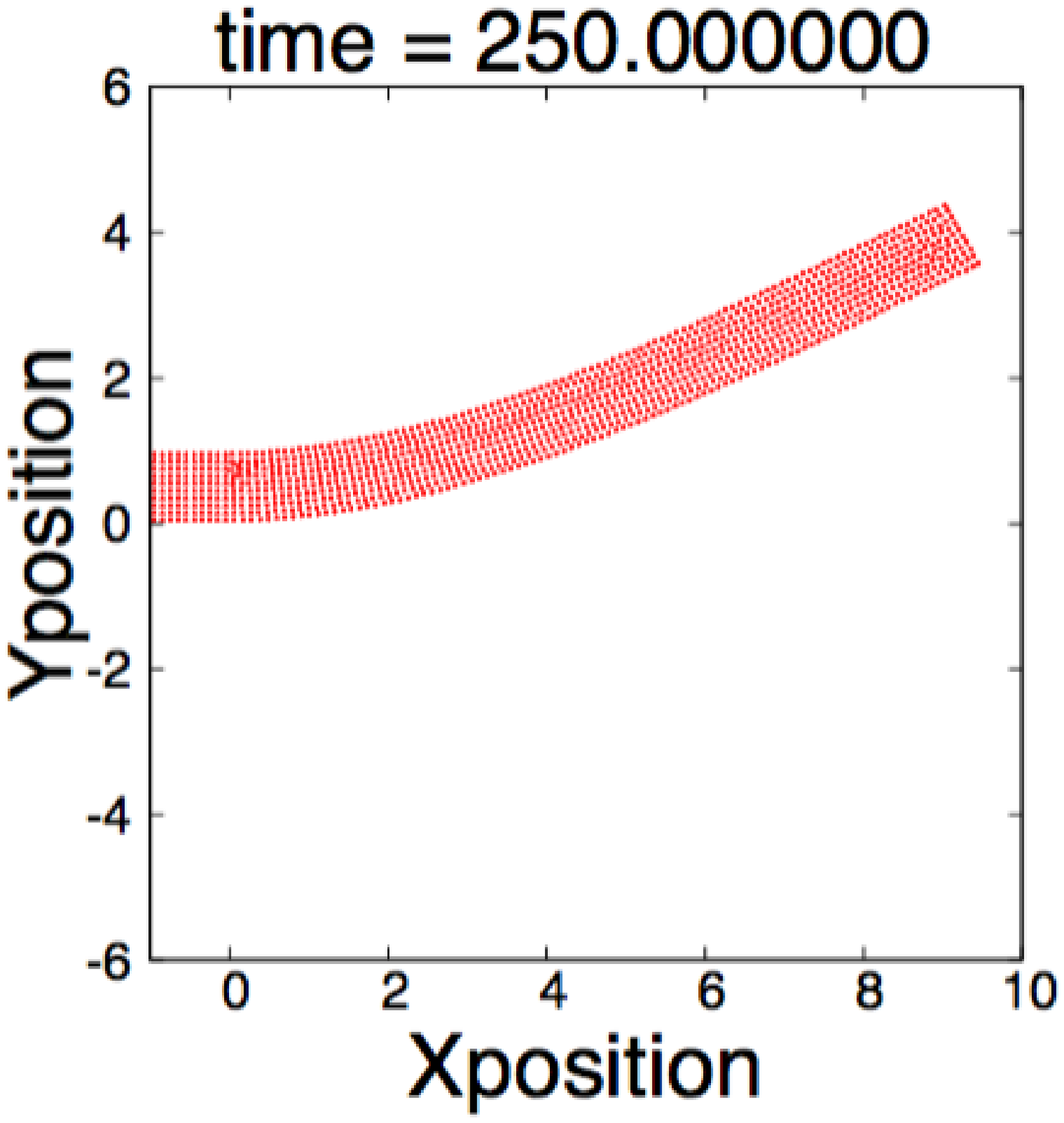}}
    \subfigure
    {\includegraphics[width=5cm,height=3.75cm]{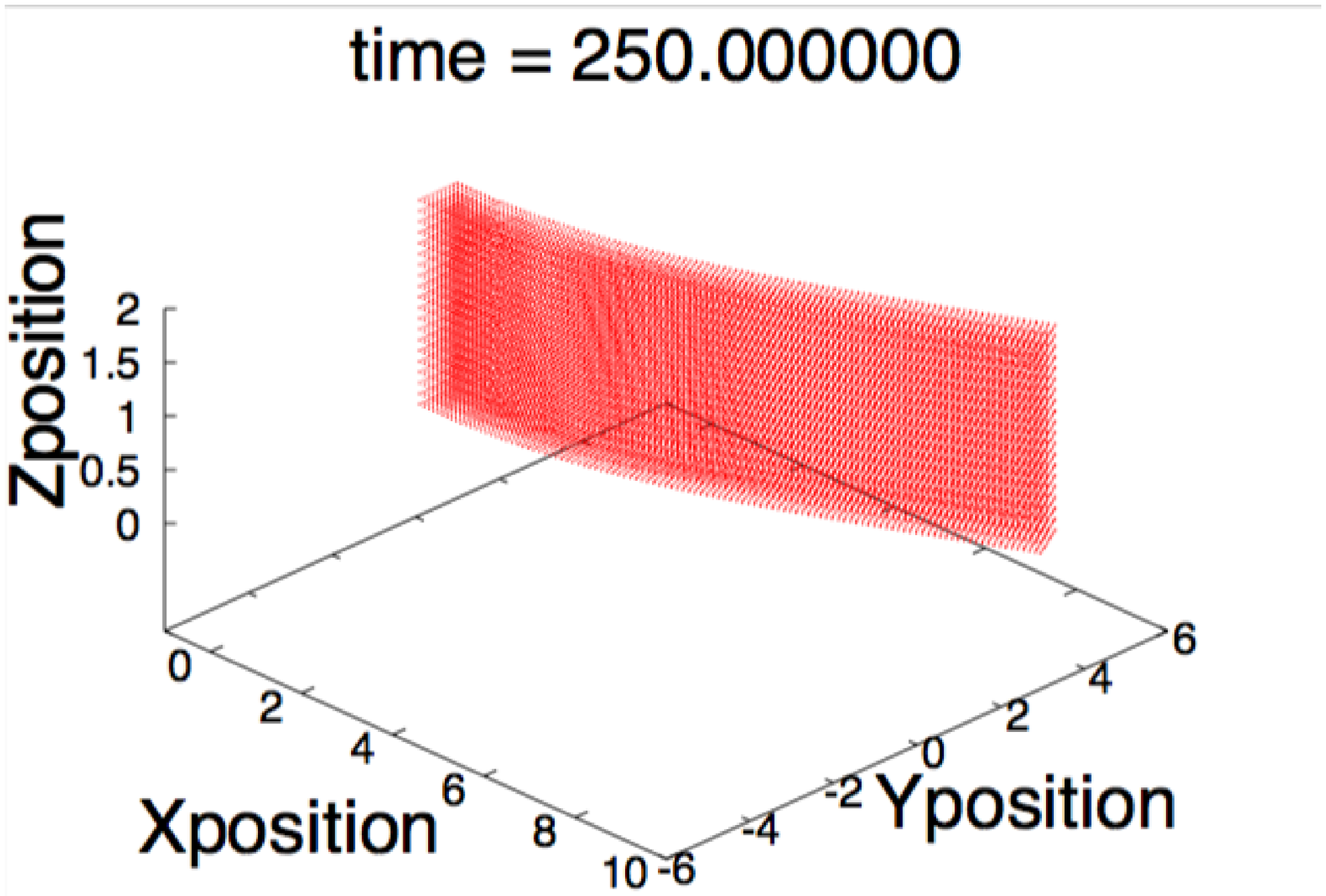}}
    \\
    \subfigure
    {\includegraphics[width=5cm,height=3.75cm]{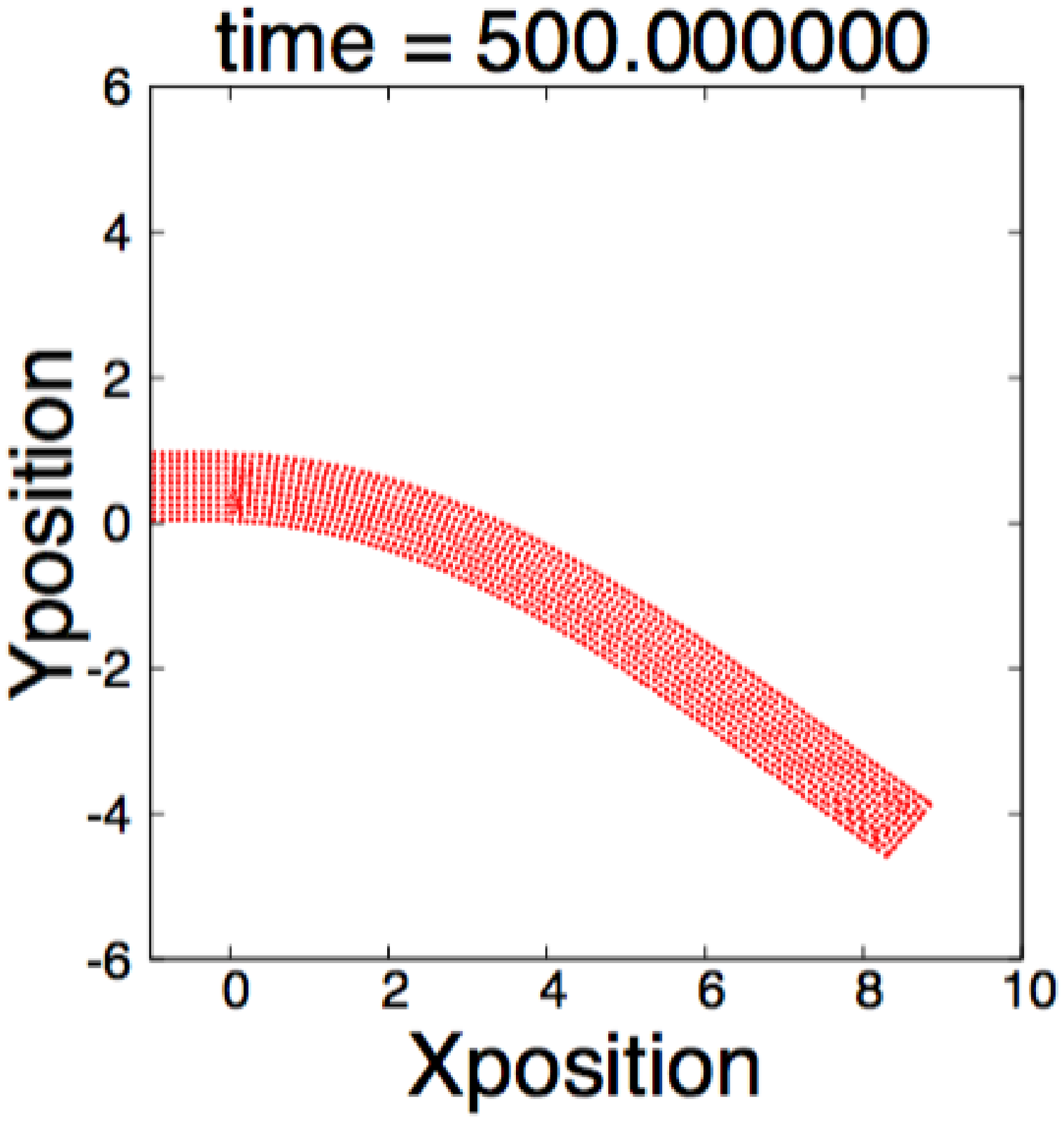}}
    \subfigure
    {\includegraphics[width=5cm,height=3.75cm]{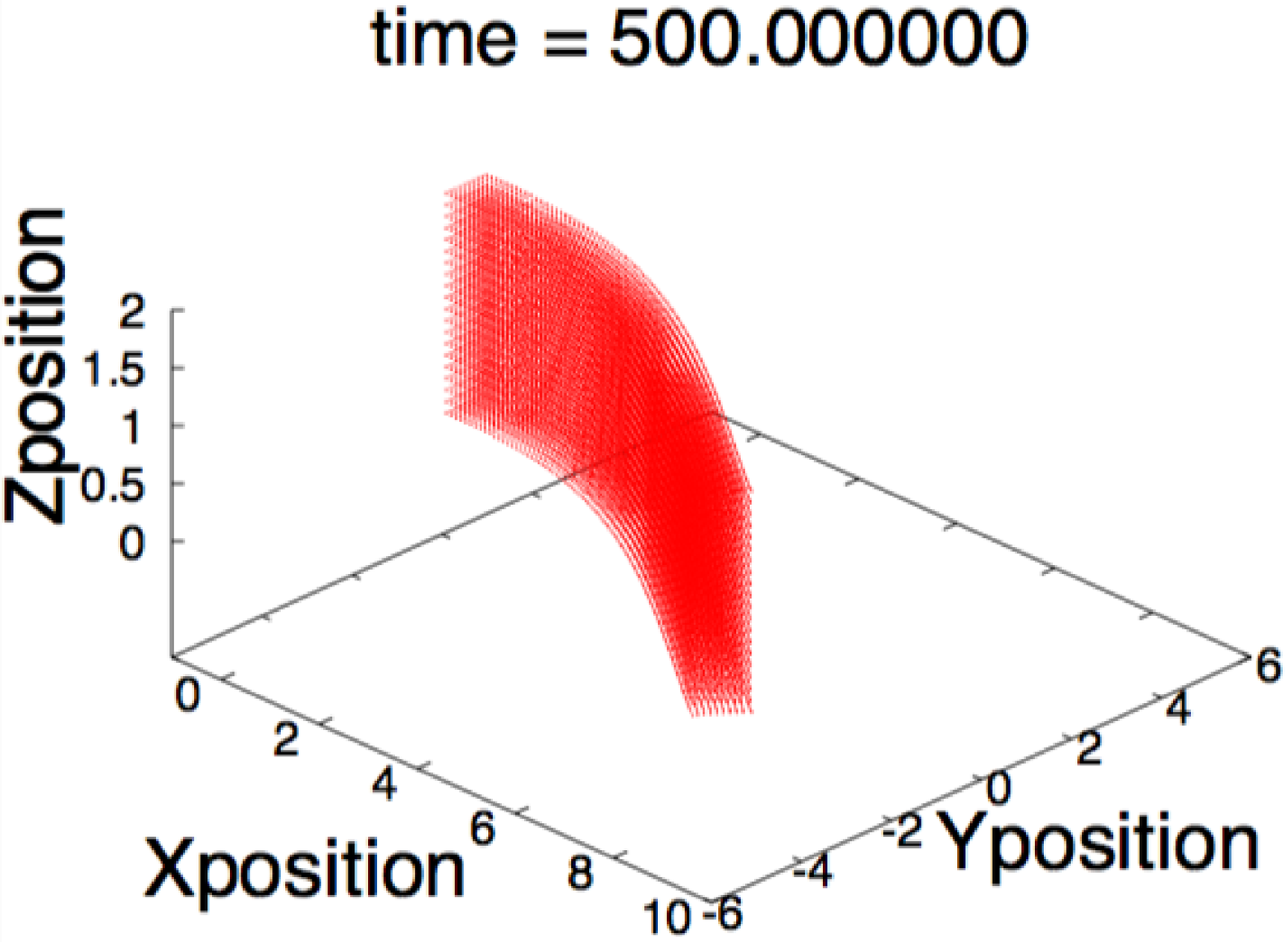}}
    \\
    \subfigure
    {\includegraphics[width=5cm,height=3.75cm]{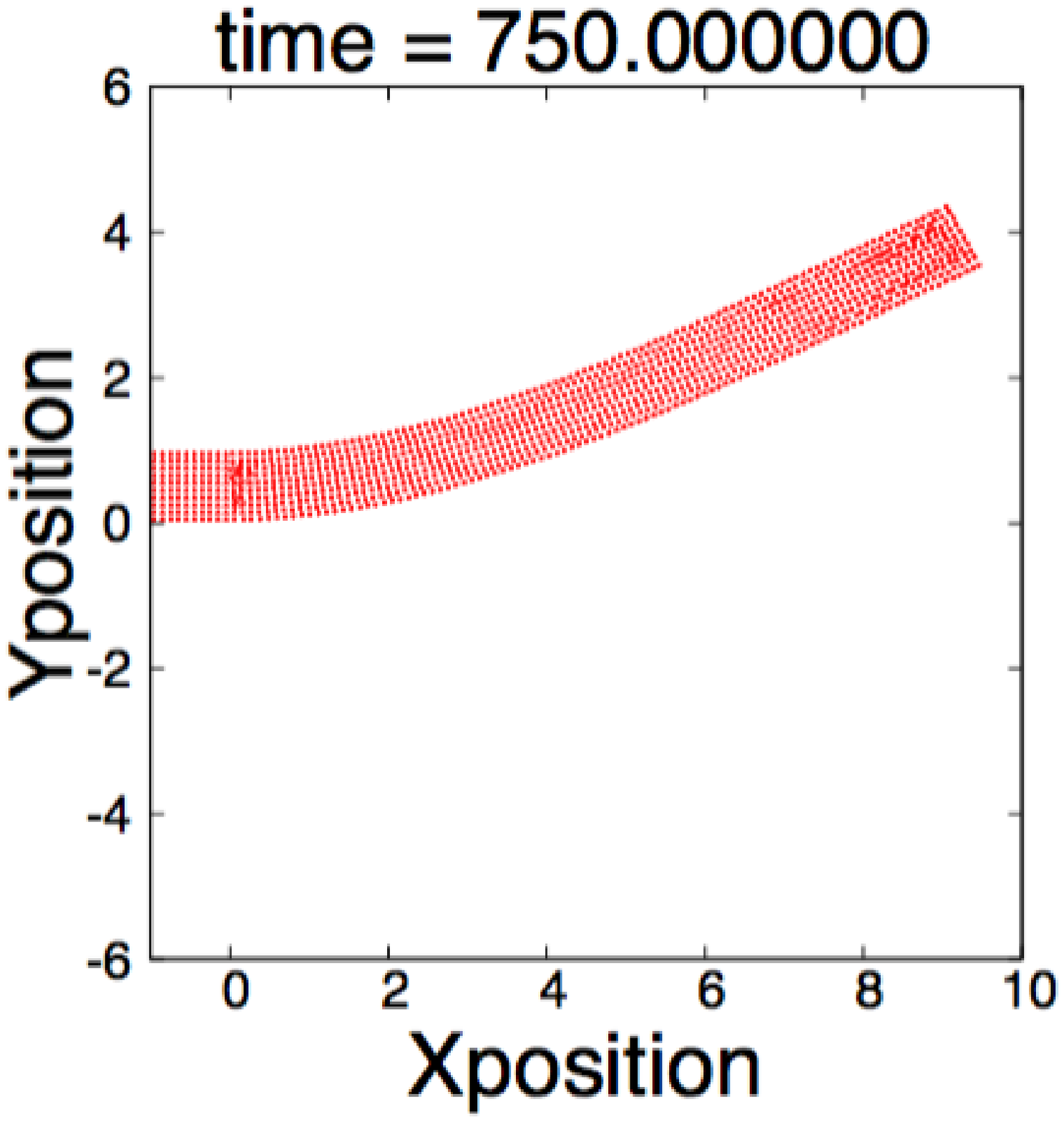}}
    \subfigure
    {\includegraphics[width=5cm,height=3.75cm]{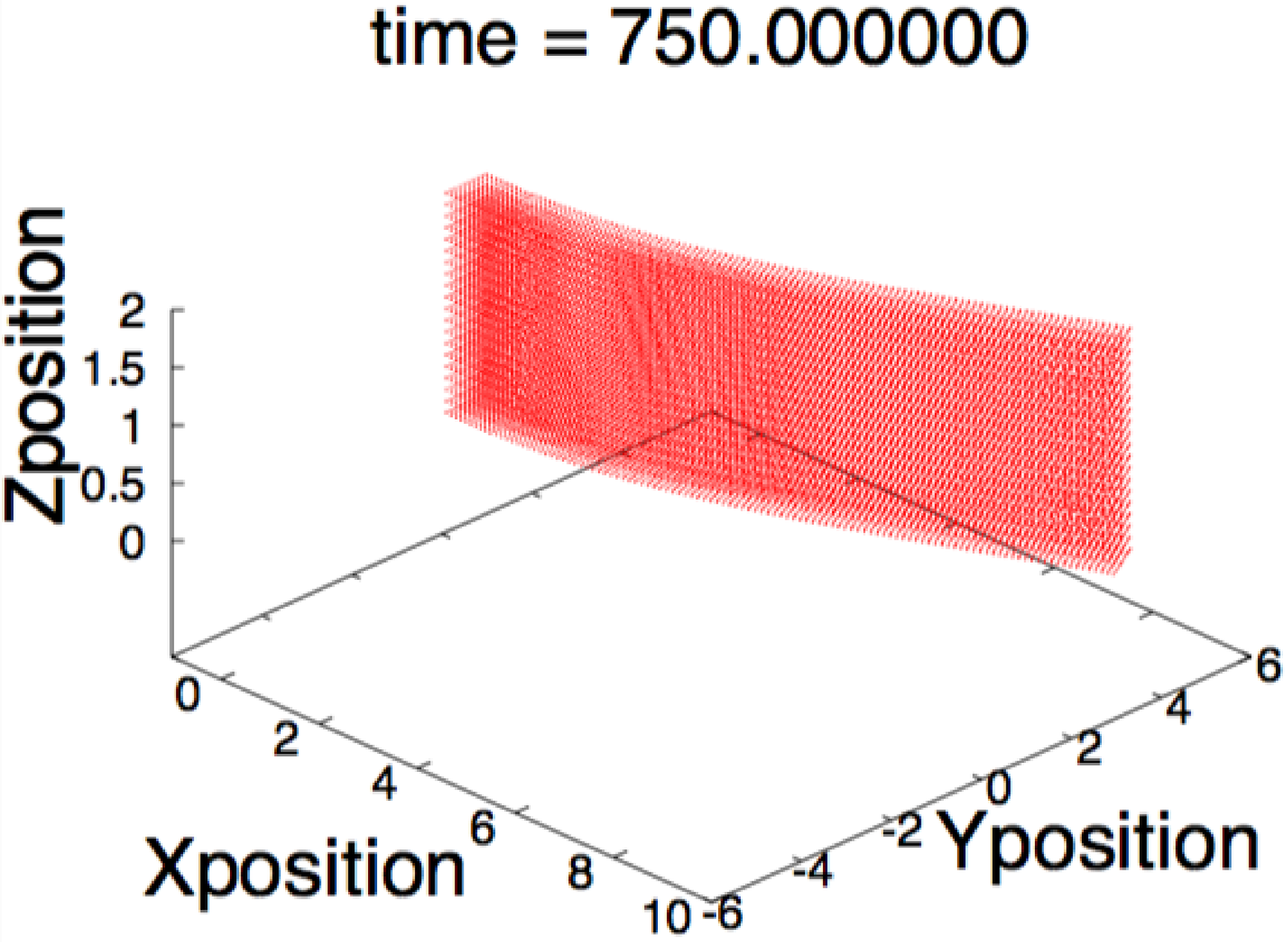}}
    \caption{Result of simulation of the oscillation plate when we select appropriate interpolation method. Left panels show the configurations observed from +$z$ direction, right panels show those observed from oblique direction.}
    \label{Oscillation-plate-3D-0123}
  \end{center}
\end{figure}

\begin{figure}[!htb]
  \begin{center}
    \leavevmode
    \subfigure
    {\includegraphics[width=5cm,height=3.75cm]{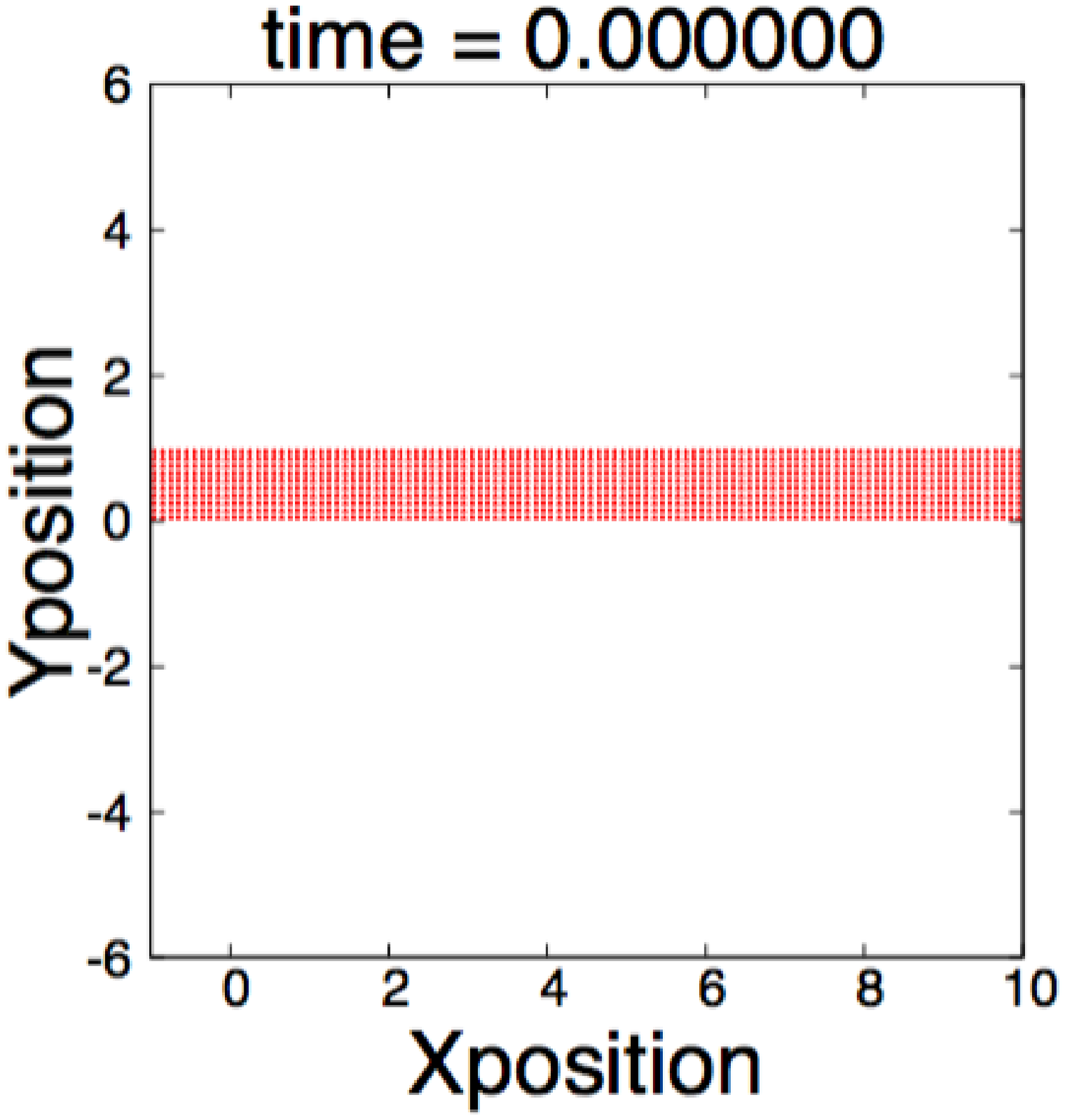}}
    \subfigure
    {\includegraphics[width=5cm,height=3.75cm]{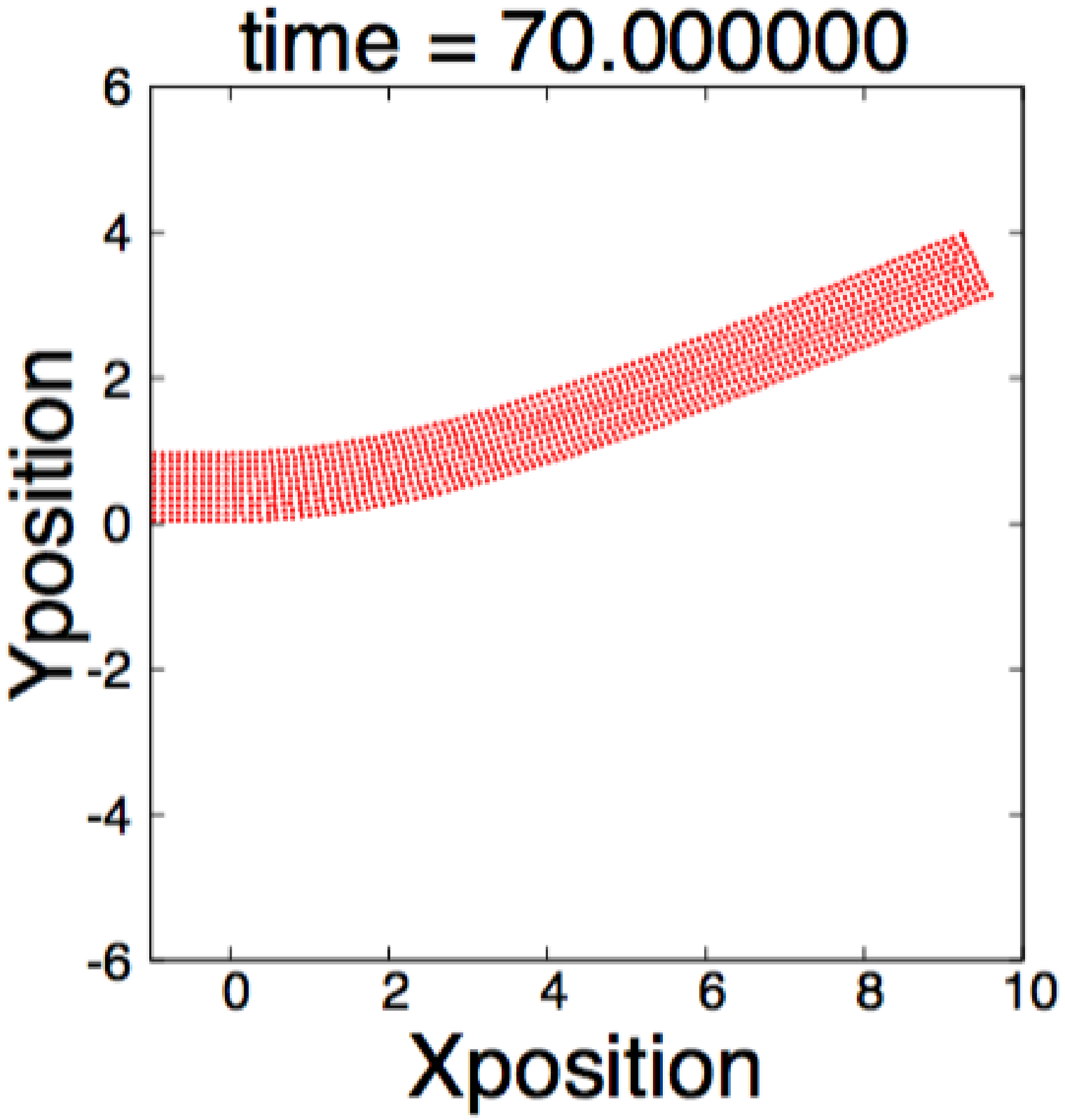}}
    \\
    \subfigure
    {\includegraphics[width=5cm,height=3.75cm]{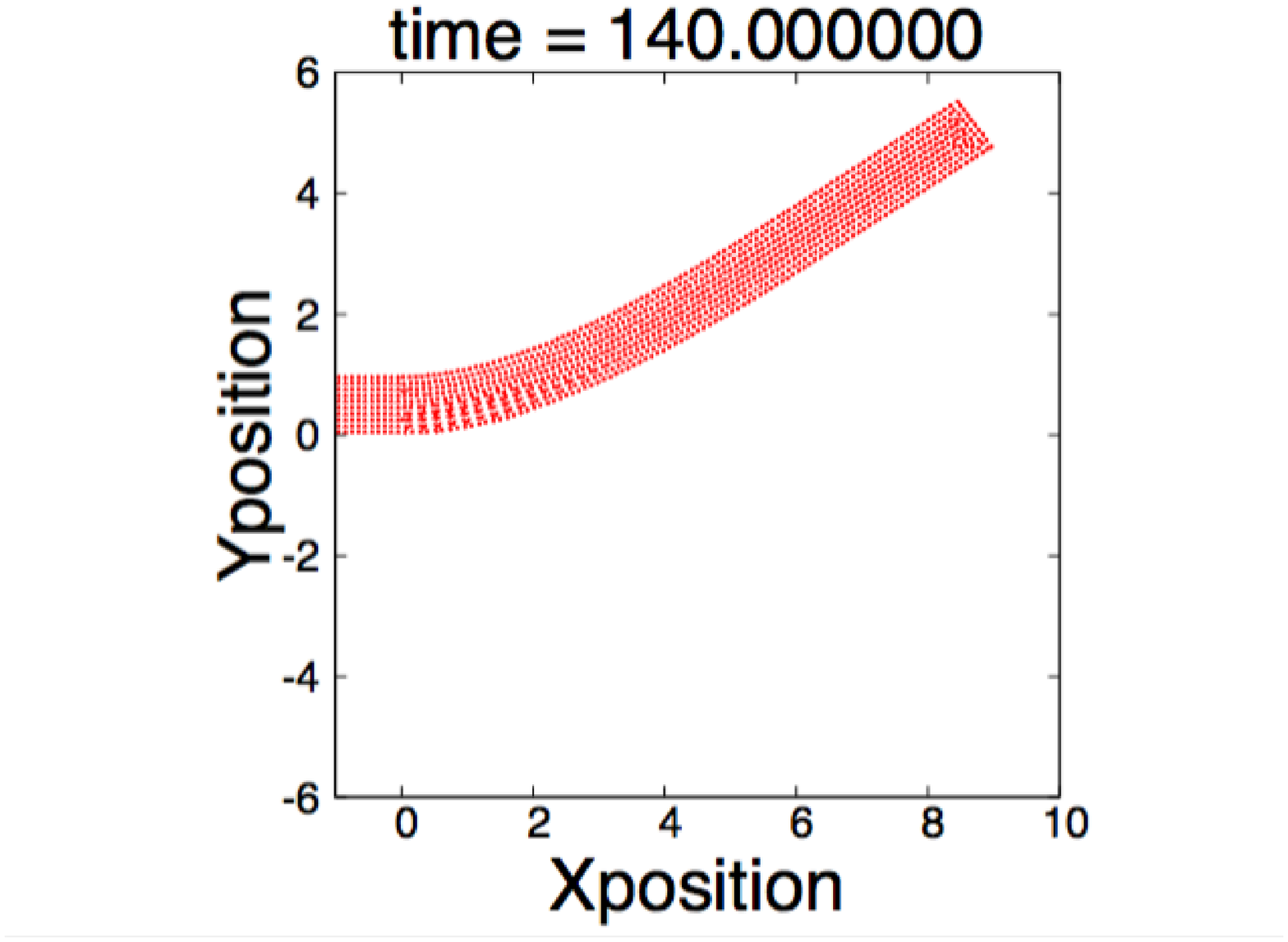}}
    \subfigure
    {\includegraphics[width=5cm,height=3.75cm]{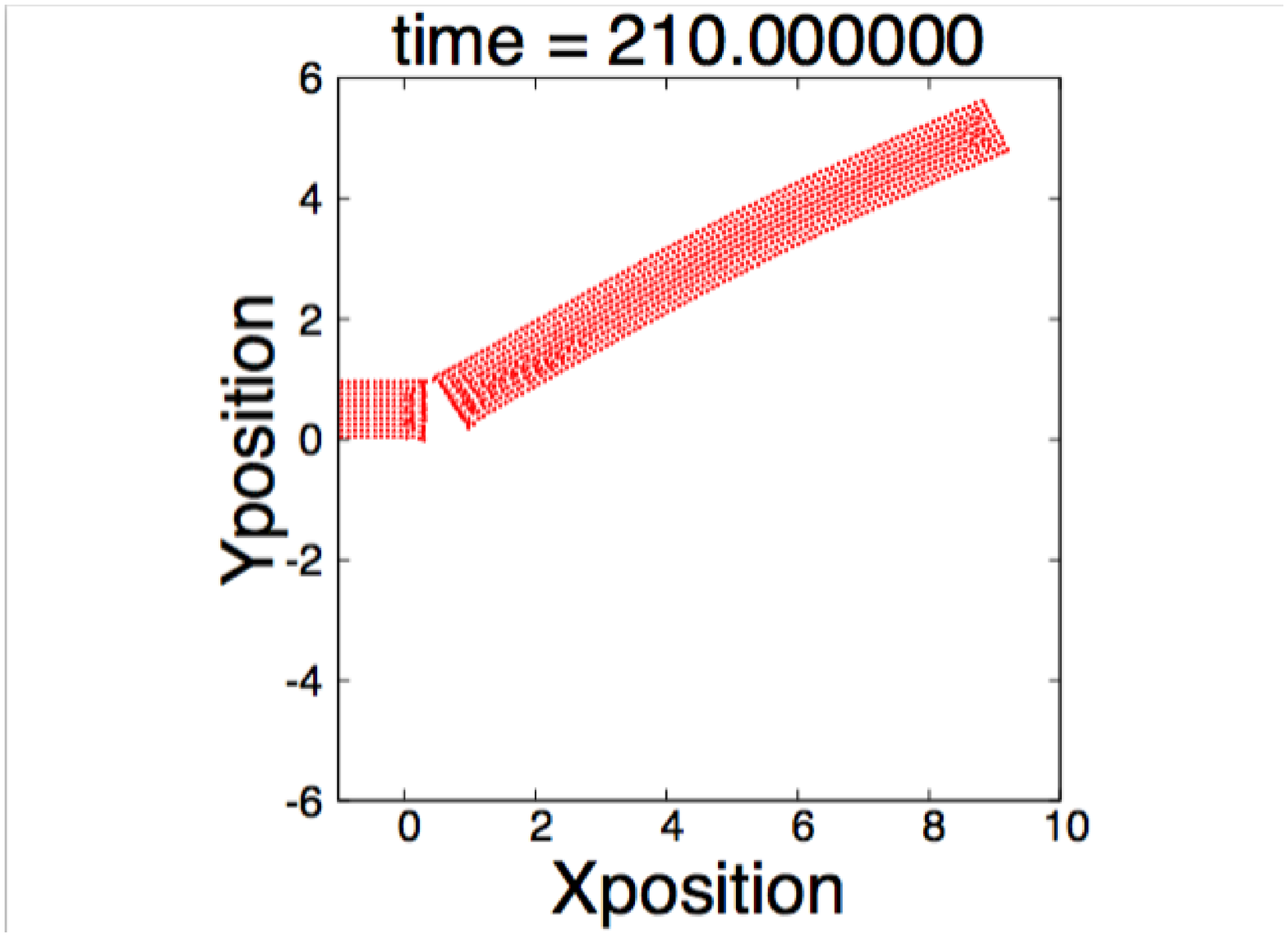}}
    \caption{Result of simulation of the oscillation plate when we use only linear interpolation.}
    \label{Oscillation-plate-3D-linear-0123}
  \end{center}
\end{figure}

Figure \ref{Oscillation-plate-3D-0123} shows the configurations of the plate when we use appropriate interpolation method, and Fig.\,\ref{Oscillation-plate-3D-linear-0123} shows the same configurations but we use only linear interpolation irrespective of the sign of pressure.

From Fig.\,\ref{Oscillation-plate-3D-linear-0123}, when we use only linear interpolation the plate breaks at stretched root, where the pressure becomes negative and linear interpolation is unstable. On the other hand, from Fig.\,\ref{Oscillation-plate-3D-0123}, we can calculate the oscillation stably if the method of this paper is applied. We confirmed that this oscillation continues stably until many periods. 

The artificial stress of \cite{Gray-et-al2001} requires the procedure as follows: first we rotate a frame of reference to diagonalize the stress tensor. Then if each diagonal part is positive (i.e. tensile stress), we added the artificial stress to that part. Finally we rotate again a frame of reference to original coordinate. In this procedure, we need to derive eigenvalue and eigenvector of the stress tensor of each particle. We can derive eigenvalue and eigenvector analytically in two-dimensional case. However, in three dimensions, to derive eigenvalue and eigenvector we have to use numerical method such as Jacobi method \cite{Mathews-and-Fink2004}. In contrast, our method does not require time consuming procedure, and we just need to select appropriate interpolation method.

According to \cite{Landau-Lifshitz1970}, the angular frequency of extremely thin plate is written as,

\begin{equation}
\omega^{2}=\frac{EH^{2}k^{4}}{12\rho (1-\nu^{2})},
\label{omega-oscillation-plate}
\end{equation}

\noindent where $E$ is Young's modulus, $\nu$ is Poisson's ratio. $E$ and $\nu$ are expressed as,

\begin{align}
&E=\frac{9K\mu}{3K+\mu}, \nonumber \\ &\nu = \frac{3K-2\mu}{2(3K+\mu )}, \label{E-and-nu}
\end{align}

\noindent where $K$ is the bulk modulus,

\begin{equation}
\frac{1}{K}=-\frac{1}{V}\Bigl( \frac{\partial V}{\partial P} \Bigr)_{T} = \frac{1}{\rho}\Bigl( \frac{1}{(\partial P / \partial \rho)} \Bigr)_{T}.
\label{K}
\end{equation}

\noindent In the case of EoS of (\ref{EoS-elastic}), $K=\rho_{0} C_{s}^{2}$. The angular frequency of this calculation is $\omega = 0.01201$. Thus analytical period of oscillation of plate in the limit of infinitesimal thickness is the following:

\begin{equation}
T_{{\rm theo}}=\frac{2\pi}{\omega} = 523.2.
\label{period-oscillation-plate}
\end{equation}

\noindent Oscillation period of our simulation is $T_{{\rm sim}} \approx 665$. We expect that the difference between the period observed in our simulations and the period of infinitesimally thin plate decreases with decreasing the ratio of the thickness to the length of plate of simulation. To show this, we additionally conduct the simulations of oscillation of plate with the length of $15H$ and $20H$, and obtain oscillation period for each case. Oscillation period of the plate with the length of $15H$ and $20H$ is $1406$ and $2353$, respectively. The error between theoretical and computed results (=$(T_{{\rm theo}}-T_{{\rm sim}})/T_{{\rm theo}}$) for $L=10H$ is 27.3\%, for $L=15H$ is 19.5\%, and for $L=20H$ is 12.5\%. The error significantly decreases with decreasing the ratio of the thickness to the length of plate.

\subsection{Impact of aluminum sphere on thin aluminum plate}
In \cite{Mehra-et-al2012}, Mehra et al. evaluate the effect of the artificial stress on the tensile instability by conducting simulations of the impact of steel sphere on aluminum plate. According to \cite{Mehra-et-al2012}, the tensile instability of these simulations makes unphysical void at the surface of collision between sphere and plate. They reported that the artificial stress of \cite{Gray-et-al2001} can not suppress the tensile instability of these simulations. Similar calculations in \cite{Mehra-and-Chaturvedi2006} (impact of aluminum sphere on aluminum plate) also observe the void. In this subsection, we conduct the simulation of impact of aluminum sphere on aluminum plate in two dimensions, and evaluate the effect of our method for the tensile instability. Similar to \cite{Mehra-and-Chaturvedi2006}, we use the same material properties as those of \cite{Howell-and-Ball2002}.

In this subsection, we use cgs unit. The radius of aluminum sphere is 0.5 [cm], and the sickness of plate is 0.2 [cm]. Initially we put the sphere and the plate with 0.1 [cm] separation. The velocity of collision is $3.1\times 10^{5} [{\rm cm/s}]$. The particles are put on the square lattice with the side length of 0.02 [cm] within the sphere and the plate. We use stiffened gas EoS of Eq.\,(\ref{stiffened-gas-equation-of-state}). Here $\gamma_{0}$ is the Gruneisen parameter, $C_{s}$ is a bulk sound speed of aluminum, $\rho_{0}$ is a reference density of aluminum. Each value is $\gamma_{0}=2.0,\ C_{s}=5.328 \times 10^{5}[{\rm cm/s}],\ \rho_{0}=2.785 [{\rm g/cm^{3}}]$. The shear modulus of aluminum is $\mu = 2.760 \times 10^{11} [{\rm dyne/cm^{2}}]$.

In this test calculation, average particle spacing largely varies due to hypervelocity impact. Thus we use the variable smoothing length. $C_{{\rm smooth}}$ is set to 2.0 to suppress the tensile instability at negative pressure region caused by variable smoothing length. As explained in Section 3.1, we select the Riemann solver for ideal gas EoS or simple EoS of elastic body using the criterion of Eq.\,(\ref{criterion-for-riemann-solver}). 

To introduce the effect of plasticity of aluminum, we adopt elastic-perfectly plastic model using von Mises yielding criterion \cite{Benz-and-Asphaug1995}. In this model, we limit the deviatoric stress tensor that is used for time evolution equations as,

\begin{equation}
S^{\alpha \beta}_{i} \Rightarrow f_{i}S^{\alpha \beta}_{i},
\label{plastic-Sab}
\end{equation}

\noindent where,

\begin{equation}
f_{i} = \min \Bigl[ \frac{Y_{0}^{2}}{3J_{2,i}}, 1 \Bigr],
\label{plastic-f}
\end{equation}

\noindent $Y_{0}$ is a yielding stress, and $J_{2,i}$ is the second invariant of the deviatoric stress tensor defined as,

\begin{equation}
J_{2,i}=\frac{1}{2}S^{\alpha \beta}_{i}S^{\alpha \beta}_{i}.
\label{plastic-J2}
\end{equation}

\noindent $Y_{0}$ is set to $3.0 \times 10^{9} [{\rm dyne/cm^{2}}]$.

\begin{figure}[!htb]
 \begin{center}
 \includegraphics[width=8cm,height=6cm]{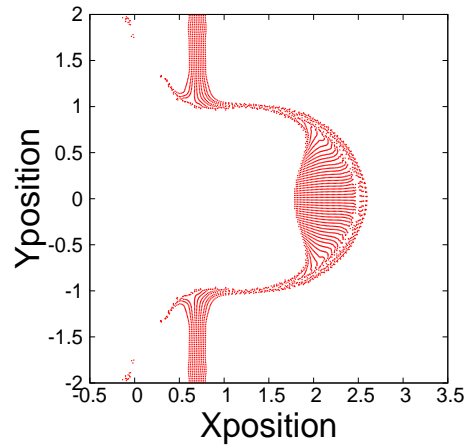}
 \caption{Configuration during simulation of impact between aluminum sphere and aluminum plate. We used appropriate interpolation method depending on the sign of pressure.}
 \label{Al-sphere-Al-plate-collision-2D}
 \end{center}
\end{figure}

\begin{figure}[!htb]
 \begin{center}
 \includegraphics[width=8cm,height=6cm]{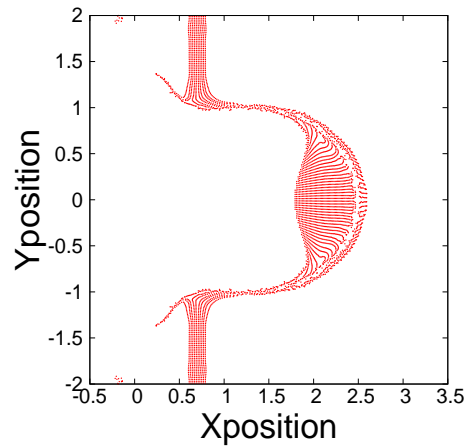}
 \caption{Same as Fig.\,\ref{Al-sphere-Al-plate-collision-2D}, but for only linear interpolation.}
 \label{Al-sphere-Al-plate-collision-2D-linear}
 \end{center}
\end{figure}

\begin{figure}[!htb]
 \begin{center}
 \includegraphics[width=8cm,height=6cm]{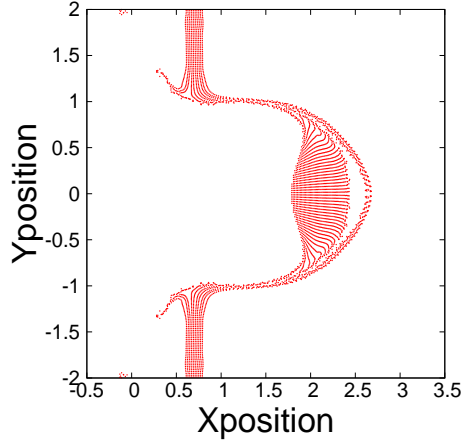}
 \caption{Same as Fig.\,\ref{Al-sphere-Al-plate-collision-2D}, but for only cubic spline interpolation.}
 \label{Al-sphere-Al-plate-collision-2D-cubic}
 \end{center}
\end{figure}

Figure \ref{Al-sphere-Al-plate-collision-2D} shows the result of calculation when we use appropriate interpolation depending on the sign of pressure. Figure \ref{Al-sphere-Al-plate-collision-2D-linear} shows the result when we use only linear interpolation independent of the sign of pressure, and Fig.\,\ref{Al-sphere-Al-plate-collision-2D-cubic} shows that when we use only cubic spline interpolation. All results are plotted at $t=8 \mu {\rm s}$.

As we can notice from Fig.\,\ref{Al-sphere-Al-plate-collision-2D} and \ref{Al-sphere-Al-plate-collision-2D-linear}, if we select interpolation or use only linear interpolation, there is no void at the surface of collision. Fig.\,\ref{Al-sphere-Al-plate-collision-2D-cubic} shows the appearance of void in the case of cubic spline interpolation. Actually voids appear in the compressed regions where the pressure is positive. This is not surprising since cubic spline interpolation in two dimensions is known to be unstable in the positive pressure regime. In \cite{Dehnen-and-Aly2012}, instability in compressed region is called pairing instability.

\begin{figure}[!htb]
 \begin{center}
 \includegraphics[width=8cm,height=6cm]{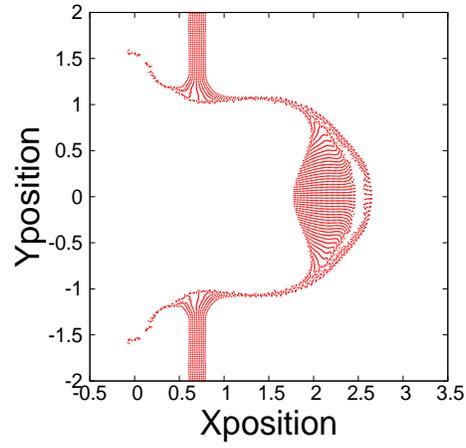}
 \caption{The same as Fig.\,\ref{Al-sphere-Al-plate-collision-2D}, but without modified monotonicity constraint of Eq.\,(\ref{monotonicity-constraint}).}
 \label{Al-sphere-Al-plate-collision-2D-org-monotonicity}
 \end{center}
\end{figure}

\begin{figure}[!htb]
 \begin{center}
 \includegraphics[width=8cm,height=6cm]{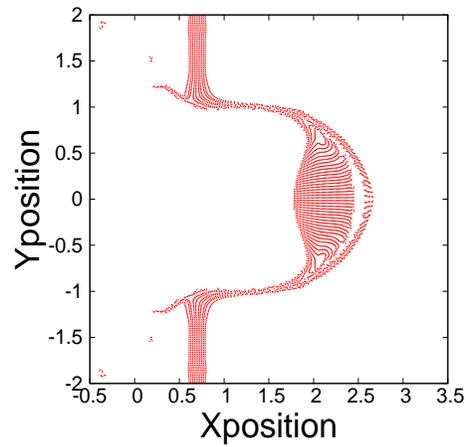}
 \caption{The same as Fig.\,\ref{Al-sphere-Al-plate-collision-2D}, but we use constant smoothing length.}
 \label{Al-sphere-Al-plate-collision-2D-const-h}
 \end{center}
\end{figure}

To do a reasonable numerical simulation with Godunov SPH method, we need not only to use appropriate interpolation, but also to use an appropriate monotonicity constraint and smoothing length. To show the importance of using an appropriate monotonicity constraint, we calculate the same simulation without the modified monotonicity constraint of Eq.\,(\ref{monotonicity-constraint}). In addition, to investigate the importance of using the appropriate smoothing length, we conduct the simulation using constant smoothing length with $h=0.02$ [cm]. Here, in both simulations, we select interpolation method depending on the sign of pressure as in Fig.\,\ref{Al-sphere-Al-plate-collision-2D}. Figure \ref{Al-sphere-Al-plate-collision-2D-org-monotonicity} shows the result without modified monotonicity constraint, and Fig.\,\ref{Al-sphere-Al-plate-collision-2D-const-h} shows that with constant smoothing length.

In both cases of Fig.\,\ref{Al-sphere-Al-plate-collision-2D-org-monotonicity} and \ref{Al-sphere-Al-plate-collision-2D-const-h}, we can see small void. The pairing instability in the positive pressure is essentially caused when the particle spacing is much smaller than the smoothing length. In that case particles can not push back each other, and result in clustering. 

According to the test calculations of \cite{Mehra-et-al2012}, the void is created in the case of the standard SPH method with general artificial viscosity. This implies that numerical dissipation due to artificial viscosity term is not sufficient to prevent the pairing instability at surface of collision. Dissipation due to the Riemann solver becomes strong depending on the strength of resultant shock wave. Therefore, as pointed out by \cite{Mehra-et-al2012}, Godunov-type scheme is effective for pairing instability at the surface of collision.

\subsection{Calculation of restitution coefficient}
Finally, to show that our Godunov SPH method for elastic dynamics can be used for describing practical experiments, we calculate the restitution coefficient in the impact of steel sphere on steel plate.

Aryaei et al. \cite{Aryaei-et-al2010} measure the restitution coefficient by dropping steel or aluminum sphere on steel or aluminum plate, and investigate the dependence of sphere diameter on the restitution coefficient. The restitution coefficient is calculated from height that spheres jump up. As a result, they find that the restitution coefficient is decreasing with increasing sphere diameter. They also analyze the restitution coefficient by Finite Element Method and show the same dependence.

In this subsection, we simulate the impact of various-size steel spheres on steel plate with the Godunov SPH method for elastic dynamics. In the experiment of \cite{Aryaei-et-al2010}, Aryaei et al. drop spheres from the height of $H=150[{\rm cm}]$, so that the impact velocity becomes $v=\sqrt{2gH}=542.2[{\rm cm/s}]$, where $g=980[{\rm cm/s^{2}}]$ is the gravitational acceleration. Thus we set the initial velocity of spheres to this value, and follow the motion of sphere from just before the impact until just after the impact. We only calculate the head-on collision between sphere and plate. We ignore the gravity of the Earth because timescale of the impact is very short. Initially we put sphere and plate with the separation of four times larger than the smoothing length, and derive the restitution coefficient by the velocity when sphere comes back to the initial position after rebound. In the case of Gaussian-type kernel function, we can ignore interactions between the pair of SPH particles that have separation larger than four times of the smoothing length. Thus the velocity of sphere sufficiently converges when sphere comes back to initial position. Here, the velocity of sphere is calculated by averaging the velocity of SPH particles that constitute sphere. 

The size of steel plate is set to $0.21[{\rm cm}]\times 1.2[{\rm cm}] \times 1.2[{\rm cm}]$. In the experiment of \cite{Aryaei-et-al2010}, the bottom of plate is fixed by frame. To reproduce this condition, we fix three layers of SPH particles from the bottom of plate.

In the calculation of Finite Element Method of \cite{Aryaei-et-al2010}, the number of element for sphere is fixed independent of the size of sphere. Thus we also use the same number of particles for every size of spheres. SPH particles are put on the square lattice with the side length of $R/20$, where $R$ represents the radius of sphere. In other words, we put twenty particles along the radial direction. 

In this subsection, we use constant smoothing length with $h=R/20$, and use EoS of Eq.\,(\ref{EoS-elastic}). We can find material density, Young's modulus $E$ and Poisson's ratio $\nu$ of steel in \cite{Aryaei-et-al2010}. Reference density for EoS $\rho_{0}$ is set to material density of steel, $\rho_{0}=7.57[{\rm g/cm^{3}}]$. Sound speed for EoS $C_{s}$ is calculated from Young's modulus and Poisson's ratio as,

\begin{align}
& C_{s}=\sqrt{\frac{K}{\rho_{0}}}, \nonumber \\ &K=\frac{E}{3(1-2\nu)}, \label{Cs-elastic-EoS-RC}
\end{align}

\noindent where $K$ is bulk modulus. The value of $C_{s}$ becomes $4.71\times 10^{5}[{\rm cm/s}]$. We also use the Riemann solver for Eq.\,(\ref{EoS-elastic}). 

Shear modulus $\mu$ is calculated from Young's modulus and Poisson's ratio as,

\begin{equation}
\mu=\frac{E}{2(1+\nu)}.
\label{shear-modulus-RC}
\end{equation}

\noindent The value of shear modulus becomes $8.00\times 10^{11}[{\rm dyne/cm^{2}}]$.

Plastic deformation plays an important role when the restitution coefficient is determined. Energy is dissipated by plastic deformation, and the restitution coefficient becomes small. In this subsection we adopt elastic-perfectly plastic model. Yielding stress $Y$ is set to $4.50\times 10^{9}[{\rm dyne/cm^{2}}]$, and we reduce deviatoric stress tensor using Eqs\,(\ref{plastic-Sab}) and (\ref{plastic-f}).

In general, tension does not work between different solids, and the same is true for shear force if we ignore friction. In this simulation tension should not work between sphere and plate. Previous test calculations ignore this point, but in this subsection we consider about this point to determine the restitution coefficient correctly. When we calculate the force between particles that consist sphere and particles that consist plate, we permit only the repulsive force along the line joining two particles. In particular, acceleration of the $i$-th particle exerted by the $j$-th particle $\bm{a}_{ij}$ is calculated as,

\begin{equation}
\bm{a}_{ij} = -2m_{j}(P_{ij}^{\ast}-S_{ij}^{ss\ast})V_{ij}^{2}(h)\frac{\partial}{\partial \bm{r}_{i}}W(\bm{r}_{i}-\bm{r}_{j},\sqrt{2}h),  
\label{acceleration-for-other-structure}
\end{equation}

\noindent if the $i$-th and $j$-th particle represent different solid (sphere or plate). Here, $S_{ij}^{ss\ast}$ is $\bm{r}_{i}-\bm{r}_{j}$ direction component of $S_{ij}^{\alpha \beta \ast}$ of Eq.\,(\ref{Sabij-ast}), and

\begin{align}
&P_{ij}^{\ast}=0 \ \ {\rm if} \ \ P_{ij}^{\ast}<0,\nonumber \\ &S_{ij}^{ss\ast}=0 \ \ {\rm if} \ \ S_{ij}^{ss\ast}>0. \label{Pijast-and-Sssijast-for-other-structure}
\end{align}

\begin{figure}[!htb]
  \begin{center}
    \includegraphics[width=10cm,height=7cm]{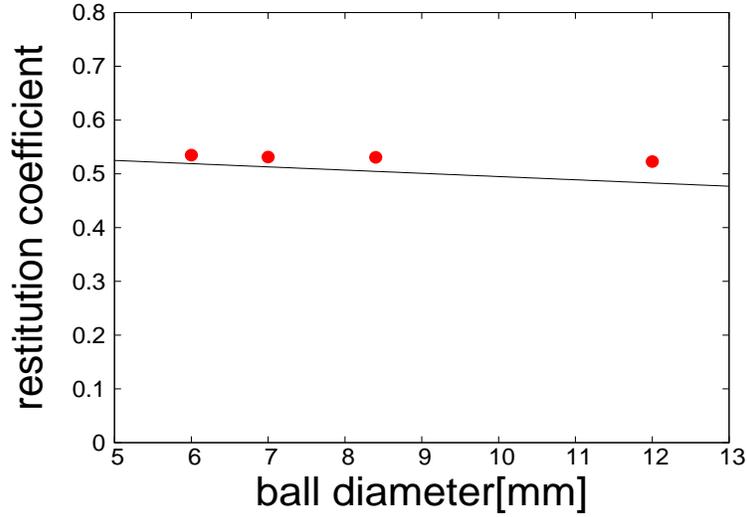}
    \caption{Ball size dependence on the restitution coefficient calculated by the Godunov SPH method for elastic dynamics. Horizontal axis shows the diameter of sphere, and the vertical axis shows the restitution coefficient. Red circles show the result of our simulation, and solid line shows the result obtained by the experiments of \cite{Aryaei-et-al2010}}
    \label{RC-summary}
  \end{center}
\end{figure}

Figure \ref{RC-summary} shows ball size dependence on the restitution coefficient. From this figure we notice that the restitution coefficient is decreasing with increasing sphere diameter even with our method. The slope is $-0.019 [{\rm mm}^{-1}]$. The solid line in Fig.\,\ref{RC-summary} shows the result obtained by experiments of \cite{Aryaei-et-al2010}. We can notice that our result agrees well with the result of experiments. In \cite{Aryaei-et-al2010}, Aryaet et al. also numerically calculate the restitution coefficient using Finite Element Method. Note that the result of their calculation does not seem to provide monotonically decreasing of the restitution coefficient while increasing the diameter of sphere.

The restitution coefficient decreases with increasing sphere diameter because the mass of sphere increases. If the mass increases, force applied to surface of collision becomes large and plastic deformation becomes large. In that case energy dissipation by plastic deformation increases, so that the restitution coefficient decreases.

Although we need to examine the validity of plastic model or parameters such as shear modulus, simulations with the Godunov SPH method for elastic dynamics seem to reproduce the result of experiments reasonably well. 

\section{Summary}
In this paper, we extended the Godunov SPH method to elastic dynamics. On the basis of the formulation of the Godunov SPH method, we formulate the equation of motion, the equation of energy and the time evolution equation of the deviatoric stress tensor. We confirmed that these formulated equations achieve the second-order accuracy in space by convergence test. Moreover, we develop the method to handle the Riemann solver for non-ideal gas equation of state. Next, we apply the stabilizing method for the tensile instability of \cite{Sugiura-and-Inutsuka2016} to elastic dynamics, and conduct several test calculations such as rubber rings collision, oscillation plate and impact of sphere on plate to evaluate the validity of our method. We confirmed that the method to suppress the tensile instability using the Godunov SPH method for  hydrodynamics equations developed by \cite{Sugiura-and-Inutsuka2016} is also valid for elastic dynamics equations. This stabilizing method is selecting appropriate interpolation method for $V_{ij}^{2}$ depending on the sign of pressure. The results show that if we select appropriate interpolation method for $V_{ij}^{2}$ we can calculate stably. To suppress the tensile instability in the calculation of hypervelocity impact, we should also consider about monotonicity constraint or the way to treat the smoothing length, and we confirmed that Godunov-type scheme is valid for such problems. We hope that we can use our method to solve various problems in elastic dynamics.

\section*{Acknowledgement}
The authors thank Hiroshi Kobayashi, Kazunari Iwasaki, Yusuke Tsukamoto for useful discussions and comments. SI is supported by Grant-in-Aid for Scientific Research (23244027, 23103005). Simulations in Section 4.6 were carried out on PC cluster at Center for Computational Astrophysics, National Astronomical Observatory of Japan.

\section*{References}

\bibliography{mybibfile}

\begin{thebibliography}{10}
\expandafter\ifx\csname url\endcsname\relax
  \def\url#1{\texttt{#1}}\fi
\expandafter\ifx\csname urlprefix\endcsname\relax\def\urlprefix{URL }\fi
\expandafter\ifx\csname href\endcsname\relax
  \def\href#1#2{#2} \def\path#1{#1}\fi

\bibitem{Lucy1977}
L.~B. Lucy, A numerical approach to the testing of the fission hypothesis, AJ
  82 (1977) 1013--1024.

\bibitem{Gingold-and-Monaghan1977}
R.~A. Gingold, J.~J. Monaghan, Smoothed particle hydrodynamics: theory and
  application to non-spherical stars, MNRAS 181 (1977) 375--389.

\bibitem{Monaghan1992}
J.~J. Monaghan, Smoothed particle hydrodynamics, Annu. Rev. Astron. Astrophys.
  30 (1992) 543--574.

\bibitem{Benz-and-Asphaug1995}
W.~Benz, E.~Asphaug, Simulations of brittle solids using smooth particle
  hydrodynamics, Computat. Phys. Comm. 87 (1995) 253--265.

\bibitem{Benz-and-Asphaug1999}
W.~Benz, E.~Asphaug, Catastrophic disruptions revisited, Icarus 142 (1999)
  5--20.

\bibitem{Swegle-et-al1995}
J.~W. Swegle, D.~L. Hicks, S.~W. Attaway, Smoothed particle hydrodynamics
  stability analysis, J. Comput. Phys 116 (1995) 123--134.

\bibitem{Morris1996}
J.~P. Morris, A study of stability properties of smoothed particle
  hydrodynamics, Publ. Astron. Soc. Aust. 13 (1996) 97--102.

\bibitem{Iwasaki2015}
K.~Iwasaki, Minimizing dispersive errors in smoothed particle
  magnetohydrodynamics for strongly magnetize medium, J. Comput. Phys. 302
  (2015) 359--373.

\bibitem{Dehnen-and-Aly2012}
W.~Dehnen, H.~Aly, Improving convergence in smoothed particle hydrodynamics
  simulations without pairing instability, MNRAS 425 (2012) 1068--1082.

\bibitem{Gray-et-al2001}
J.~P. Gray, J.~J. Monaghan, R.~P. Swift, {SPH} elastic dynamics, Methods Appl.
  Engrg. 190 (2001) 6641--6662.

\bibitem{Randles-and-Libersky1996}
P.~W. Randles, L.~D. Libersky, Smoothed particle hydrodynamics: Some recent
  improvements and applications, Comput. Meth. Appl. Mech. Eng. 139 (1996)
  375--408.

\bibitem{Johnson-and-Beissel1996}
G.~R. Johnson, S.~R. Beissel, Normalized smoothing functions for {SPH} impact
  computations, Int. J. Num. Methods. Eng. 39 (1996) 2725--2741.

\bibitem{Monaghan1999}
J.~J. Monaghan, {SPH} without a tensile instability, J. Comput. Phys. 159
  (1999) 290--311.

\bibitem{Mehra-et-al2012}
V.~Mehra, C.~D. Sijoy, V.~Mishra, S.~Chaturvedi, Tensile instability and
  artificial stresses in impact problems in {SPH}, Journal of Physics:
  Conference Series 377 (2012) 012102.

\bibitem{Sugiura-and-Inutsuka2016}
K.~Sugiura, S.~Inutsuka, An extension of {G}odunov {SPH}: {A}pplication to
  negative pressure media, J. Comput. Phys. 308 (2016) 171--197.

\bibitem{Inutsuka2002}
S.~Inutsuka, Reformulation of smoothed particle hydrodynamics with riemann
  solver, J. Comput. Phys. 179 (2002) 238--267.

\bibitem{Genda-et-al2015}
H.~Genda, T.~Fujita, H.~Kobayashi, H.~Tanaka, T.~Abe, Resolution dependence of
  disruptive collisions between planetesimals in the gravity regime, Icarus 262
  (2015) 58--66.

\bibitem{Mehra-and-Chaturvedi2006}
V.~Mehra, S.~Chaturvedi, High velocity impact of metal sphere on thin metallic
  plates: a comparative smooth particle hydrodynamics study, J. Comput. Phys.
  212 (2006) 318--337.

\bibitem{Tillotson1962}
J.~H. Tillotson, Metallic equations of state for hypervelocity impact, General
  Atomic Report GA-3216.

\bibitem{Iwasaki-and-Inutsuka2011}
K.~Iwasaki, S.~Inutsuka, Smoothed particle magnetohydrodynamics with {R}iemann
  solver and the method of characteristics, Mon. Not. R. Astron. Soc. 418
  (2011) 1668--1688.

\bibitem{Monaghan1988}
J.~J. Monaghan, An introduction to {SPH}, Comput. Phys. Comm. 48 (1988) 89--96.

\bibitem{Price-and-Monaghan2004}
D.~J. Price, J.~J. Monaghan, Smoothed particle magnetohydrodynamics - {II}.
  {V}ariational principles and variable smoothing-length terms, Mon. Not. R.
  Astron. Soc. 348 (2004) 139--152.

\bibitem{Bonet-and-Lok1999}
J.~Bonet, T.~S.~L. Lok, Variational and momentum preservation aspects of smooth
  particle hydrodynamics formulations, Comput. Methods Appl. Mech. Engrg. 180
  (1999) 97--115.

\bibitem{Leer1978}
B.~V. Leer, Towards the ultimate conservative difference scheme. {V}. a
  second-order sequel to godunov's method, J. Comput. Phys 32 (1978) 101--136.

\bibitem{Puri-and-Ramachandran2014}
K.~Puri, P.~Ramachandran, Approximate {R}iemann solvers for the {G}odunov {SPH}
  ({GSPH}), J. Comput. Phys. 270 (2014) 432--458.

\bibitem{Einfeldt-et-al1991}
B.~Einfeldt, C.~D. Munz, P.~L. Roe, B.~Sj{\"o}green, On {G}odunov type-methods
  near low densities, J. Comput. Phys. 92 (1991) 273--295.

\bibitem{Hernquist-and-Katz1989}
B.~V. Leer, {TREE-SPH} - a unification of {SPH} with the hierarchical tree
  method, Astrophys. J. Supple. 70 (1989) 419--446.

\bibitem{Price2012}
D.~J. Price, Smoothed particle hydrodynamics and magnetohydrodynamics, J.
  Comput. Phys. 231 (2012) 759--794.

\bibitem{Landau-Lifshitz1970}
L.~D. Landau, E.~M. Lifshitz, Elasticity, course of theoretical physics, vol.
  7, Pergamon Press, Oxford.

\bibitem{Mathews-and-Fink2004}
J.~H. Mathews, K.~K. Fink, Numerical methods using matlab, 4th edition, chap.
  11, Prentice-Hall Inc.

\bibitem{Howell-and-Ball2002}
B.~P. Howell, G.~J. Ball, A free-lagrange augmented godunov method for the
  simulation of elastic-plastic solids, J. Comput. Phys. 175 (2002) 128--167.

\bibitem{Aryaei-et-al2010}
A.~Aryaei, K.~Hashemnia, K.~Jafarpur, Experimental and numerical study of ball
  size effect on restitution coefficient in low velocity impacts, Int. J.
  Impact Eng. 37 (2010) 1037--1044.

\end{thebibliography}

\section*{Appendix A}
In this Appendix, we conduct linear stability analysis of the Godunov SPH method for elastic dynamics. Particle spacing is affected by longitudinal wave, and instability of longitudinal wave causes the tensile instability. Thus we conduct the linear stability analysis for longitudinal wave. We neglect discretization in the direction of time and assume infinitely-accurate time integration, because the tensile instability does not depend on time integration method. We assume that the mass of each particle $m$ is the same for all particles. Constant smoothing length is considered. We conduct linear stability analysis for variable smoothing length in Appendix B. To separate the effect of viscosity and the tensile instability, we do not use the Riemann solver for $P_{ij}^{\ast}$, but assume $P_{ij}^{\ast}=(P_{i}+P_{j})/2$.

In unperturbed state, the particles are put on the square lattice with the side length of $\Delta x$. This unperturbed position is expressed as,

\begin{equation}
\overline{\bm{r}}_{i}=(\overline{x}_{i},\overline{y}_{i},\overline{z}_{i}). \tag{A1}
\label{eq-A1}
\end{equation}

\noindent We add the perturbation to the component of $x$-direction. Perturbed positions of particles are written as,

\begin{align}
&\bm{r}_{i}=(\overline{x}_{i}+\delta x_{i},\overline{y}_{i},\overline{z}_{i}), \nonumber \\ &\delta x_{i}=\epsilon_{x} \exp [i(k\overline{x}_{i}-\omega t)], \tag{A2} \label{eq-A2}
\end{align}

\noindent where $\epsilon_{x}$ is infinitesimal constant, $k$ and $\omega$ represents wave number and angular frequency of perturbation respectively, $i$ that is not subscript shows imaginary unit. Hereafter, $\epsilon$ represents infinitesimal constant, and we neglect second or higher order of infinitesimal values. 

From Eq.\,(\ref{eq-A2}) and $\dot{\bm{r}}_{i}=\bm{v}_{i}$, the velocity of the $i$-th particle becomes,

\begin{equation}
\bm{v}_{i}=(-i\omega \delta x_{i},0,0). \tag{A3}
\label{eq-A3}
\end{equation}

We define the density in unperturbed state as $\overline{\rho}$, and we write the density of the $i$-th particle as,

\begin{align}
&\rho_{i}=\overline{\rho}+\delta \rho_{i},\nonumber \\ &\delta \rho_{i}=\epsilon_{\rho}\exp [i(k\overline{x}_{i}-\omega t)]. \tag{A4} \label{eq-A4}
\end{align}

\noindent From Eq.\,(\ref{Godunov-EoC}), we can write $\delta \rho_{i}$ using $\delta x_{i}$ as,

\begin{align}
&\delta \rho_{i}=-i\overline{\rho}D\delta x_{i}, \nonumber \\ & D\equiv \sum_{j}-\sin [k(\overline{x}_{i}-\overline{x}_{j})]\frac{\partial}{\partial \overline{x}_{i}}W(\overline{\bm{r}}_{i}-\overline{\bm{r}}_{j},h)\frac{m}{\overline{\rho}}. \tag{A5} \label{eq-A5}
\end{align}

\noindent From Eqs.\,(\ref{eq-A4}) and (\ref{eq-A5}), density is represented as $\rho_{i}=\overline{\rho}(1-iD\delta x_{i})$. Note that this representation of density is the same as that is calculated by Eq.\,(\ref{SPH-density}) as shown in Appendix B of \cite{Sugiura-and-Inutsuka2016}. Therefore, the stability does not change even if we calculate density by Eq.\,(\ref{SPH-density}) or we use time-evolved density by Eq.\,(\ref{Godunov-EoC}).

The pressure of the $i$-th particle is represented as $P_{i}=\overline{P}+\delta P_{i}=\overline{P}+\overline{C_{s}}^{2}\delta \rho_{i}=\overline{P}-i\overline{C_{s}}^{2}\overline{\rho}D\delta x_{i}$, where $\overline{P}$ and $\overline{C_{s}}$ represents the pressure and the sound speed in unperturbed state respectively. 

Only $x$-direction component of the acceleration is not $0$. Thus we have to consider only the $x$ component. As shown in Eq.\,(\ref{Godunov-EoM-2}), the equation of motion of the Godunov SPH method for elastic dynamics includes $\sigma_{ij}^{x y\ast},\sigma_{ij}^{x z\ast}$. However, we only have to focus on $\sigma_{ij}^{xx\ast}$ because the terms that include $\sigma_{ij}^{x y\ast},\sigma_{ij}^{x z\ast}$ vanish if we take the summation over $y$- or $z$-direction. Moreover, only $xx$ component of $\dot{\epsilon}_{\rho ,ij}^{\alpha \beta}$ exist and all components of $R_{\rho ,ij}^{\alpha \beta}$ are $0$. Linearized $\dot{\epsilon}_{\rho ,ij}^{xx}$ is written as,

\begin{equation}
\dot{\epsilon}_{\rho ,ij}^{xx} = \frac{m}{\overline{\rho}^{2}}(v_{i}^{x}-v_{j}^{x})\frac{\partial}{\partial \overline{x}_{i}}W(\overline{\bm{r}}_{i}-\overline{\bm{r}}_{j},\sqrt{2}h). \tag{A6}
\label{eq-A6}
\end{equation}

We define $xx$ component of the deviatoric stress tensor in unperturbed state as $\overline{S^{xx}}$, and we write $S^{xx}/\rho$ of the $i$-th particle as,

\begin{align}
&\Bigl( \frac{S^{xx}}{\rho} \Bigr)_{i}=\frac{\overline{S^{xx}}}{\overline{\rho}} +\delta \Bigl( \frac{S^{xx}}{\rho} \Bigr)_{i},\nonumber \\ &\delta \Bigl( \frac{S^{xx}}{\rho} \Bigr)_{i} = \epsilon_{S/\rho}\exp [i(k\overline{x}_{i}-\omega t)]. \tag{A7} \label{eq-A7}
\end{align}

\noindent We substitute Eqs.\,(\ref{eq-A6}) and (\ref{eq-A7}) into Eq.\,(\ref{Godunov-dSab_rho_dt-2}), and then we obtain,

\begin{align}
&\delta \Bigl( \frac{S^{xx}}{\rho} \Bigr)_{i}=-i\frac{(4/3)\mu+\overline{S^{xx}}}{\overline{\rho}}a\delta x_{i},\nonumber \\ &a\equiv \sum_{j}\sin [k(\overline{x}_{i}-\overline{x}_{j})]\frac{\partial}{\partial \overline{x}_{i}}W(\overline{\bm{r}}_{i}-\overline{\bm{r}}_{j},\sqrt{2}h)\frac{m}{\overline{\rho}}.\tag{A8} \label{eq-A8}
\end{align}

Using Eqs.\,(\ref{eq-A8}) and (\ref{way-to-express-Sab}), the deviatoric stress tensor of the $i$-th particle can be written as,

\begin{equation}
S_{i}^{xx}=\overline{S^{xx}}+\delta \Bigl( \frac{S^{xx}}{\rho} \Bigr)_{i} \overline{\rho}+\frac{\overline{S^{xx}}}{\overline{\rho}}\delta \rho_{i} = \overline{S^{xx}} -i\Bigl(\frac{4}{3}\mu a + \overline{S^{xx}}(a+D)\Bigr)\delta x_{i}. \tag{A9}
\label{eq-A9}
\end{equation}

Finally, substituting linearized density, pressure and $xx$ component of deviatoric stress tensor into the equation of motion of the Godunov SPH method for elastic dynamics (\ref{Godunov-EoM-2}), we can derive the dispersion relation because the left hand side of the equation of motion becomes $-\omega^{2}\delta x_{i}$. For example, the dispersion relation in the case of linear interpolation for $V_{ij}^{2}$ becomes,

\begin{align}
&\omega^{2}_{{\rm linear}}=-\Bigl[ \overline{C_{s}}^{2}D-\frac{4\mu a}{3\overline{\rho}} -\frac{\overline{S^{xx}}(a+D)}{\overline{\rho}} \Bigr] a + \frac{2(\overline{P}-\overline{S^{xx}})}{\overline{\rho}}a + \frac{2(\overline{P}-\overline{S^{xx}})}{\overline{\rho}}b,\nonumber \\ &b\equiv \sum_{j}(1-\cos [k(\overline{x}_{i}-\overline{x}_{j})])\frac{\partial^{2}}{\partial \overline{x}_{i}^{2}}W(\overline{\bm{r}}_{i}-\overline{\bm{r}}_{j},\sqrt{2}h)\frac{m}{\overline{\rho}}.\tag{A10} \label{eq-A10}
\end{align}

\noindent Here, for the perturbations of long wavelength $D\sim k$ and $a\sim -k$. Thus the third term in square brackets of Eq.\,(\ref{eq-A10}) is almost $0$. If we compare Eq.\,(\ref{eq-A10}) with the dispersion relation of the Godunov SPH method for hydrodynamics, we notice that the dispersion relations for hydrodynamics become that for elastic dynamics if $\overline{C_{s}}^{2}Da \rightarrow [\overline{C_{s}}^{2}D-(4\mu a/3\overline{\rho})]a$ and $\overline{P} \rightarrow \overline{P}-\overline{S^{xx}}$. This is the same for all interpolation methods. Therefore, the stability depends on the sign of $\overline{P}-\overline{S^{xx}}$. In usual simulation, $\overline{P}>0$,  $\overline{S^{xx}}<0$ for compressed region and $\overline{P}<0$, $\overline{S^{xx}}>0$ for tensile region. Therefore, it is sufficient that we select appropriate interpolation method depending only on the sign of pressure.

We may expect, in principle, even if pressure is positive, a region becomes effectively tensile dominant due to strong side slip force, and criterion of the sign of pressure may not be sufficient. In that case, using $\bm{r}_{i}-\bm{r}_{j}$ direction component of the deviatoric stress tensor $S_{i}^{ss}$ and $S_{j}^{ss}$, criterion of the sign of $P_{i}-S_{i}^{ss}+P_{j}-S_{j}^{ss}$ may be effective. According to our experience on test calculations, however, this criterion does not seem to be required.

\section*{Appendix B}
In Appendix B, we conduct the linear stability analysis of equations for variable smoothing length. For simplicity, we use the equations for hydrodynamics of the Godunov SPH method, and we use $\eta =1$. We treat smoothing length as constant when we linearize density, because density distribution in the case of variable smoothing length is almost the same as that in the case of constant smoothing length. The positions of particles are the same as those of Appendix A, and we also neglect the second or higher order of infinitesimal values. 

We write the smoothing length of the $i$-th particle as,

\begin{align}
& h_{i}=\overline{h}+\delta h_{i}, \nonumber \\ &\delta h_{i}=\epsilon_{h}\exp [i(k\overline{x}_{i}-\omega t)]. \tag{B1} \label{eq-B1}
\end{align}

\noindent From Eq.\,(\ref{variable-h}), we can express $\rho_{i}^{\ast}$ as,

\begin{align}
\rho_{i}^{\ast} &= \sum_{j} m \Bigl[ W(\overline{\bm{r}}_{i}-\overline{\bm{r}}_{j},C_{{\rm smooth}}\overline{h}) + (\delta x_{i} - \delta x_{j})\frac{\partial}{\partial \overline{x}_{i}}W(\overline{\bm{r}}_{i}-\overline{\bm{r}}_{j},C_{{\rm smooth}}\overline{h}) \nonumber \\ &  \ \ \ \ \ + \delta h_{i}\frac{\partial}{\partial \overline{h}}W(\overline{\bm{r}}_{i}-\overline{\bm{r}}_{j},C_{{\rm smooth}}\overline{h}) \Bigl] \nonumber \\ & = \overline{\rho}^{\ast} ( 1 + ia_{s}\delta x_{i} + b_{s} \delta h_{i}), \tag{B2} \label{eq-B2}
\end{align}

\noindent where,

\begin{align}
& \overline{\rho}^{\ast} \equiv \sum_{j} m W(\overline{\bm{r}}_{i}-\overline{\bm{r}}_{j},C_{{\rm smooth}}\overline{h}), \nonumber \\ & a_{s} \equiv \sum_{j} \sin [k(\overline{x}_{i}-\overline{x}_{j})] \frac{\partial}{\partial \overline{x}_{i}}W(\overline{\bm{r}}_{i}-\overline{\bm{r}}_{j},C_{{\rm smooth}}\overline{h}) \frac{m}{\overline{\rho}^{\ast}}, \nonumber \\ & b_{s} \equiv \sum_{j}\frac{\partial}{\partial \overline{h}}W(\overline{\bm{r}}_{i}-\overline{\bm{r}}_{j},C_{{\rm smooth}}\overline{h})\frac{m}{\overline{\rho}^{\ast}}. \tag{B3} \label{eq-B3}
\end{align}

\noindent Then we can express $h_{i}$ using Eq.\,(\ref{variable-h}) as,

\begin{align}
& h_{i}=\Bigl[ \frac{m}{\overline{\rho}^{\ast}(1+ia_{s}\delta x_{i}+b_{s}\delta h_{i})} \Bigr]^{1/d} \nonumber \\ & \approx \Bigl[ \frac{m}{\overline{\rho}^{\ast}} \Bigr] \Bigl( 1 - i\frac{a_{s}}{d}\delta x_{i} - \frac{b_{s}}{d}\delta h_{i} \Bigr). \tag{B4} \label{eq-B4}
\end{align} 

\noindent $m/\overline{\rho}^{\ast}$ means the smoothing length in unperturbed state. Thus $\overline{h}=m/\overline{\rho}^{\ast}$. From Eq.\,(\ref{eq-B4}), $\delta h_{i}$ can be expressed using $\delta x_{i}$ as,

\begin{equation}
\delta h_{i} = -i \Bigl( \frac{\overline{h}a_{s}}{d+\overline{h}b_{s}} \Bigr) \delta x_{i}. \tag{B5}
\label{eq-B5}
\end{equation}

The equation of motion of the Godunov SPH method for hydrodynamics in the case of variable smoothing length is,

\begin{equation}
\dot{v}^{\alpha}_{i}=-\sum_{j}m_{j}P_{ij}^{\ast}\Bigl[ V_{ij}^{2}(h_{i})\frac{\partial}{\partial x^{\alpha}_{i}}W(\bm{r}_{i}-\bm{r}_{j},\sqrt{2}h_{i})+V_{ij}^{2}(h_{j})\frac{\partial}{\partial x^{\alpha}_{i}}W(\bm{r}_{i}-\bm{r}_{j},\sqrt{2}h_{j}) \Bigr]. \tag{B6}
\label{eq-B6}
\end{equation}

\noindent Substituting linearized density, pressure and smoothing length into Eq.\,(\ref{eq-B6}), we obtain,

\begin{align}
&\omega^{2}_{{\rm variable \ h}}=\omega^{2}_{{\rm constant \ h}} - \Bigl( \frac{\overline{h}a_{s}}{d+\overline{h}b_{s}} \Bigr) \frac{\overline{P}}{\overline{\rho}}c_{s}, \nonumber \\ &c_{s} \equiv \sum_{j}\sin [k(\overline{x}_{i}-\overline{x}_{j})]\frac{\partial}{\partial \overline{h}}\frac{\partial}{\partial \overline{x}_{i}}W(\overline{\bm{r}}_{i}-\overline{\bm{r}}_{j},\sqrt{2}\overline{h})\frac{m}{\overline{\rho}}, \tag{B7} \label{eq-B7}
\end{align}

\noindent where $\omega^{2}_{{\rm constant \ h}}$ is $\omega^{2}$ in the case of constant smoothing length, which is written in \cite{Sugiura-and-Inutsuka2016}. The formula of $\omega^{2}_{{\rm constant \ h}}$ is different for linear interpolation, cubic spline interpolation and quintic spline interpolation.

For perturbations with any frequency lower than Nyquist frequency, $a_{s}<0$, $c_{s}>0$ and $b_{s}$ is positive constant that does not depend on wave number. Thus, in the case of negative pressure, the term of variable smoothing length makes $\omega^{2}$ negative and the method in unstable. At Nyquist frequency $a_{s}=0$. For perturbations with smaller wavelength than $C_{{\rm smooth}}\overline{h}$, $a_{s}$ becomes almost 0. In consequence, extension to variable smoothing length can make perturbations of longer wavelength than Nyquist frequency unstable even if this perturbation is stable in the case of constant smoothing length. However, if we make the value of $C_{{\rm smooth}}$ larger, $a_{s}$ becomes smaller and we can make this perturbation stable again. 

According to \cite{Sugiura-and-Inutsuka2016}, $\omega^{2}_{{\rm constant \ h}}$ can be decomposed into the term that becomes $\overline{C_{s}}^{2}k^{2}$ at long wavelength and the other error terms,

\begin{equation}
\omega^{2}_{{\rm constant \ h}} = -\overline{C_{s}}^{2}D a + \frac{\overline{P}}{\overline{\rho}} \times {\rm other \ terms}. \tag{B8}
\label{eq-B8}
\end{equation}

\noindent As we can notice from Eqs.\,(\ref{eq-B7}) and (\ref{eq-B8}), only the first term of $\omega^{2}_{{\rm variable \ h}}$ is proportional to $\overline{C_{s}}^{2}$, and all the other terms are proportional to $\overline{P}/\overline{\rho}$. Thus we can evaluate whether arbitrary state (including spatial dimension, interpolation method and $C_{{\rm smooth}}$) is stable or not only by $(\overline{P}/\overline{\rho})/\overline{C_{s}}^{2}$. Conversely, for arbitrary spatial dimension, interpolation method and value of $(\overline{P}/\overline{\rho})/\overline{C_{s}}^{2}$, we can evaluate the minimum $C_{{\rm smooth}}$ to achieve stable simulation.

\begin{figure}[!htb]
 \begin{center}
 \includegraphics[width=7cm,height=5cm]{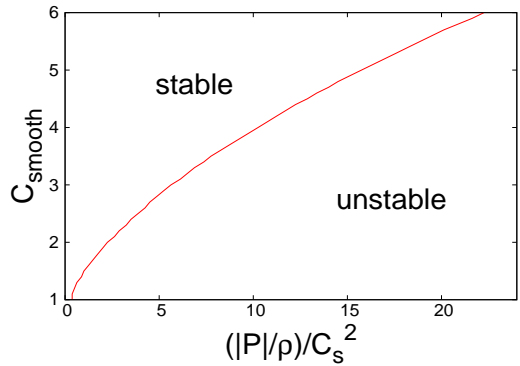}
 \caption{This figure shows which pair of $(\overline{P}/\overline{\rho})/\overline{C_{s}}^{2}$ and $C_{{\rm smooth}}$ provides stable calculation in the case of negative pressure, one dimension and quintic spline interpolation. The horizontal axis shows $(\overline{|P|}/\overline{\rho})/\overline{C_{s}}^{2}$, the vertical axis shows $C_{{\rm smooth}}$. In this parameter space, left hand side of curve provides stable simulation.}
 \label{fig-B1}
 \end{center}
\end{figure}

\begin{figure}[!htb]
 \begin{center}
 \includegraphics[width=7cm,height=5cm]{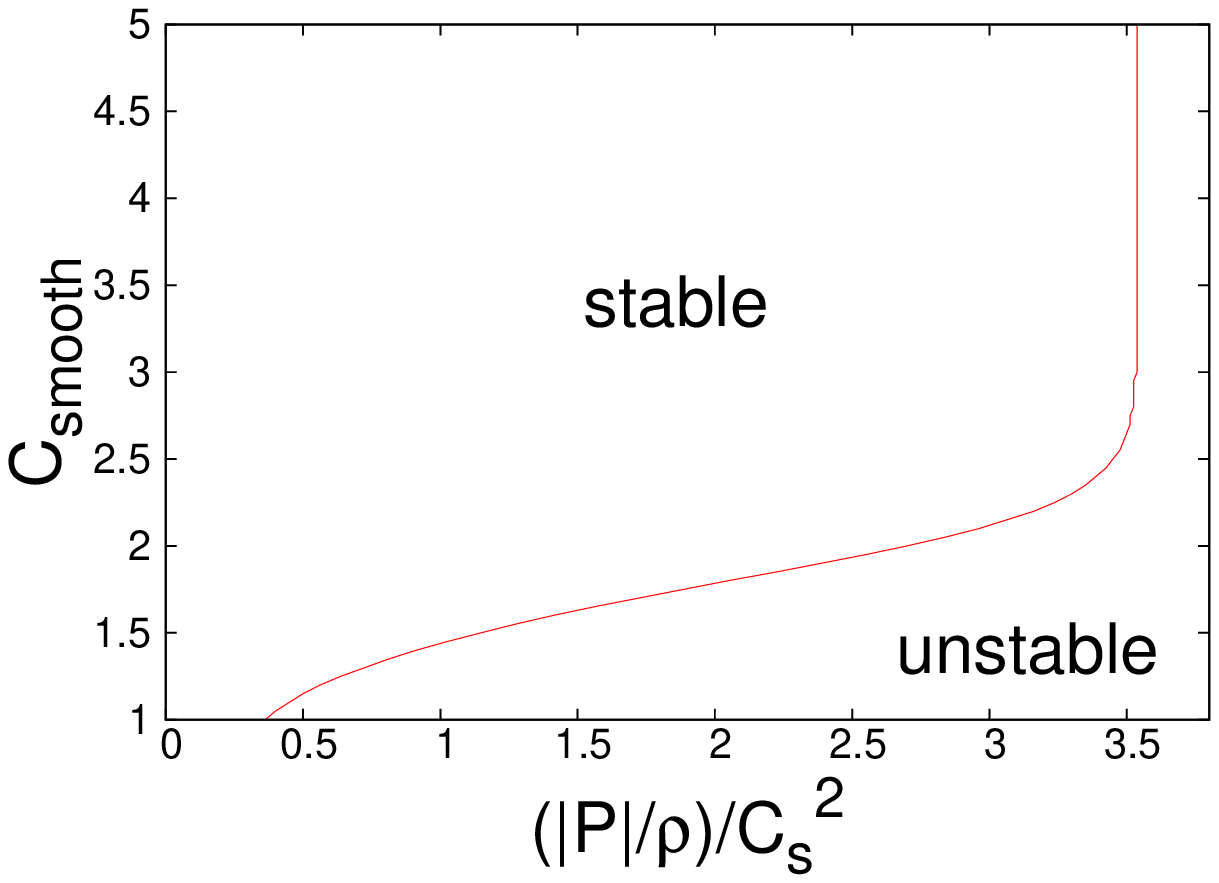}
 \caption{Same as Fig.\,\ref{fig-B1}, but for two dimensions and cubic spline interpolation.}
 \label{fig-B2}
 \end{center}
\end{figure}

\begin{figure}[!htb]
 \begin{center}
 \includegraphics[width=7cm,height=5cm]{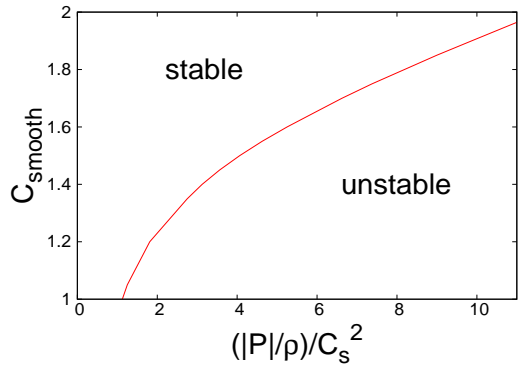}
 \caption{Same as Fig.\,\ref{fig-B1}, but for three dimensions and cubic spline interpolation.}
 \label{fig-B3}
 \end{center}
\end{figure}

In \cite{Sugiura-and-Inutsuka2016}, in the negative pressure region, Sugiura and Inutsuka (2016) use quintic spline interpolation for one dimension, cubic spline interpolation for two dimensions, and cubic spline interpolation for three dimensions. Thus, we investigate which pair of $(\overline{P}/\overline{\rho})/\overline{C_{s}}^{2}$ and $C_{{\rm smooth}}$ provides stable calculation for these three cases. Figure \ref{fig-B1}, \ref{fig-B2} and \ref{fig-B3} show the results of this investigation for quintic spline interpolation in one dimension, cubic spline interpolation in two dimensions, and cubic spline interpolation in three dimensions respectively.

In the Fig.\,\ref{fig-B2}, curve extends vertically around $(\overline{|P|}/\overline{\rho})/\overline{C_{s}}^{2}\sim 3.5$. This is owing to constant smoothing length term, and if $(\overline{P}/\overline{\rho})/\overline{C_{s}}^{2}$ is smaller than -3.5, calculation becomes unstable even with constant smoothing length. However, $(\overline{P}/\overline{\rho})/\overline{C_{s}}^{2}\sim -3.5$ can not be realized in usual calculation. If we assume the equation of state of $P=C_{s}^{2}(\rho -\rho_{0})$, the density of $\rho \sim 0.22\rho_{0}$ is required to achieve $(\overline{P}/\overline{\rho})/\overline{C_{s}}^{2}\sim -3.5$. In other words, material should be stretched until the density becomes five times smaller than the average density. In that case ordinary material should break up. 

We express $C_{{\rm smooth}}$ on the curve of figures as $C_{{\rm smooth,crit}}$. In the region of negative pressure, the calculation is stable if $C_{{\rm smooth}}$ is larger than $C_{{\rm smooth,crit}}$. For convenience, we made fitting formula for this $C_{{\rm smooth,crit}}$. Fitting formula is expressed as,

\begin{align}
&C_{{\rm smooth,crit}}=A \ln [ B(X-C)],\nonumber \\ & X\equiv (\overline{|P|}/\overline{\rho} )/\overline{C_{s}}^{2}. \tag{B9} \label{eq-B9}
\end{align}

\noindent In the case of one dimension and quintic spline interpolation,

\begin{equation}
A=3.96448, \ \ B=0.143576, \ \ C=-9.07397. \tag{B10}
\label{eq-B10}
\end{equation}

\noindent In the case of two dimensions and cubic spline interpolation,

\begin{equation}
A=0.926887, \ \ B=2.37512, \ \ C=-0.89341. \tag{B11}
\label{eq-B11}
\end{equation}

\noindent In the case of three dimensions and cubic spline interpolation,

\begin{equation}
A=0.495342, \ \ B=4.37086, \ \ C=-0.673217. \tag{B12}
\label{eq-B12}
\end{equation}

\noindent Here, we use data point of $(\overline{|P|}/\overline{\rho})/\overline{C_{s}}^{2}< 3.5$ for two dimensions and cubic spline interpolation. Large computational cost is required if $C_{{\rm smooth}}$ is large. Thus, in practical calculation, we just make $C_{{\rm smooth}}$ larger in negative pressure region locally, and for positive pressure region $C_{{\rm smooth}}=1.0$ is sufficient. We can calculate $C_{{\rm smooth,crit}}$ of the $i$-th particle using physical quantities of this particle as,

\begin{align}
&C_{{\rm smooth,crit},i}=A \ln [ B(X_{i}-C)],\nonumber \\ & X_{i}\equiv (|P_{i}|/\rho_{i} )/C_{s,i}^{2}, \tag{B13} \label{eq-B13}
\end{align}

\noindent and $C_{{\rm smooth}}$ of the $i$-th particle can be calculated as,

\begin{align}
&C_{{\rm smooth},i}=\left\{ \begin{array}{ll} \max [C_{{\rm smooth,crit},i}+\epsilon_{{\rm margin}},1.0] & {\rm if} \ P<0, \\ 1.0 & {\rm if} \ P>0, \\ \end{array} \right. \tag{B14} \label{eq-B14}
\end{align}

\noindent where $\epsilon_{{\rm margin}}$ is small value for safety. $\epsilon_{{\rm margin}}=0.1$ is sufficient. In this case, we can obtain smoothing length of the $i$-th particle by substituting $C_{{\rm smooth},i}$ for $C_{{\rm smooth}}$ in Eq.\,(\ref{variable-h}).

\end{document}